\documentclass[lettersize,journal]{IEEEtran}
\pdfoutput=1
\usepackage{amsmath,amsfonts}
\usepackage{array}
\usepackage[caption=false,font=normalsize,labelfont=sf,textfont=sf]{subfig}
\usepackage{textcomp}
\usepackage{stfloats}
\usepackage{url}
\usepackage{verbatim}
\usepackage{graphicx}
\usepackage{cite}
\usepackage{lscape}
\usepackage[utf8]{inputenc} 
\usepackage[T1]{fontenc}    
\usepackage{hyperref}       
\usepackage{url}            
\usepackage{booktabs}       
\usepackage{amsfonts}       
\usepackage{nicefrac}       
\usepackage{microtype}      
\usepackage{xcolor}         
\usepackage{multicol}
\usepackage{multirow}
\usepackage{tikz}
\usetikzlibrary{snakes}
\usetikzlibrary{trees}
\usepackage{wrapfig}
\usepackage{makecell}
\usepackage{algorithmic}
\usepackage[boxed,ruled,lined]{algorithm2e}
\usepackage{bigstrut}
\usepackage{wasysym}
\usepackage{enumitem}
\usepackage{enumitem}
\usepackage{threeparttable}
\usepackage{rotating}
\usepackage{color, colortbl}
\usepackage{xr-hyper}
\usepackage{float}
\usepackage{mathtools}

\usepackage{titlesec}

\titlespacing\subsection{0pt}{5pt plus 2pt minus 2pt}{3pt plus 2pt minus 2pt}
\definecolor{mygray}{gray}{0.9}

\long\def\comment#1{}
\def\ie{$i.e.$}
\def\eg{$e.g.$}
\def\etc{$etc. $}
\def\wrt{$w.r.t.$}

\def\blue#1{\textcolor{black}{#1}}

\def\bluetwo#1{\textcolor{black}{#1}}

\begin{document}

\title{BlackboxBench: A Comprehensive Benchmark of Black-box Adversarial Attacks}

\author{
Meixi Zheng\textsuperscript{*}, Xuanchen Yan\textsuperscript{*}, Zihao Zhu, Hongrui Chen, Baoyuan Wu\textsuperscript{\dag}, \IEEEmembership{Senior Member,~IEEE}
\\
\IEEEcompsocitemizethanks{
\IEEEcompsocthanksitem 
All authors are with the School of Data Science, The Chinese University of Hong Kong, Shenzhen, Guangdong, 518172, P.R. China, email: meixizheng1@link.cuhk.edu.cn, xuanchenyan@link.cuhk.edu.cn, zihaozhu@link.cuhk.edu.cn, 
hongruichen@\\link.cuhk.edu.cn, wubaoyuan@cuhk.edu.cn.
\IEEEcompsocthanksitem
\textsuperscript{*}These two authors contribute equally.
\IEEEcompsocthanksitem
\textsuperscript{\dag} Corresponding author: Baoyuan Wu (wubaoyuan@cuhk.edu.cn).
}
}



\maketitle

\begin{abstract}
Adversarial examples are well-known tools to evaluate the vulnerability of deep neural networks (DNNs). Although lots of adversarial attack algorithms have been developed, it's still challenging in the practical scenario that the model's parameters and architectures are inaccessible to the attacker/evaluator, \ie, black-box adversarial attacks. 
Due to the practical importance, there has been rapid progress from recent algorithms, reflected by the quick increase in attack success rate and quick decrease in query numbers to the target model. However, there lacks thorough evaluations and comparisons among these algorithms, causing difficulties in tracking the real progress, analyzing advantages and disadvantages of different technical routes, as well as designing future development roadmap of this field. 
Thus, we aim at building a comprehensive benchmark of black-box adversarial attacks, called \textit{BlackboxBench}. It mainly provides: 1) a unified, extensible and modular-based codebase, implementing \blue{29} query-based attack algorithms and \blue{30} transfer-based attack algorithms; 2) comprehensive evaluations: we evaluate the implemented algorithms against several mainstreaming model architectures on 2 widely used datasets (CIFAR-10 and a subset of ImageNet), leading to \blue{14,950} evaluations\footref{fn: evaluation} in total; 3) thorough analysis and new insights, as well analytical tools. 
The website and source codes of BlackboxBench are available at \url{https://blackboxbenchmark.github.io/} and \url{https://github.com/SCLBD/BlackboxBench/}, respectively. 

\end{abstract}

\begin{IEEEkeywords}
Black-box adversarial attacks, benchmark, query-based adversarial attacks, transfer-based adversarial attacks.
\end{IEEEkeywords}


\begin{figure*}[t]
 \centering
\includegraphics[width=0.87\linewidth]{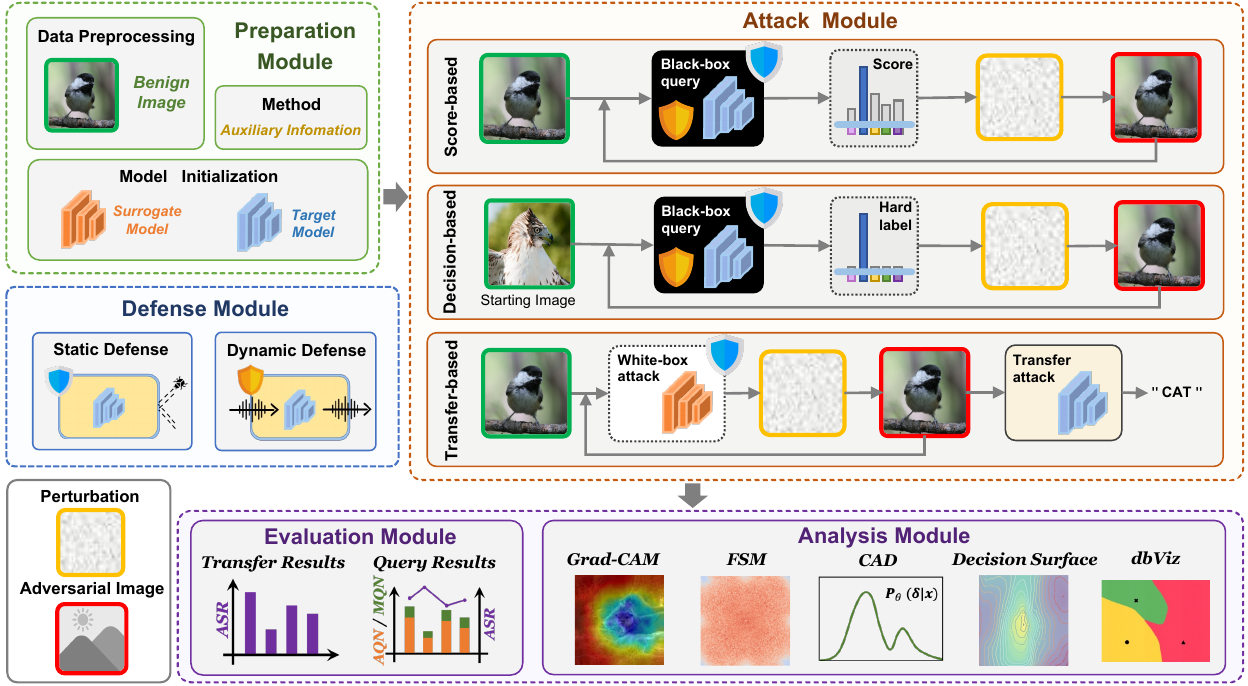}
 \caption{The general structure of the modular-based codebase of BlackboxBench.}
 \label{modules}
 \vspace{-15pt}
\end{figure*}

\section{Introduction}
\IEEEPARstart{D}{eep} neural networks (DNNs) have been densely applied to safety-critical applications such as autonomous driving and face recognition due to their remarkable success. Worryingly, adversarial examples, which add human-imperceptible noises to benign inputs, capable of deceiving well-trained models, pose a significant threat to DNNs. In real-world situations, attackers typically have limited access to information about the target model. \blue{This makes the study of adversarial attacks with in a black-box scenario crucial for understanding the robustness of models against more realistic attack conditions.}

The field of machine learning security has witnessed a rapid evolution of black-box adversarial attacks. \blue{Due to the lack of access to target model's complete information, attackers must rely on either external queries or surrogate models, corresponding to query-based and transfer-based attacks, respectively. As a result, significant efforts have been made in these two areas to enhance the state-of-the-art (SOTA) attack performance \cite{prgf, cisa, sia, bayesian}.} These endeavors are also inherently double as robustness assessments to develop stronger defenses and more resilient models \cite{pamidefense1, pamidefense2, pamidefense3, pamidefense4}. 
\blue{
However, the proliferation of diverse attack methods, evaluated in different model architectures and datasets, compared to limited previous methods, perplexes researchers regarding the actual threat posed by them, as well as the actual progress in this field. Meanwhile, variations in code styles and lack of a unified attack pipeline make it hard to use them to examine the actual protection provided by robust models or defense mechanisms.
}As a result, there is a growing demand for building a comprehensive platform for black-box adversarial attacks to facilitate fair comparisons within the community.

To this end, we build a comprehensive benchmark of black-box adversarial attacks, called \textbf{BlackboxBench}. 
To date, we have included \blue{29} query-based attacks and \blue{30} transfer-based attacks, the largest collection of black-box adversarial attacks to our best knowledge. 
\blue{
BlackboxBench adopts a modular and unified implementation. As the core module of BlackboxBench, \ie, the attack module, is implemented based on a unified attack pipeline composed of several functional blocks (see Fig. \ref{pipeline}), BlackboxBench allows fair and standard evaluation, illuminates promising technical routes, facilitates the involvement of new contributors, and is user-friendly.
}Using BlackboxBench, we conduct evaluations of all these query-based attacks and transfer-based attacks on comprehensive attack scenarios\bluetwo{—varying attack settings, target models, and datasets—}resulting in \blue{14,950} evaluations\footnote{\blue{
One evaluation is conducted by generating adversarial examples on a dataset using a specific attack method against a particular target model under a defined attack setting.
}\label{fn: evaluation}} in total. These standardized evaluations establish a leaderboard to identify the most effective methods and track actual progress in this field. 
\bluetwo{
Our results indicate that black-box adversarial attacks have evolved significantly over the years, with CISA emerging as the SOTA method for decision-based attacks, MCG for score-based attacks, and SIA and Bayesian Attack for untargeted and targeted transfer-based attacks, respectively. Additional observations addressing benchmark-level questions (see the first four entries in Tab. \ref{tab:questions}) are briefly recapped at the end of each subsection in Sec. \ref{sec: result overview}.
}

\bluetwo{
Beyond benchmarking, we further leverage BlackboxBench to foster a deeper understanding of black-box adversarial vulnerability. In addition to the evaluations above, we extend our experimental scope to systematically assess the influence of critical factors-including the data used for attacking, the architecture of surrogate and target models, constraints on attack budget, deployed defenses, and attack strategies—on the black-box adversarial vulnerability of models. Unlike the benchmark evaluations, which focus on tracking progress, these additional analyses aim to uncover underlying factors behind black-box adversarial vulnerability.
These analyses are supported by 10 analytical tools (see Tab. \ref{tab:tools} in the Supplemental Material for details on tools and their functionalities), and are organized around 11 key analytical questions (see the latter entries in Tab. \ref{tab:questions}) from the perspectives of data, model architecture, budget, defense, and attack procedure, 
providing several insightful takeaways, including (but not limited to) the following:
\begin{itemize}[leftmargin=15pt,itemsep=1pt,topsep=0pt]
\item \textbf{Data} We find that samples far from the model’s decision boundary would take higher query costs for successful attacks across various query-based methods, as they make it harder for attackers to efficiently find the attack direction.
\item \textbf{Model architecture} We show that complex models like ConvNeXt-T and ViT-B/16, with smoother decision boundaries, exhibit stronger robustness to adversarial queries, while simpler models like ResNet-50 and VGG-19$^{\dagger}$, with more fragmented decision boundaries, are more vulnerable. We also propose a quantitative metric to measure the transferability of adversarial examples between models.
\item \textbf{Attack budget} We find that surrogate-assisted score-based methods (\eg, MCG, PRGF, BASES, $\mathcal{CG}$-attack) are more efficient in query and norm budgets. Transfer-based methods aiming for semantically meaningful perturbations (\eg, FIA, NAA, DRA, IAA) generally require higher norm budgets to maintain their strong performance.
\item \textbf{Defense} We observe that adversarial examples transfer more effectively between models with the same training strategy, due to their shared attention in the frequency domain.
\item \textbf{Attack procedure} We visualize successful adversarial attacks would manage to shift the model’s attention from discriminative areas to trivial areas, thus resulting in misclassification.
\end{itemize}
Further insights into other key questions are presented at the beginning of each analytical section (\ie, Sec. \ref{sec: effect of data}, \ref{sec: effect of model architecture} and \ref{sec: effect of methods}, Supplementary \ref{sec: effect of defense} and \ref{sec: visualization analysis}).
}

This paper is organized as follows.  Sec. \ref{sec: related work} introduces the related benchmarks and highlights the uniqueness of BlackboxBench. Sec. \ref{our benchmark} provides the preliminary and an overview of BlackboxBench architecture and library modules. Sec. \ref{sec:evaluation and analysis} presents the evaluation results, systematic analyses and our new insights, followed by conclusions in Sec. \ref{sec: conclusion}.

\section{Related Work}
\label{sec: related work}

ML models have demonstrated vulnerability to imperceptive deliberate perturbations. Some researchers demonstrate these vulnerabilities by designing new attacks, while others pay ongoing efforts toward new defenses to enhance robustness \cite{wei2022physically}. 
Within this research community, many benchmarks have emerged for standardized evaluations. The evaluated perturbations are of various types (\textit{e.g.}, patch-based\cite{reap}, norm-bounded\cite{carben, multirobustbench}) to fool models in different ML tasks (\textit{e.g.}, computer vision\cite{smartbox}, natural language understanding\cite{advglue}, \cite{graph}).


BlackboxBench is designed for norm-bounded noise perturbed vision tasks, currently the most extensively studied ML setting. Moreover, unlike some excellent benchmarks specialized in defenses (\textit{e.g.}, RobustBench \cite{robustbench}, \href{https://www.robust-ml.org/}{RobustML}), our aim is to track the real progress in black-box adversarial attacks. Various libraries including CleverHans\cite{papernot2016technical}, FoolBox\cite{rauber2017foolbox}, ART\cite{nicolae2018adversarial}, DEEPSEC \cite{deepsec} AdverTorch\cite{ding2019advertorch}, SecML\cite{pintor2019secml}, RealSafe \cite{realsafe}, AdvBox\cite{goodman2020advbox}, DeepRobust\cite{li2020deeprobust}, TransferAttackEval \cite{eval}, \blue{Torchattacks\cite{torchattacks} and TA-Bench\cite{tabench}} have all provided evaluation for popular adversarial attacks. \textbf{Nevertheless, BlackboxBench sets itself apart from existing benchmarks from four key perspectives, as detailed in Tab. \ref{tab:difference} of the Supplemental Material.}
\begin{itemize}[leftmargin=15pt,itemsep=1pt,topsep=0pt]
    \item 
    \bluetwo{
    \textbf{Programming characteristic} Most libraries employ object-oriented programming (OOP) to define attacks as separate classes. In contrast, BlackboxBench features a modular codebase with an attack module that integrates various attacks into a unified attack pipeline composed of several functional blocks. This procedural structure allows easy switching between attacks or integrating multiple attacks into a more powerful one by modifying or combining functional blocks within the unified attack pipeline, without the need to redefine the entire optimization process as required in OOP-style benchmarks. Note that while DEEPSEC\cite{deepsec} and SecML\cite{pintor2019secml} also feature modular architectures, they lack a unified attack module compared to BlackboxBench.
    }
    \item \textbf{Algorithm coverage} Taking a look at the number and published years of white-box (WA) and black-box (BA) adversarial attacks\footnotemark{} that each library covers, BlackboxBench stands out by supporting \blue{59} black-box attack methods from 2017 to 2023, diverging from other benchmarks that mainly encompass white-box or early black-box techniques.
    \footnotetext{White-box attacks can tackle black-box tasks owing to the transferability of adversarial examples. Without specifications in original papers, we categorize white-box attacks evaluated within a black-box setting as black-box attacks.}
    \item \textbf{Leaderboard} BlackboxBench is the first platform that furnishes a leaderboard for monitoring advancements in black-box adversarial attacks.
    \item \textbf{Tools \& Analyses} Besides only analyzing quantitative results like DEEPSEC, RealSafe, TransferAttackEval and \blue{TA-Bench}, \bluetwo{BlackboxBench also offers analytical tools to facilitate in-depth analyses of underlying factors behind black-box adversarial vulnerability}, aiming at provide fresh perspectives for future research. Note that although SecML provides tools, only one analysis has been conducted.
    
\end{itemize}

\section{Our Benchmark}
\label{our benchmark}
\subsection{Preliminary}
\label{Preliminary}

\begin{figure*}[t]
 \centering
\includegraphics[width=0.85\linewidth]{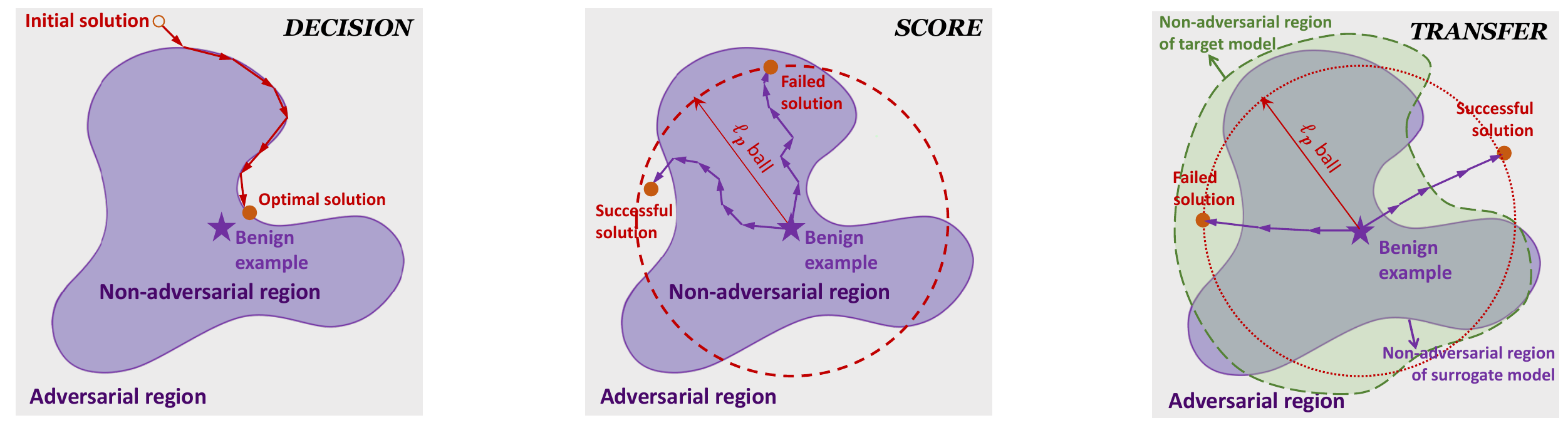}
 \caption{Graphical illustrations of score-base attack (\textbf{left}), decision-based attack (\textbf{middle}) and transfer-based attack (\textbf{right}), under the untargeted attack setting.}
 \label{illustrate_of_attack}
 \vspace{-13pt}
\end{figure*}

BlackboxBench focuses on black-box adversarial attacks against deep learning models. This section seeks to provide novice readers with a complete grasp of this domain.

Given target neural network $f: \mathcal{X} \rightarrow \mathcal{P}(\mathcal{Y})$, parameterized by $\theta$, where $\mathcal{X} \subset \mathbb{R}^d$, and $\mathcal{P}(\mathcal{Y})$ is the $K$-dimensional probability simplex over $\mathcal{Y}$. For benign example $\boldsymbol{x} \in\mathcal{X}$ whose true label is $y \in\mathcal{Y}$, adversarial attacks aim at misleading, with imperceptible perturbation $\boldsymbol{\delta}_{\boldsymbol{x}}$, the prediction of $\boldsymbol{x}$ from $f$, which lead to the adversarial example $\boldsymbol{x}^*=\boldsymbol{x}+\boldsymbol{\delta}_{\boldsymbol{x}}$. Formally, $\boldsymbol{x}^*$ satisfies two properties:
\begin{itemize}[leftmargin=15pt,itemsep=1pt,topsep=0pt]
    \item To make sure $\boldsymbol{x^*}$ is perceptually similar to $\boldsymbol{x}$, $\boldsymbol{x^*}$ is generated within a $\ell_p$ norm ball centered at $\boldsymbol{x}$ with radius $\epsilon$, \ie, $\left\|\boldsymbol{x^*}-\boldsymbol{x}\right\|_{p} \leq \epsilon$. $\ell_\infty$ and $\ell_2$ are the most commonly used norms. The perturbation norm budget $\epsilon$ is small enough so as to remain undetectability.
    \item $\boldsymbol{x^*}$ can be untargeted, in which case the goal of the adversary is to classify $\boldsymbol{x}^*$ as an arbitrary class than the correct one, \ie, $f(\boldsymbol{x^*}) \neq y$. $\boldsymbol{x^*}$ also can be targeted, the targeted $\boldsymbol{x^*}$ is similar but is required to induce a prediction for the target class $t$, \ie, $f(\boldsymbol{x^*}) = t$.
\end{itemize}

Depending on the full access to the $f$ or not, adversarial attacks can be divided into \textit{white-box} and \textit{black-box}. The former is deemed easier but less realistic because attackers could access parameters of the target models. The latter is more practical and challenging because no information is known about the target model, and is the focus of BlackboxBench. As background, we introduce different types of black-box attacks: \textit{query-based attacks}, which query the target model repeatedly and can be categorized into score-based and decision-based according to the model feedback, and \textit{transfer-based attacks}, which work offline without interacting with the target model\footnotemark{}. Their graphical illustrations are shown in Fig. \ref{illustrate_of_attack}.

\footnotetext{In the latest survey \cite{wusurvey}, adversarial attacks are categorized as white-box, black-box (specifically referring to query-based attacks) and transfer-based attacks. Here we incorporate transfer-based attacks into black-box category to emphasize the importance of adversarial transferability in enhancing black-box attack ability.}

\subsubsection{Decision-based query attacks} Decision-based query attacks primarily work with only the final decision of a model (e.g., the classification output) without access to the model's gradients, logits, or any other internal states. 
The general formulation is:
\begin{equation}\label{eq: decision}
    \min _{\boldsymbol{x^*}} \mathcal{D}\left(\boldsymbol{x}, \boldsymbol{x^*}\right) \text{ s.t. } \Omega\left(f\left(\boldsymbol{x^*}\right),y\right)=1.
\end{equation}
The perturbation aims to be as imperceptible as possible (corresponding to minimizing the distance, represented by $\mathcal{D}$, between the initial input $\boldsymbol{x}$ and its perturbed counterpart $\boldsymbol{x^*}$), while keeping misleading the model into incorrect classifications (corresponding to the hard constraint introduced by $\Omega$). 
$\Omega$ forces the perturbed input to meet the attack goal, which takes 1 if the constraint is satisfied, otherwise 0. According to the type of attack, the $\Omega$ is specified as follows:
\begin{equation}\label{eq: decision_constraint}
\begin{cases}\text { Untargeted: } & \Omega\left(f\left(\boldsymbol{x}^*\right), y\right) \coloneqq \mathbb{I}\left(f\left(\boldsymbol{x}^*\right) \neq y\right) ; \\ \text { Targeted: } & \Omega\left(f\left(\boldsymbol{x}^*\right), t\right) \coloneqq \mathbb{I}\left(f\left(\boldsymbol{x}^*\right)=t\right). \end{cases}
\end{equation}
$\mathbb{I}(\cdot)$ is the indicator function, and  $\mathbb{I}(c)=1$ if $c$ is true and $\mathbb{I}(c)=0$ otherwise.
In essence, attackers are navigating the adversarial space defined by the hard constraints. Their goal is to find a point that is both adversarial (misclassified) and as close as possible to the original input.

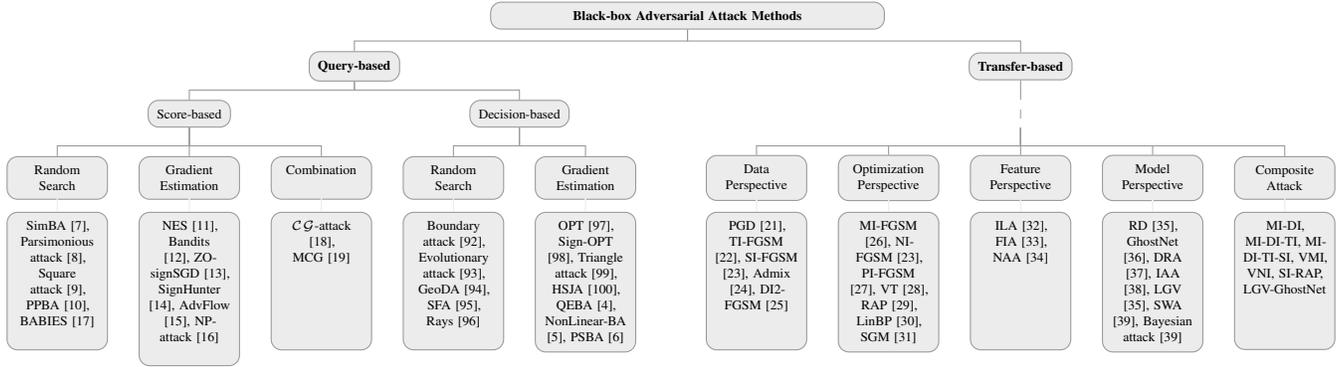
\begin{figure*}[t]
 \centering
\tikzset{global scale/.style={
    scale=#1,
    every node/.append style={scale=#1}
  }
}
\tikzstyle{every node}=[scale=1, font=\tiny]
\begin{tikzpicture}[scale=0.42]
\begin{scope}[
  gluon1/.style={decorate, draw=gray!80},
  gluon/.style={decorate, draw=gray!15},
  man/.style={rectangle, draw=gray!80, fill=gray!15,text width=1.1cm,
		text centered, rounded corners},
  par/.style={rectangle, draw=gray!80,fill=gray!15,text width=1.1cm,
		text centered,anchor=north, rounded corners},
  woman/.style={rectangle, draw=gray!80,fill=gray!15,text width=5cm,
		text centered, rounded corners},
  method/.style={rectangle, draw=gray!80,fill=gray!15, rounded corners},
  grandchild/.style={grow=down,xshift=1em,anchor=west,
    edge from parent path={(\tikzparentnode.south) |- (\tikzchildnode.west)}},
  grandchild_left/.style={grow=down,xshift=1em,anchor=east,
    edge from parent path={(\tikzparentnode.south) -| (\tikzchildnode.east)}},
  first/.style={level distance=8ex},
  second/.style={level distance=14ex},
  third/.style={level distance=20ex},
  fourth/.style={level distance=26ex},
  level 1/.style={sibling distance=56em, level distance=1.5cm},
  level 2/.style={sibling distance=1.85em, level distance=1.4cm},
  level 3/.style={sibling distance=1.85em, level distance=1.9cm},
  level 4/.style={sibling distance=5.5em, level distance=1.0cm},
  ]
    \node [woman]{\textbf{\tiny{Black-box Adversarial Attack Methods}}}
    [edge from parent fork down]
    child{node[method] {\textbf{\tiny{Query-based}}}
      edge from parent [gluon1]
      child{node[method] {Score-based}
        edge from parent [gluon1]
        child{node[man] {Random Search}
            edge from parent [gluon1]
            child{node[par]{
            SimBA\cite{guo2019simple}, Parsimonious attack\cite{moon2019parsimonious}, Square attack\cite{andriushchenko2020square}, PPBA\cite{li2020projection}, BABIES\cite{tran2022exploiting} \textcolor{gray!15}{111111111} \textcolor{gray!15}{111111111} }
            edge from parent [gluon]}
        }
        child [missing] {}
        child [missing] {}
        child [missing] {}
        child [missing] {}
        child [missing] {}
        child{node[man] {Gradient Estimation}
            edge from parent [gluon1]
            child{node[par]{NES\cite{ilyas2018black}, Bandits\cite{ilyas2018prior}, ZO-signSGD\cite{liu2019signsgd}, SignHunter\cite{al2020sign}, AdvFlow\cite{mohaghegh2020advflow}, NP-attack\cite{bai2020improving} \textcolor{gray!15}{111111111} }
            edge from parent [gluon]}
        }
        child [missing] {}
        child [missing] {}
        child [missing] {}
        child [missing] {}
        child [missing] {}
        child{node[man] {Combination \textcolor{gray!15}{111111111}}
            edge from parent [gluon1]
            child{node[par]{$\mathcal{CG}$-attack\cite{feng2022boosting}, MCG\cite{yin2023generalizable}, PRGF\cite{prgf}, Subspace attack\cite{subspaceattack}, BASES\cite{bases}  \textcolor{gray!15}{111111111} \textcolor{gray!15}{111111111}  \textcolor{gray!15}{111111111}}
            edge from parent [gluon]}
        }
      }
      child [missing] {}
      child [missing] {}
      child [missing] {}
      child [missing] {}
      child [missing] {}
      child [missing] {}
      child [missing] {}
      child [missing] {}
      child [missing] {}
      child [missing] {}
      child [missing] {}
      child [missing] {}
      child [missing] {}
      child [missing] {}
      child{node[method] {Decision-based}
        edge from parent [gluon1]
        child{node[man] {Random Search}
            edge from parent [gluon1]
            child{node[par]{Boundary attack\cite{brendel2017decision}, Evolutionary attack\cite{dong2019efficient}, GeoDA\cite{rahmati2020geoda}, SFA\cite{chen2020boosting}, Rays\cite{chen2020rays} \textcolor{gray!15}{111111111} \textcolor{gray!15}{111111111}}
            edge from parent [gluon]}
        }
        child [missing] {}
        child [missing] {}
        child [missing] {}
        child [missing] {}
        child [missing] {}
        child{node[man] {Gradient Estimation}
            edge from parent [gluon1]
            child{node[par]{OPT\cite{cheng2018query}, Sign-OPT\cite{cheng2019sign}, Triangle 
            attack\cite{wang2022triangle}, HSJA\cite{chen2020hopskipjumpattack}, QEBA\cite{li2020qeba}, NonLinear-BA\cite{li2021nonlinear}, PSBA\cite{zhang2021progressive}, CISA\cite{cisa}} 
            edge from parent [gluon]}
        }
      }
    }
    child{node[method] {\textbf{\tiny{Transfer-based}}}
    edge from parent [gluon1]
        child{ node {\textcolor{gray!85}{|}}
            child{node[man] {Data Perspective}
                edge from parent [gluon1]
                child{node[par]{PGD\cite{pgd}, TI-FGSM\cite{tifgsm},  SI-FGSM\cite{nisifgsm}, Admix\cite{admix}, DI2-FGSM\cite{difgsm}, SIA\cite{sia}
                \textcolor{gray!15}{111111111}
                \textcolor{gray!15}{111111111} \textcolor{gray!15}{111111111}}
                edge from parent [gluon]}
            }
            child [missing] {}
            child [missing] {}
            child [missing] {}
            child [missing] {}
            child [missing] {}
            child{node[man] {Optimization Perspective}
                edge from parent [gluon1]
                child{node[par]{MI-FGSM\cite{mifgsm}, NI-FGSM\cite{nisifgsm}, PI-FGSM\cite{pifgsm}, VT\cite{vt}, RAP\cite{rap}, LinBP\cite{linbp}, SGM\cite{sgm}, PGN\cite{pgn} \textcolor{gray!15}{111111111}}
                edge from parent [gluon]}
            }
            child [missing] {}
            child [missing] {}
            child [missing] {}
            child [missing] {}
            child [missing] {}
            child{node[man] {Feature Perspective}
                edge from parent [gluon1]
                child{node[par]{ILA\cite{ila}, FIA\cite{fia}, NAA\cite{naa} \textcolor{gray!15}{111111111} \textcolor{gray!15}{111111111} \textcolor{gray!15}{111111111} \textcolor{gray!15}{111111111} \textcolor{gray!15}{111111111} \textcolor{gray!15}{111111111}}
                edge from parent [gluon]}
            }
            child [missing] {}
            child [missing] {}
            child [missing] {}
            child [missing] {}
            child [missing] {}
            child [missing] {}
            child [missing] {}
            child [missing] {}
            child{node[man] {Model Perspective}
                edge from parent [gluon1]
                child{node[par]{\textcolor{gray!15}{11}\textit{(Tuning)}\textcolor{gray!15}{11} RD\cite{lgv}, GhostNet\cite{ghostnet}, DRA\cite{dra}, IAA\cite{iaa}, LGV\cite{lgv}, SWA\cite{bayesian}, Bayesian attack\cite{bayesian}}
                edge from parent [gluon]}
                child [missing] {}
                child{node[par]{\textcolor{gray!15}{111}\textit{(Fusion)}\textcolor{gray!15}{111} Logit ensemble\cite{mifgsm}, Loss ensemble\cite{mifgsm}, Longitudinal ensemble\cite{ghostnet}, CWA\cite{cwa}, AdaEA\cite{adaea}
                }
                edge from parent [gluon]}
            }
            child [missing] {}
            }
      };
\end{scope}
\end{tikzpicture}
\caption{Taxonomy of black-box adversarial attacks, and implemented methods in each category.}\label{listmethods}
 \vspace{-13pt}
\end{figure*}

\subsubsection{Score-based query attacks} Score-based query attacks primarily exploit the confidence scores outputted by models, rather than relying solely on the final decision, to create adversarial examples. Their general formulation can be viewed from the perspective of margin maximization. 
\begin{equation}
    \min _{\boldsymbol{x^*}} \xi\left(\boldsymbol{x^*} \in \mathbb{B}_{\boldsymbol{x}, \epsilon}\right)+\max (0, \Delta).
\end{equation}
The term $\mathbb{B}_{\boldsymbol{x}, \epsilon}$ denotes a $\ell_p$ norm ball centered at $\boldsymbol{x}$ with radius $\epsilon$. 
$\xi$ denotes the distance function which returns 0 if $x^*$ is located in the ball $\mathbb{B}_{\boldsymbol{x}, \epsilon}$ otherwise returns $\infty$. It ensures the imperceptibility of adversarial perturbations.
The variable $\Delta$ differentiates based on the type of attack:
\begin{equation}
\begin{cases}{ } \text { Untargeted: } & \Delta \coloneqq f\left(\boldsymbol{x}^*, y\right)-\max _{j \neq y} f\left(\boldsymbol{x}^*, j\right); \\ \text { Targeted: } & \Delta \coloneqq \max _{j \neq t} f\left(\boldsymbol{x}^*, j\right)-f\left(\boldsymbol{x}^*, t\right).\end{cases}
\end{equation}
Note that the margin loss, captured by the value of $\Delta$, should always be non-negative. A value of 0 for this loss means the attack successfully found its adversarial example, suggesting that the attack process terminates.
The primary objective of the attack is to find successful adversarial perturbations while maintaining them within the $\ell_p$ norm ball $\mathbb{B}_{\boldsymbol{x}, \varepsilon}$.

\subsubsection{Transfer-based attacks} Unlike query-based attacks that demand repeated queries, transfer-based attacks are inspired by the phenomenon that adversarial samples produced to mislead a surrogate model $f^{\prime}$ can mislead target models $f$, even if their architectures are significantly different \cite{transferability}, as long as $f$ and $f^{\prime}$ solve the same tasks. Let $\mathcal{L}$ be the loss function (\textit{e.g.} cross-entropy). To fool the classifier $f$, transfer-based attacks use gradient-based methods to craft $\ell_p$-norm bounded adversarial perturbation $\boldsymbol{\delta}_{\boldsymbol{x}}$ against the white-box models $f^{\prime}$,
\begin{equation}\label{equ: white}
    \boldsymbol{x}^*=\boldsymbol{x}+\boldsymbol{\delta}_{\boldsymbol{x}}(f^{\prime}) \text {, where }\boldsymbol{\delta}_{\boldsymbol{x}}(f^{\prime})=\underset{\|\boldsymbol{z}\|_{p \leq \epsilon}}{\arg \max } \mathcal{L}(f^{\prime}(\boldsymbol{x}+\mathbf{z}), y),
\end{equation}
and then directly utilize the transferability $\mathcal{T}_{\boldsymbol{x}}\left(f^{\prime}, f\right)$ of the adversarial examples to attack the black-box model $f$,
\begin{equation}\label{equ: transferbaility}
    \mathcal{T}_{\boldsymbol{x}}\left(f^{\prime}, f\right)=\left|\left\{f(\boldsymbol{x}) \neq f\left(\boldsymbol{x}+\boldsymbol{\delta}_{\boldsymbol{x}}(f^{\prime})\right)\right\}\right|.
\end{equation}

In BlackboxBench, as depicted in Fig. \ref{listmethods}, we have implemented a broad range of algorithms for black-box adversarial attacks, spanning the above three types: decision-based query attacks, score-based query attacks, and transfer-based attacks. Each of these types further encompasses different technical routes. We have also implemented various defense methods, including training-time and inference-time approaches.
\textbf{We strongly recommend that readers refer to the Supplementary \ref{sec:implemented algorithm} for an elaborate taxonomy of the specific algorithms implemented for each category.}

\subsection{Codebase}
\label{sec:codebase}

We build a unified, extensible, modular-based codebase as the basis of BlackboxBench. As shown in Fig. \ref{modules}, it consists of five modules: preparation module (preparing clean data, pretrained model and auxiliary information), attack module, defense module, evaluation module, and analysis module. 
\blue{Attack module, the core component of the entire codebase, is responsible for generating adversarial samples. Defense module offers options for users to apply defense strategies against attacks. Evaluation module provides metrics to assess the performance of attack methods. Lastly, analysis module incorporates various analytical tools to enhance our understanding of black-box adversarial attacks.
Here we only detail the attack module. \textbf{Supplementary} \ref{app: other modules} includes more information about the analysis module, defense module and evaluation module.
}

\begin{figure}[t]
 \setlength{\abovecaptionskip}{-0.1cm}
\setlength{\belowcaptionskip}{-0.1cm}
 \centering
\includegraphics[width=8.5cm]{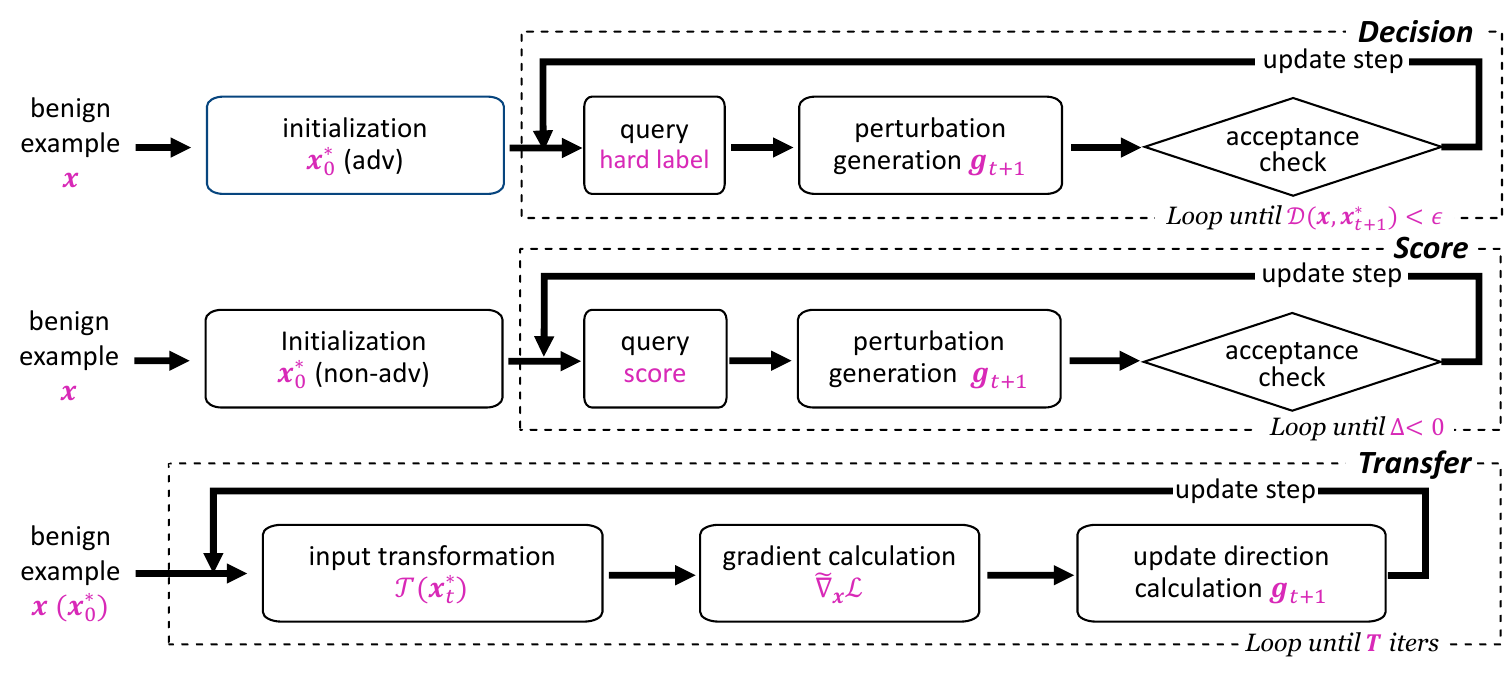}
 \caption{Unified pipelines of attack modules in decision-based attack methods (\textbf{top}), score-based attack methods (\textbf{middle}), and transfer-based attack methods (\textbf{bottom}).}\label{pipeline}
 \vspace{-15pt}
\end{figure}

\subsubsection{Attack module}\label{sec: Attack module}
The attack module takes preprocessed benign examples, pretrained models, and auxiliary information from the preparation module to generate adversarial images. \blue{As schematically illustrated }in Fig. \ref{pipeline}, we summarize the generation process of various algorithms in decision-based, score-based and transfer-based attacks into three common pipelines. More detailed unified algorithms can be found in \textbf{Supplementary} \ref{Unified algorithms}. 
\blue{
The attack module of BlackboxBench is constructed based on the unified pipeline. Fig. \ref{uml} in the \textbf{Supplemental Material} describes the attack module from an implementation perspective by showing the corresponding component diagrams in Unified Modeling Language (UML).
}

\noindent \textbf{Query-based attack module} For query-based attacks, both score-based and decision-based attack flows can be made up of the following four blocks. Nevertheless, they differ significantly inside each block, leading to two independent submodules.
\begin{itemize}[leftmargin=15pt,itemsep=1pt,topsep=0pt]
\item {\textbf{Initialization block}} The decision-based (score-based) attack initiation phase begins with choosing a starting point $\boldsymbol{x}_0^*$ within the adversarial (non-adversarial) region.
\item {\textbf{Query block}} In this phase, the adversarial example $\boldsymbol{x}_t^*$ queries the target model and receives feedback---a label for the decision-based or a posterior probability for the score-based.
\item {\textbf{Perturbation-generation block}} In the decision-based attack submodule, this block computes a perturbation $\textbf{g}_{t+1}$ based on hard-label feedback to narrow the distance between $\boldsymbol{x}_t^*$ and $\boldsymbol{x}$. In the score-based attack sub-module, it finds $\textbf{g}_{t+1}$ based on the probability feedback, aiming to cross the decision boundary within an $\ell_p$ norm ball. The employed perturbation generation function is user-optional. 
\item {\textbf{Acceptance-check block}} This part checks whether the perturbation from last step is acceptable. Upon a true check, the perturbation is added to $\boldsymbol{x}_t^*$ through an update step.
\end{itemize}
The latter three blocks are iterated until the objective of decision-based attacks or score-based attacks is achieved.

\noindent \textbf{Transfer-based attack module} According to the basic structure constituted by I-FGSM, BlackboxBench proposes a unified transfer-based attack pipeline, consisting of the input-transformation block, gradient-calculation block, update-direction-calculation block and update step. Other methods developed on the top of I-FGSM from different perspectives are implemented by modifying corresponding blocks along this pipeline, as follows: 
\begin{itemize}[leftmargin=15pt,itemsep=1pt,topsep=0pt]
\item {\textbf{Input-transformation block}} This block contains several image transformation implementations. Methods from data perspective conducted different transformations $\mathcal{T}\left(\cdot\right)$ on adversarial samples $\boldsymbol{x}_t^*$ in this block.
\item {\textbf{Gradient-calculation block}} Gradient-calculation block takes transformed samples $\mathcal{T}\left(\boldsymbol{x}_t^*\right)$ and outputs the gradients $\widetilde{\nabla}_{\boldsymbol{x}} \mathcal{L}$ \textit{w.r.t.} the adversarial loss $\mathcal{L}$. Here, methods from feature perspective defined novel loss functions $\mathcal{L}\left(\cdot\right)$ based on the intermediate layer features to generate more transferable adversarial examples. Also part of the methods from optimization perspective modifies the way to calculate the gradient in this block.
\item {\textbf{Update-direction-calculation block}} In this block, based on the gradients $\widetilde{\nabla}_{\boldsymbol{x}} \mathcal{L}$, methods from optimization perspective derived effective update direction $\boldsymbol{g}_{t+1}$, \textit{i.e.,} adversarial perturbation.
\item {\textbf{Update step}} Passing through the above three blocks, all methods generate adversarial examples $\boldsymbol{x}_{t+1}^*$ by a one-step update---moving $\boldsymbol{x}_{t}^*$ in the direction of $\boldsymbol{g}_{t+1}$.
\end{itemize}
The above procedure will be iterated ${T}$ times. Note that methods from model perspective refine surrogate models to improve adversarial transferability during the preparation stage rather than in the iterative attack module.


\blue{
By building a unified attack module, BlackboxBench offers several advantages:  
 ~\textbf{1)} From the perspective of a fair and standard evaluation, a unified attack module allows differences between methods to exist only in the core function's code, while the rest of the process remains consistent across all methods. This allows evaluating the method's actual attack performance.  
 ~\textbf{2)} From a researcher's perspective, BlackboxBench offers inspiration and promising research directions by organizing methods into improvements from different functional blocks within a unified attack process. Meanwhile, this flat procedural programming structure avoids using a large number of inheritance classes, unlike other benchmarks. In this way, code readability is prioritized, ensuring that researchers can quickly understand and begin their work. 
~\textbf{3)} For engineers, should they wish to implement a more powerful attack by combining multiple attacks, BlackboxBench simplifies integration of each method's core functional module through defining configuration files. This eliminates need for engineers to redefine the entire optimization process.  
~\textbf{4)} For contributors, adding a new attack method to BlackboxBench is easy. They simply need to design the core function and register it as the appropriate functional block to integrate it into the unified attack process.
}


\begin{table*}[t]
  \centering
  \caption{A summary of analyses in Sec. \ref{sec:evaluation and analysis}. \CIRCLE \ denotes the analyzed question is  properly examined in this attack, while \Circle \ indicates it's not applicable to this attack. \textit{Used tool} indicates the employed tool from analysis module to facilitate this analysis.}
    \begingroup
    \scalebox{0.73}{
    \begin{threeparttable}
    \begin{tabular}{cp{34.085em}cccc}
    \toprule
    \multirow{2}[4]{*}{\textbf{Analysis object}} & 
    \multicolumn{1}{c}{\multirow{2}[4]{*}{\textbf{Analysis content}}} & \multicolumn{2}{c}{\textbf{Query-based}} & 
    \multicolumn{1}{c}{\multirow{2}[4]{*}{\textbf{Transfer-based}}}  & \multicolumn{1}{c}{\multirow{2}[4]{*}{\textbf{Used tool}\tnote{1}}} \bigstrut\\
\cline{3-4}          & \multicolumn{1}{c}{} & \multicolumn{1}{p{4.085em}}{\textbf{Score-based}} & \multicolumn{1}{p{4.085em}}{\textbf{Decision-based}} &  \bigstrut\\
    \hline \hline
    \multirow{4}[1]{*}{\textbf{Result overview} (Sec. \ref{sec: result overview})} &     1) What is the progress in the field over the past years? & \CIRCLE     & \CIRCLE     & \CIRCLE & AEV \bigstrut[t]\\
          &     2) Which category presents the most promising avenue?  & \CIRCLE     & \CIRCLE     & \CIRCLE & AEV \\
          &     3) Under each attack setting, what is the best-performing method in each category and what is the SOTA method?   & \CIRCLE     & \CIRCLE     & \CIRCLE & AEV \\
          &     4) Can individual attacks be combined to form a stronger attack? & \Circle     & \Circle     & \CIRCLE & AEV \bigstrut[b]\\
    \hline
    \multirow{2}[1]{*}{\textbf{Effect of data} (Sec. \ref{sec: effect of data})} 
    & {1) Does the evaluation results generalize across different datasets?} & \CIRCLE     & \CIRCLE     & \CIRCLE & AEV\bigstrut[t]\\
    & {2) Whether there are samples harder to be attacked by all attack methods?} & \CIRCLE     & \CIRCLE     & \Circle & AEV+DBV-B\bigstrut[t]\\
          &     3) How does the dimensionality the adversary restricted to affect its performance?  & \CIRCLE     & \CIRCLE     & \CIRCLE & AEV+DBV-A  \bigstrut[b]\\
    \hline
    \multirow{2}[1]{*}{\textbf{Effect of model architecture} (Sec. \ref{sec: effect of model architecture})} &     1) Which model architecture is most resistant to malicious queries? & \CIRCLE     & \CIRCLE     & \Circle & AEV+DBV-B \bigstrut[t]\\
          &     2) What kind of misalignment between surrogate models and target models then undermines transferability? & \Circle     & \Circle     & \CIRCLE & AEV+ADM \bigstrut[b]\\
    \hline
    \multirow{2}[1]{*}{\textbf{Effect of attack budget} (Sec. \ref{sec: effect of methods})} &     1) How about the methods' reliance on the perturbation norm budget? & \CIRCLE     & \CIRCLE     & \CIRCLE      & AEV+MAV-A \bigstrut[t]\\
          &     2) How about the methods' reliance on the query number budget? & \CIRCLE     & \CIRCLE     & \CIRCLE      & AEV \bigstrut[b]\\
    \hline
    \multirow{3}[1]{*}{\textbf{Effect of defense} (\textbf{Supplementary} \ref{sec: effect of defense})} &    1) How different attacks perform against various defenses?  & \CIRCLE     & \CIRCLE     & \Circle      & AEV \bigstrut[t]\\
          & 2) How different is training-time defense versus inference-time defense? & \CIRCLE     & \CIRCLE     & \Circle     & AEV \\
          &     3) Are adversarial images transferable between normally trained models and adversarially trained models? & \Circle     & \Circle     & \CIRCLE     & AEV+MAV-B \bigstrut[b]\\
    \hline
    \multirow{1}[1]{*}{\textbf{Attack procedure analysis} (\textbf{Supplementary} \ref{sec: visualization analysis})} &     1) How does the attacker traverse from the benign example to its perturbed counterpart? & \CIRCLE     & \CIRCLE     & \CIRCLE     & MAV-A \bigstrut[b]\\
    \bottomrule
    \end{tabular}
    \begin{tablenotes}
    \item[1] AEV $\to$ attack-component effect visualization (5 tools); MAV-A $\to$ model attention visualization - FullGrad; MAV-B $\to$ model attention visualization - FSM; DBV-A $\to$ decision boundary visualization - Decision Surface; DBV-B $\to$ decision boundary visualization - dbViz; ADM $\to$ adversarial divergence measurement.
    \end{tablenotes}
    \end{threeparttable}}
    \endgroup
    \label{tab:questions}
    \vspace{-10pt}
\end{table*}%

\section{Evaluation and Analysis}
\label{sec:evaluation and analysis}

Based on the benchmark shown in Sec. \ref{our benchmark}, we further provide thorough evaluations and detailed analyses of implemented black-box adversarial attacks. 
In this section, we first analyze the overview results, then study the effects of data, model architecture, and attack budget. \blue{We also analyze the influence caused by deployed defense and the attack procedure, please refer to \textbf{Supplementary} \ref{sec: effect of defense} and \ref{sec: visualization analysis}.}
Our analyses are broken into several related questions, summarized in Tab. \ref{tab:questions}. 

\subsection{Experimental Setup}
\label{sec: experimental setup}

\textbf{Datasets} We evaluate our benchmark on CIFAR-10\cite{cifar10} and ImageNet\cite{imagenet}. For CIFAR-10, the test set with 10,000 images is evaluated. For ImageNet, considering the popularity in different fields, query-based attacks use randomly chosen 1000 images belonging to the 1,000 categories (1 image in 1 class) from ILSVRC 2012 validation set, transfer-based attacks use the ImageNet-compatible dataset\footnote{\url{https://github.com/tensorflow/cleverhans/tree/master/examples/nips17\_adversarial\_competition/dataset}} in the NIPS 2017 adversarial competition, which contains 1,000 images with a resolution of $299 \times 299 \times 3$.


\blue{
\textbf{Models} In query-based attacks, we attack 4 normally trained target models and 2 adversarially trained target models on the CIFAR-10 dataset, and 5 normally trained target models and 3 adversarially trained target models on the ImageNet dataset. In transfer-based attacks, following prior works \cite{linbp, bayesian, lgv, rap, robustbench}, we choose 4 surrogate models, evaluate 8 normally trained target models and 1 adversarially trained target model on the CIFAR-10 dataset, and choose 5 surrogate models, evaluate 14 normally trained target models and 3 adversarially trained target models on the ImageNet dataset. The model architectures are diverse and belong to different families to study the transferability. The clean accuracies and checkpoints of adopted models are provided in the \textbf{Supplementary} \ref{app:pretrained model download links}.
}


\subsection{Result Overview}
\label{sec: result overview}


We first exhibit the attack results of all attack methods in BlackboxBench. Each method is evaluated under four attack settings. An attack setting $\mathcal{A}$ is defined by a combination of attack targets $t \in\{$targeted, untargeted$\}$ and norms $p \in\{2, \infty\}$, \textit{i.e.}, $\mathcal{A}= t \times p$.

In query-based attacks, we evaluate \blue{29} methods outlined in \textbf{Supplementary} \ref{sec:implemented algorithm} using datasets and models detailed in Sec. \ref{sec: experimental setup} across all four attack settings. $\ell_2$ and $\ell_{\infty}$ are implemented for most attacks, even though some attacks are explicitly designed for one certain norm. There are a total of 630 pairs of evaluations for decision-based black-box attacks and 988 pairs of evaluations for score-based black-box attacks. Each method's performance in attacking a target model is assessed using three metrics: average query number (AQN), median query number (MQN), and attack success rate (ASR). A better attacker should report a higher ASR and a lower AQN and MQN simultaneously.
In transfer-based attacks, we evaluate \blue{30} methods mentioned in \textbf{Supplementary} \ref{sec:implemented algorithm} on datasets and models listed in Sec. \ref{sec: experimental setup} under all four attack settings, leading to 13,332 evaluations in total. The efficacy of adversarial perturbations crafted on a surrogate model by each method is measured by averaging ASR on all targeted models. A superior attack method is expected to yield a higher averaged ASR. For the context of transfer-based attacks, unless explicitly specified otherwise, "ASR" refers to the averaged ASR herein.

Due to space limitations, in Fig. \ref{overview_query} and Fig. \ref{overview_transfer}, we exclusively offer a condensed summary of results specifically for the ImageNet dataset alongside one model architecture---ResNet-50, \ie, we only exhibit results where ResNet50 acts as a surrogate model in transfer-based attacks and as target model in query-based attacks. For comprehensive tables including additional models and the CIFAR-10 dataset, please refer to \href{https://blackboxbench.github.io/}{BlackboxBench LeaderBoard}.
Based on the results in Fig. \ref{overview_query} and Fig. \ref{overview_transfer}, for each type of attack, our overview in this section will be organized from \blue{the following perspectives: 1) what is the progress in the field over the past years? 2) which category presents the most promising avenue? 3) under each attack setting, what is the best-performing method in each category and what is the SOTA method? 4) can individual attacks be combined to form a stronger attack? If so, what is the most effective guideline for combination?\\
\textbf{Note:} For question 4), we will omit discussions on decision-based and score-based attacks, focusing solely on the case of transfer-based attacks for following reasons. Since decision-based and score-based attacks rely on limited information from the target model (only hard labels or confidence scores) and are constrained by query limits, existing works mainly focus on designing the search strategy to achieve effective attacks with minimal queries. In contrast, transfer-based attacks leverage full information from surrogate models (such as gradients) and the advantage of being unrestricted by query limits, making it possible to combine multiple strategies to significantly enhance attack success rates and transferability. }

\subsubsection{Overview of decision-based attacks}
\label{sec: Overview for decision-based attacks}

\begin{figure*}[t]
\vspace{-2mm}
    \centering
\includegraphics[width=0.9\linewidth]{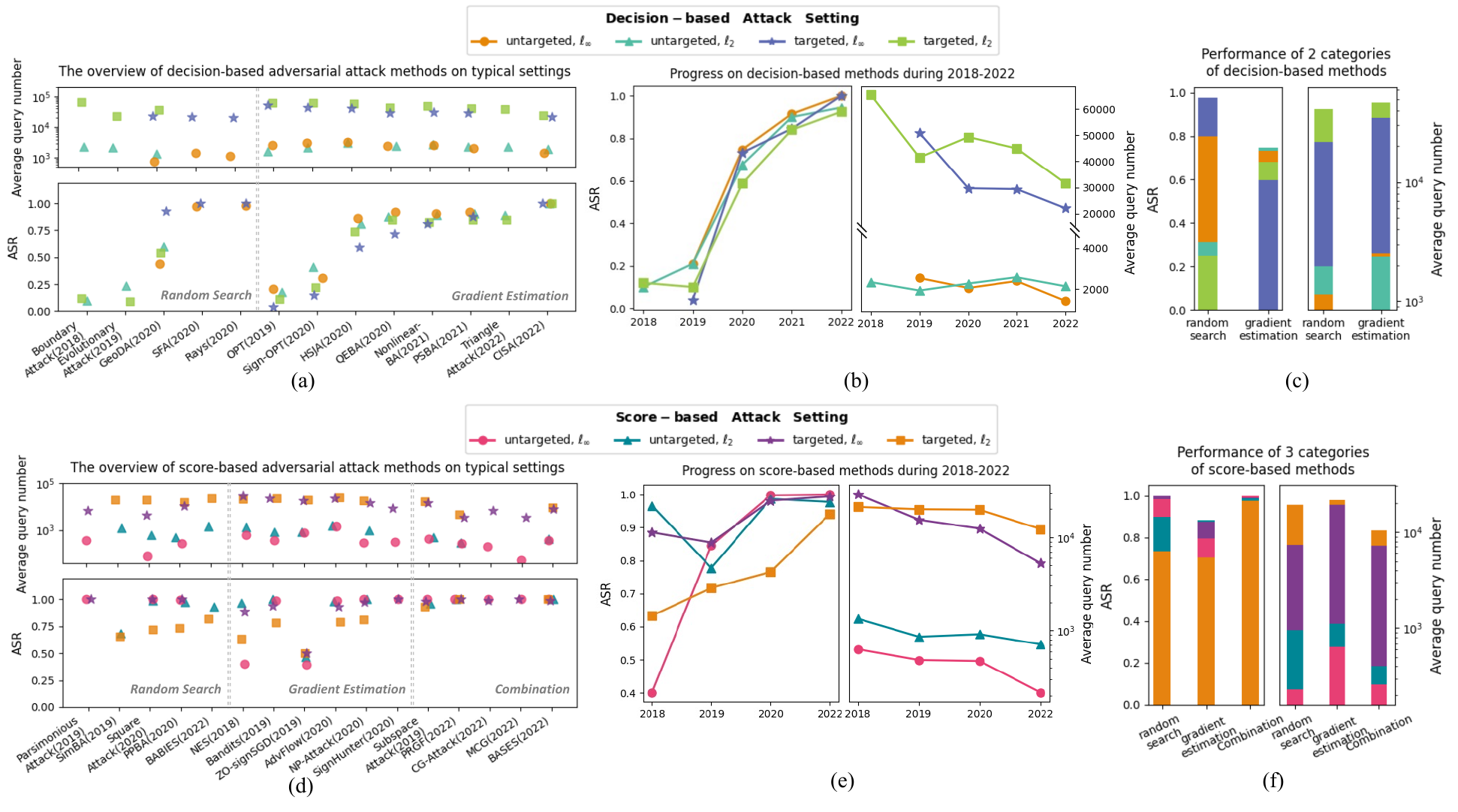}
\vspace{-5pt}
 \caption{\textbf{Result overview of query-based attacks.} The overviews of various decision-based (\textbf{top row}) and score-based (\textbf{bottom row}) black-box adversarial attacks implemented in BlackboxBench. The first column shows attack performances, measured using ASR and AQN metrics. The second and third columns present performance summaries \textit{w.r.t.} years and attack categories, respectively. Each color-mark pattern denotes an attack setting.} \label{overview_query}
 \vspace{-14pt}
\end{figure*}

\begin{itemize}[leftmargin=10pt,itemsep=1pt,topsep=0pt]
    \item \blue{\textit{What is the progress in the field over the past years?}}\\
    \blue{As shown in Fig. \ref{overview_query}(a), various methods have achieved notable advancements in improving query efficiency and attack success rates, leveraging techniques like gradient estimation and random search strategies. Fig. \ref{overview_query}(a) is categorized \wrt~ both year (see Fig. \ref{overview_query}(b)) and type of attack strategies (see Fig. \ref{overview_query}(c)). As illustrated in Fig. \ref{overview_query}(b), there has been an upward trend in both attack success and efficiency from 2017 to 2022, demonstrating sustained progress in this field.}
    \item \blue{\textit{Which category presents the most promising avenue?}}\\
    \blue{Different performance is observed under various attack settings in Fig. \ref{overview_query}(c). Upon closer inspection of \ref{overview_query}(a), CISA\cite{cisa} demonstrates the most outstanding performance across all four attack settings, achieving not only the highest ASR but also maintaining the lowest AQN. Thus, the most promising avenue might be the strategy adopted by CISA\cite{cisa}, \ie, leveraging surrogate model to simulate the behavior of target model and generating approximate gradient information to effectively attack target model. In contrast, other works mainly focused on improving query efficiency through designing better search and gradient estimation strategies. 
    }
    \item \blue{\textit{Under each attack setting, what is the best-performing method in each category and what is the SOTA method?}}\\
    \blue{ From Fig. \ref{overview_query}(a), 
    among all approaches in the \textit{gradient estimation} category for decision-based attacks, CISA~\cite{cisa} demonstrates the best performance under all four settings. 
    In the \textit{random search} category, RayS~\cite{chen2020rays} beats others in both untargeted attacks and targeted attacks. Viewing from all methods, we can find CISA~\cite{cisa} stands out as the SOTA attack method in both untargeted and targeted attacks.}
\end{itemize}

\noindent \blue{\textbf{Summary}: Decision-based attacks have evolved significantly over the past few years. 
CISA~\cite{cisa} represents the current SOTA method, showcasing the potential of the integration of prior knowledge and surrogate models with the query feedback to estimate gradients more effectively, resulting in more powerful and efficient attacks.}

\subsubsection{Overview of score-based attacks}
\label{sec: Overview for score-based attacks}

\begin{itemize}[leftmargin=10pt,itemsep=1pt,topsep=0pt]
    \item \blue{\textit{What is the progress in the field over the past years?}}\\
    \blue{As shown in Fig. \ref{overview_query}(d), various methods have steadily improved query efficiency and attack success rates by leveraging multiple types of attack strategies. Fig. \ref{overview_query}(d) is categorized \wrt~ both year (see Fig. \ref{overview_query}(e)) and attack strategies (see Fig. \ref{overview_query}(f)). We can see there is a marked improvement in either efficiency or ASR since 2018. Unfortunately, most methods before 2022 could not simultaneously improve attack efficiency and success rate. This situation changed since 2022, with the newly emerged attack strategy, which is \textit{combination}, presenting remarkable improvement in both efficiency and ASR.}
    \item \blue{\textit{Which category presents the most promising avenue?}}\\
    \blue{From Fig. \ref{overview_query}(f), we can see the most promising avenue lies in \textit{combination} category, which leverages surrogate models when estimating gradients, as shown by Subspace Attack\cite{subspaceattack}, $\mathcal{CG}$-attack\cite{feng2022boosting}, BASES\cite{bases}, PRGF\cite{prgf} and MCG\cite{yin2023generalizable}. These methods either perform transfer attacks directly on the surrogate models or leverage the transferability of adversarial examples to assist in gradient estimation. By extracting gradient-related information from the surrogate model, these approaches significantly reduce the number of queries required and improve the attack success rate. This strategy of utilizing surrogate models allows them to achieve higher efficiency and effectiveness in black-box attack scenarios.}
    \item \blue{\textit{Under each attack setting, what is the best-performing method in each category and what is the SOTA method?}}\\
    \blue{Across different categories of score-based black-box attacks, the best-performing methods are as follows: Square Attack \cite{andriushchenko2020square} performs best in the \textit{random search} category; SignHunter \cite{al2020sign} leads in the \textit{gradient estimation} category; PRGF \cite{prgf} and MCG \cite{yin2023generalizable} excel in the \textit{combination} category by leveraging extra information from surrogate models. Overall, the current SOTA method among all categories is MCG \cite{yin2023generalizable} in both untargeted and targeted attacks, while PRGF\cite{prgf} also presents remarkably comparable attack performance compared with MCG\cite{yin2023generalizable}.
    }
\end{itemize}

\noindent \blue{\textbf{Summary}: Score-based attacks have seen steady improvements over years, focusing on balancing query efficiency and attack success rates. Recent methods, including Subspace Attack \cite{subspaceattack}, PRGF \cite{prgf}, and MCG \cite{yin2023generalizable}, which belong to \textit{combination} category, have advanced the field by leveraging surrogate models and incorporating techniques like adaptive sampling, model fusion, and projected gradient estimation. MCG\cite{yin2023generalizable} stands as the current SOTA method in both untargeted attacks and targeted attacks, showcasing the potential of assisting gradient estimation with example-level and model-level transferability to implement more efficient and powerful attacks.}

\subsubsection{Overview for transfer-based attacks}
\label{sec: Overview for transfer-based attacks}

\begin{figure*}[t]
\vspace{-2mm}
 \setlength{\abovecaptionskip}{-0.1cm}
\setlength{\belowcaptionskip}{-0.1cm}
 \centering
\includegraphics[width=0.94\linewidth]{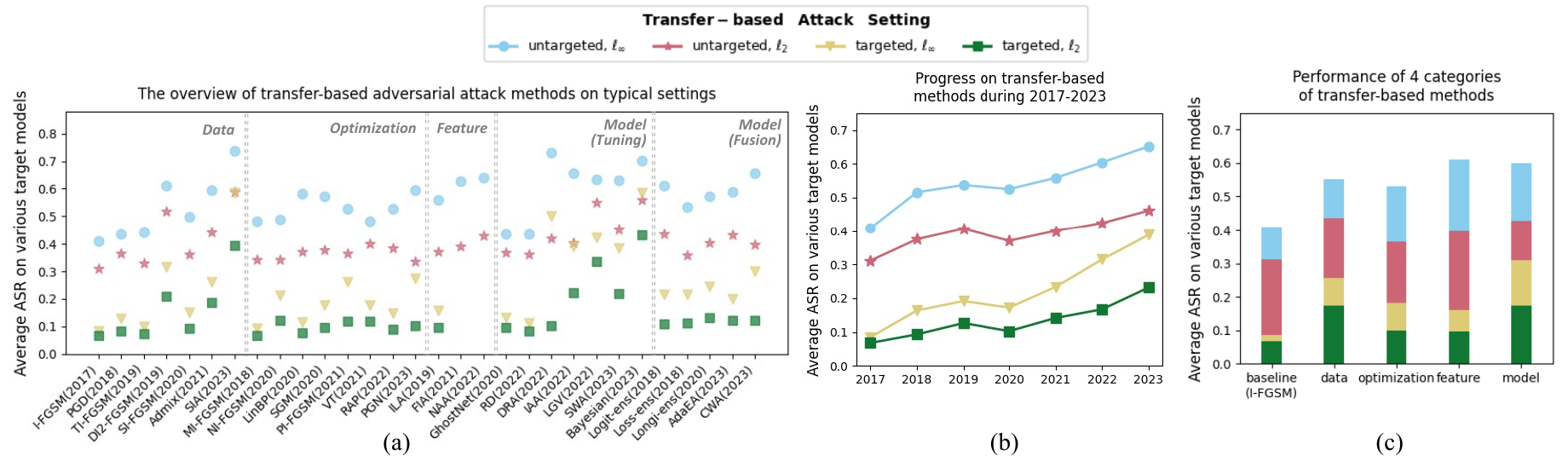}
 \caption{\textbf{Result overview of transfer-based attack.} \textbf{(a)} The performance of various transfer-based black-box adversarial attacks implemented in BlackboxBench. For in-depth analysis, the overview is further summarized according to \textbf{(b)} years and \textbf{(c)} categories. Each color-mark pattern represents an attack setting.} \label{overview_transfer}
 \vspace{-12pt}
\end{figure*}

\begin{figure}[t]
 \setlength{\abovecaptionskip}{-0.1cm}
\setlength{\belowcaptionskip}{-0.1cm}
 \centering
\includegraphics[width= 8.3cm]{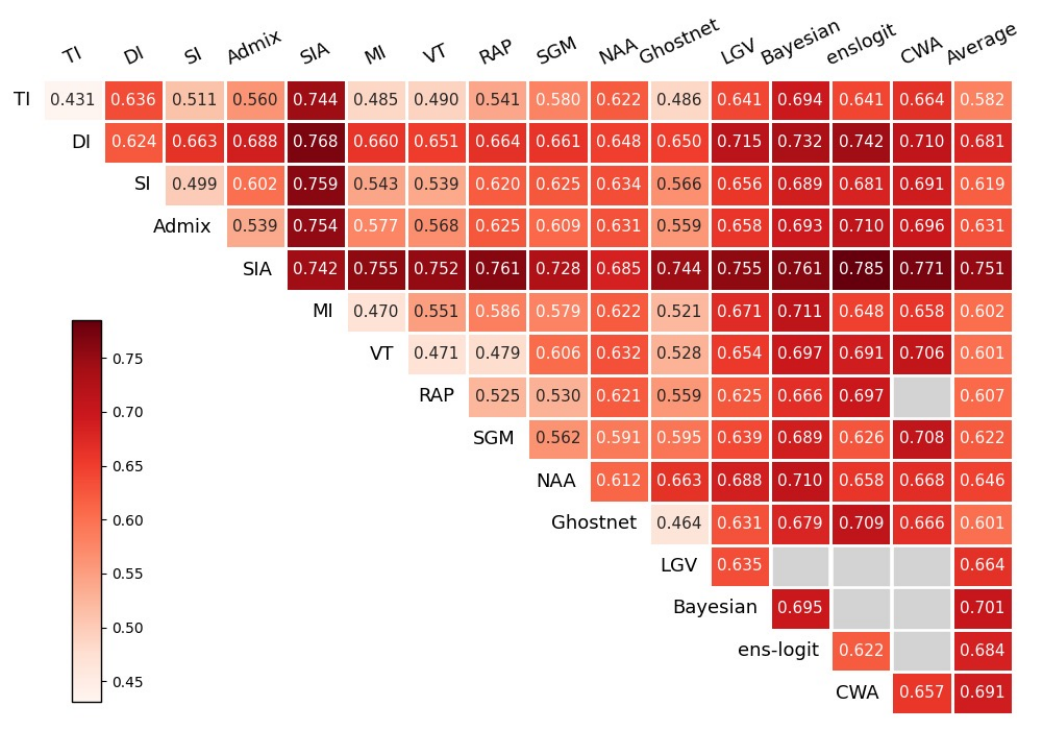}
 \caption{\blue{\textbf{Results of composite transfer-based attacks.} The composite attacks are conducted under untargeted, $\ell_{\infty}$ setting. Each cell represent an pairwise integration of methods in x-axis and y-axis. Cells in gray means these two attacks are incompatible. The last column, 'average,' represents the average attack results for the combinations formed by the corresponding method with all other methods.}}\label{fig:composite}
 \vspace{-15pt}
\end{figure}

\begin{figure*}[t]
    \centering
    \includegraphics[width=0.94\linewidth]{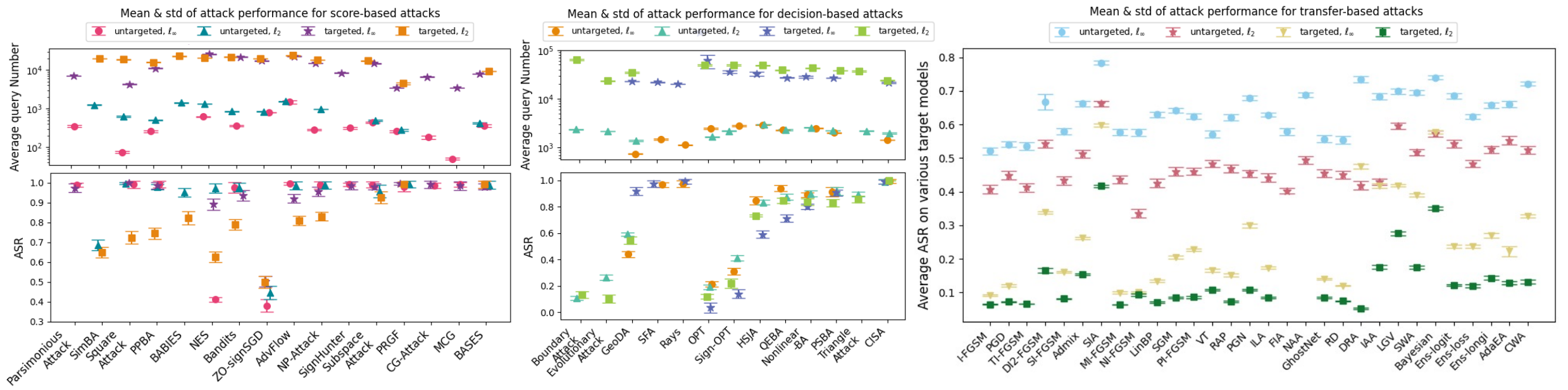}
    \caption{\blue{\textbf{Generalizability of evaluation results.}
    Let random variable $X_c \in \mathcal{X}$ be the input image from class $c$ with $\mathcal{X} \subset \mathbb{R}^d$, and $x_c^i$ the realizations. Let random variable $A_c$ be the attack result on $X_c$, and $a_c^i$ the realizations. $A_c \subset \mathbb{Z}_0^+$ takes values in $[0,1]$ and $[0, M]$ when the attack result represents ASR and AQN, respectively, where $M$ is the maximum query number.
    This figure shows the mean $\pm$ std of dataset-level attack result over 5 random subsets, where  each subset contains one image per class across 1000 classes, sampled from ImageNet validation set.
    Here, mean is defined as $\mu=\sum_{i=1}^n \mu^i/n$, and std is defined as $\sigma=\sqrt{\sum_{i=1}^n\left(\mu^i-\mu\right)^2/(n-1)}$, where $\mu^i=\sum_{c=1}^C a_c^i/C$ is the dataset-level attack result of $i$-th randomly sampled dataset. In our trials, $n=5, C=1000$.}}
    \label{fig:subimagenet_overview}
    \vspace{-15pt}
\end{figure*}

\begin{itemize}[leftmargin=10pt,itemsep=1pt,topsep=0pt]
    \item \blue{\textit{What is the progress in the field over the past years?}}\\
    \blue{From Fig. \ref{overview_transfer}(a), all methods have, to different extent, achieved improvements in the transferability of crafted adversarial examples based on I-FGSM\cite{ifgsm}. Fig. \ref{overview_transfer}(a) is summarized \wrt~ both year (also see Fig. \ref{overview_transfer}(b)) and category (also see Fig. \ref{overview_transfer}(c)).
    As depicted in Fig. \ref{overview_transfer}(b), there is a clear upward trend in attack performance over past seven years, indicating sustained efforts from the active research community.}
    \item \blue{\textit{Which category presents the most promising avenue?}}\\
    \blue{At the earliest stage of transfer-based attacks, the prevailing concept was the parallelism between the transferability of adversarial examples and the generalization ability of trained models \cite{mifgsm}. Accordingly, attempts were primarily made on input augmentation and gradient optimization, marking the birth of attacks developed from data and optimization perspectives. Some of them, although initially proposed, remains strikingly effective even today. For example, as shown in Fig.  \ref{overview_transfer}(a), DI2-FGSM\cite{difgsm} introduced in 2019, which applies random resizing and padding to the input with a probability, still outperforms some methods proposed later. The concept of integrating momentum into the iterative process, proposed by MI-FGSM\cite{mifgsm} in 2018, are still widely adopted in recent method like CWA from 2023\cite{cwa}. Moreover, further breakthroughs have been made along this path, as seen with SIA \cite{sia}, which significantly outperforms others.
    Later, inspired by the fact that different DNN models share similar features, transfer-based attacks from a feature perspective have shown promise in relieving the overfitting to the white-box model and the low transferability to black-box model by performing attacks in the intermediate layers. Notable examples include FIA\cite{fia}, which aims at disrupting the important object-related features. More recently, the focus has shifted towards development from model perspective, either through the tuning on the base surrogate model or the fusion of multiple models' outputs.
    In summary, as depicted in Fig. \ref{overview_transfer}(c), methods from data, optimization, feature and model perspectives have shown similar overall progress in enhancing transfer-based attacks.
    However, a closer look at Fig. \ref{overview_transfer}(a) reveals that, within the model category---specifically in the subcategory of model-tuning-based methods---most approaches, except for two baselines GhostNet\cite{ghostnet} and RD\cite{lgv}, the rest such as DRA\cite{dra}, Bayesian attack\cite{bayesian}, and LGV\cite{lgv}, have demonstrated substantial performance improvements, even in challenging targeted scenarios. This best-performing subcategory implies a promising avenue for future research. Nonetheless, it should be noted that these approaches typically involve higher computational costs due to the necessity of additional model fine-tuning.}
    \item \blue{\textit{Under each attack setting, what is the best-performing method in each category and what is the SOTA method?}}\\
    \blue{Among all approaches aimied at creating diverse input in \textit{data} category, SIA\cite{sia} beats others by constructing the most diverse inputs through applying multiple transformations on a single image. In \textit{optimization} category, PGN\cite{pgn} demonstrates the best results under $l_\infty$ constrain, while under $l_2$ constrain, VT\cite{vt} is the best untargeted attack and NI-FGSM\cite{nisifgsm} is the best targeted attack. In \textit{feature} category, NAA\cite{naa} exhibits best performance. In \textit{model} category, Bayesian attack\cite{bayesian} is inferior to DRA\cite{dra} under untargeted, $l_\infty$ setting but demonstrates best results in other three scenarios. Overall, SIA emerges as a SOTA untargeted attack and Bayesian attack as a SOTA targeted attack. Their performance difference is marginal, but they outperform others by a significant margin under all four settings.}
    \item \blue{
    \textit{Can individual attacks be combined to form a stronger attack? If so, what is the most effective guideline for combination?}\\
    As most transfer-based attacks are orthogonal to each other, seamless integration for improved attack performance is possible. In Fig. \ref{fig:composite}, 15 individual attacks from all four categories were selected and combined to form 98 pairwise composite attacks, and evaluated under untargeted, $l_\infty$ setting. 
    The results indicate that, in general, the combination of individual attacks results in more powerful ones.
    With a few exceptions, the composites involving the SOTA untargeted attack method, SIA, demonstrates superior attack results (on average 75.1\%) than other composites without SIA.
    It is noteworthy that the best composite attack are obtained through incorporating SIA with model-fusion-based methods, \ie, SIA\&Enslogit\cite{mifgsm} with ASR 78.5\% and SIA\&CWA\cite{cwa} with ASR 77.1\%, rather than with the best performing category that includes model-tuning-based methods (LGV\cite{lgv} and Bayesian attack\cite{bayesian}). Recall that the surrogate ensemble in these model-tuning-based methods is randomly sampled from the base surrogate model’s weight distribution, while the surrogate ensemble in model-fusion-based methods consists of well-trained models. Thus, a possible explanation may be that the highly transformed images, crafted by various transformations in SIA, tend to fall within the lower probability regions of the sampled models’ data distribution, compared to those of well-trained models, thereby exhibiting less compatibility with model-tuning-based methods. To sum up, the most effective guideline for combination might be to incorporate the SOTA method into the composite attacks. In addition, the complement requires strong synergy with the SOTA method. 
    Beyond pairwise composite attacks, we suppose that better attack performance can be achieved by triplets or even quartets, and leave it as future work.
    }
\end{itemize}

\noindent \blue{\textbf{Summary}: In transfer-based black-box adversarial attacks, there is a clear upward trend in attack performance over the past seven years. The best performing category comprises model-tuning-based methods, which is a promising research paths for future works. The SOTA untargeted attack is SIA\cite{sia} and SOTA targeted attack is Bayesian attack\cite{bayesian}. To achieve the most powerful pairwise composite attack, prioritize including and synergizing with the SOTA method.
}


\subsection{Effect of Data}
\label{sec: effect of data}

\blue{In this section, we investigate the effect of data in black-box adversarial attacks by following questions: Firstly, \textit{do evaluation results generalize across different datasets?}
While evaluation on a one-image-per-class subset containing 1000 classes is commonly used in the community, we examine its generalizability across different subsets. Our results show that \textbf{this subset is sufficiently diverse to reliably reflect the performance of evaluated attacks}.
Secondly, \textit{whether exists samples harder to be attacked by all methods?}
We find that \textbf{there are indeed some samples that are difficult to attack across all methods}, and one important reason is that these samples are far from the model's decision boundary. This means that generating adversarial perturbations large enough to misclassify these samples requires more efforts, especially in black-box attacks, where often involves more queries.
Beyond aforementioned questions, the influence of input dimensionality on black-box adversarial attack performance is also investigated. Please refer to \textbf{Supplementary} \ref{sec: Input dimensionality} for more discussions.}

\subsubsection{Generalizability of evaluation results}
\label{sec: Generalizability of evaluation results}

\blue{
When evaluated on ImageNet dataset, most attack methods employ a small subset of its validation dataset, typically sampling one image from each of the 1000 classes, as well as BlackboxBench. To analyze the generalizability of the evaluation results, we randomly sample 1000 images (one image per class), creating multiple random datasets, and observe the fluctuation of attack results, \ie, ASR and AQN, at the dataset level.  
Fig. \ref{fig:subimagenet_overview} plots the observed mean $\pm$ standard deviation (std) of dataset-level attack result over 5 random trials. 
The relative performance rankings of the various methods, as indicated by means, maintain consistency with the results overviewed in Sec. \ref{sec: result overview}. Meanwhile, the low stds, depicted by short error bars, suggest a minimal variation in dataset-level attack results during the switch of subsets. To sum up, a one-image-per-class subset containing 1000 classes is enough to reflect the actual performance of evaluated attacks.
}

\blue{
Note that the observed stability in attack results considered on the entire dataset does not necessarily imply a low variation in attack results of single samples within a class. In contrast, we observe a large fluctuation in attack results among individual samples (see \textbf{Supplementary} \ref{add: Fluctuation among samples} for details). Unlike transfer-based attacks, where the difference between surrogate and target models contribute significantly to the attack performance, the effectiveness and efficiency of query-based attacks are predominantly affected by the initial point of their attack path, that is, the benign sample's position on loss surface of the target model. Therefore, in the consecutive subsection, we will marginalize all the methods and delve into reasons behind attack performance fluctuations among samples.
}


\subsubsection{Characteristic of hard samples} 
\label{sec: Characteristic of hard samples} 
\begin{figure*}[t]
    \centering
    \includegraphics[width=0.9\linewidth]{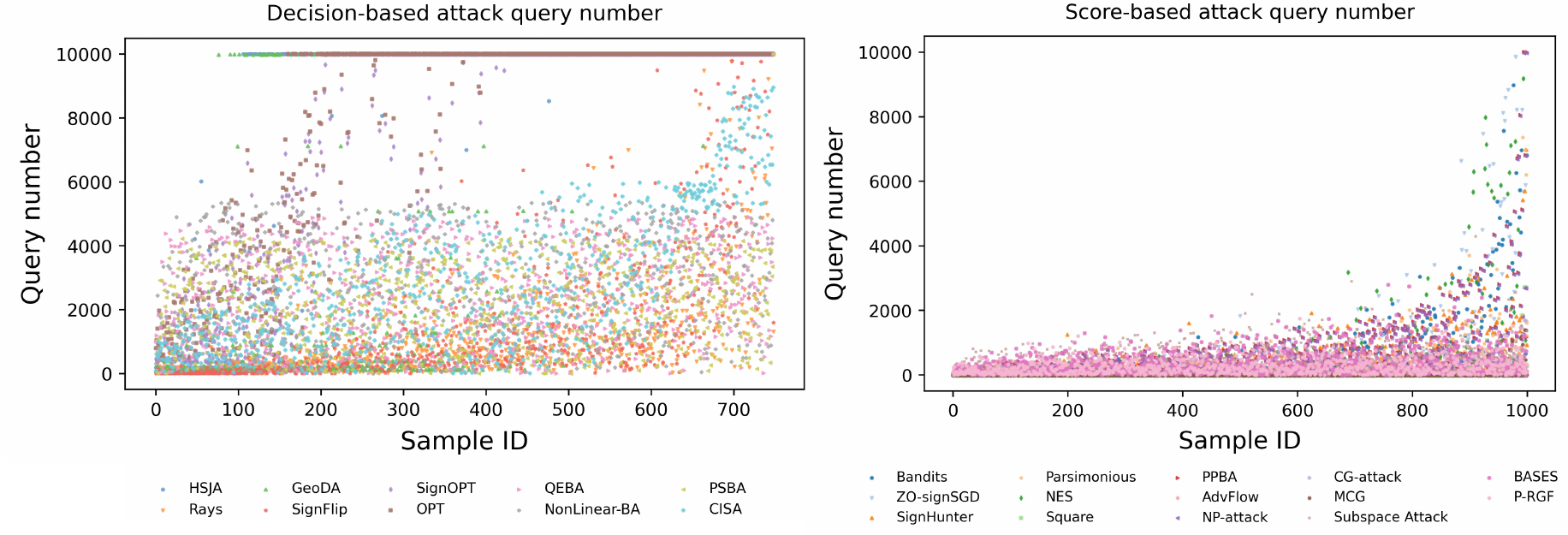}
     \vspace{-5pt}
    \caption{\textbf{Characteristic of hard samples}. Query number of each example for untargeted decision-based black-box attacks (\textbf{left}) and untargeted score-based black-box attacks (\textbf{right}). We test all query-based attack methods on ResNet-50. Note that we only recorded the query numbers of those samples which are successfully initialized at the beginning of decision-based attacks. As for score-based attacks, we recorded the query numbers of the samples which failed to be initialized to be 0 and therefore the number of recorded samples in score-based attacks is 1,000.}
    \label{fig:query-numbers}
     \vspace{-14pt}
\end{figure*}

Fig. \ref{fig:query-numbers} presents the performance of decision-based black-box attacks and score-based black-box attacks on the test dataset. The y-axis represents the number of queries required for a successful attack, while the x-axis indicates the index of clean examples sorted according to the query cost. In the case of decision-based attacks, when the number of queries exceeds 10,000 for untargeted attacks, the attack is considered a failure, leading to a cluster of data points at the top of Fig. \ref{fig:query-numbers} (left). 
Firstly, the plot shows two distinct regions corresponding to gradient-estimation-based attacks and random search-based attacks, indicating the superiority of the latter when performing $\ell_\infty$ untargeted attack. 
The results also reveal that there exist samples for which all attack algorithms fail or require a substantial number of queries to succeed. A similar phenomenon can be observed in score-based black-box attacks (see Fig. \ref{fig:query-numbers} (right)).
This discovery motivates the exploration of which samples are more difficult or easier to attack. 

\begin{figure*}[t]
\vspace{-2mm}
    \centering
\includegraphics[width=17cm]{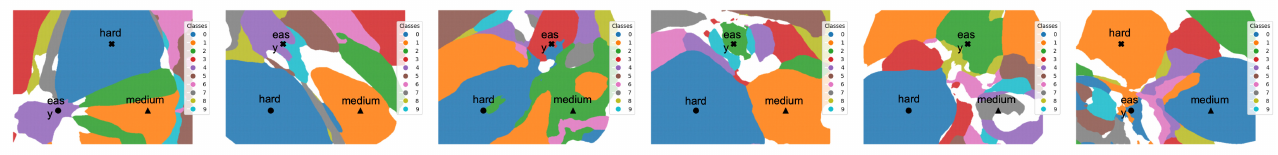}
 \caption{For the ImageNet dataset on which we deploy our attacks, we select images that are difficult to attack (\ie, those requiring a large number of queries or reaching the query limit) and those that are easily to attack, and visualize the decision boundaries around them. The difficulty level is categorized as ``\textit{hard}", ``\textit{medium}", ``\textit{easy}", corresponding to top 30\%, 30\%-60\%, 60\%-90\% in terms of the required query numbers. Considering the 1,000 categories of ImageNet, in the figure above, we only display the decision boundaries for the top 10 categories with the highest confidence for the model. Each point in the figure represents the benign images we select, and the chosen model is ResNet-50.} \label{decision-boundary-visualization}
 \vspace{-7pt}
\end{figure*}

Based on the quantitative results, our objective was to explore the reasons behind the resilience exhibited by certain samples against untargeted attacks. To this end, we conducted extensive experiments using the dbViz visualization tool to analyze the decision boundaries surrounding hard-to-attack and easy-to-attack samples in Fig. \ref{decision-boundary-visualization}. Our analysis tool will help verify our hypothesis: we believe that samples vulnerable to untargeted attacks are proximate or surrounded by complex, visually fragmented decision boundaries, while those resilient samples are distanced from these boundaries. Recall the nature of untargeted attacks, where the goal is to introduce adversarial perturbations to a clean sample and misclassify it as a label other than its ground truth. Samples near intricate decision boundaries, featuring multiple categories nearby, are more susceptible to this kind of manipulation. Conversely, upon visualizing the decision boundaries around the hard-to-attack samples, we observed their substantial distance from these boundaries, as presented in Fig. \ref{decision-boundary-visualization}. This observation suggests that a longer, more intensive attack process might be necessary to shift these perturbed samples away from their original ground-truth category. 
Our analysis tool provides a more intuitive comprehension of the quantitative results and can serve as a compass for future endeavors aimed at refining attack and defense strategies.

\begin{figure*}[t]
 \setlength{\abovecaptionskip}{-0.1cm}
\setlength{\belowcaptionskip}{-0.1cm}
 \centering
\includegraphics[width=17.5cm]{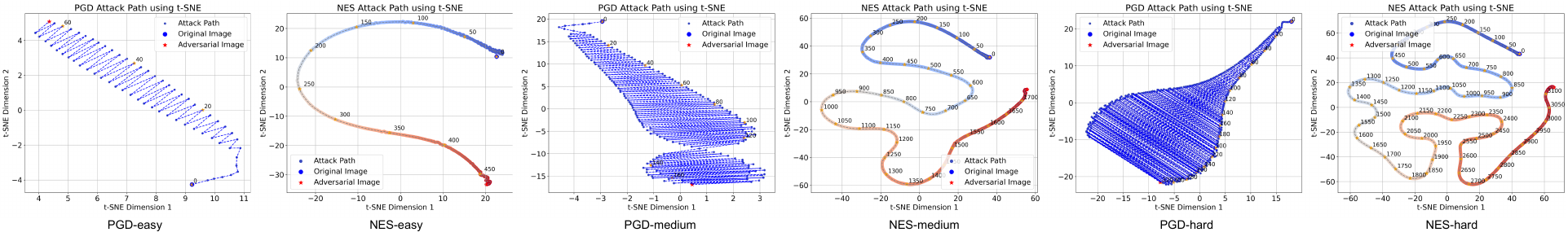}
 \caption{T-SNE visualizations of the attack paths of PGD (white-box attack) and NES (score-based black-box attack), on easy samples (the left two), medium samples (the middle two), and hard samples (the right two), respectively. The attacks are conducted on ResNet-50 model and ImageNet dataset.}\label{attack_trajetory}
 \vspace{-12pt}
\end{figure*}

\blue{We come to the conclusion based on the discussion above that those samples lead to failure attacks or take huge query costs and stay away from the decision boundary. It leads us to wonder why, for the same sample, a white-box attack (PGD\cite{pgd}) can succeed, while a black-box attack (NES\cite{ilyas2018black}) might fail, or even if it succeeds, it requires significantly more queries.}

\blue{In the experiment, the attack paths of NES and PGD are visualized using t-SNE dimensionality reduction as shown in Fig. \ref{attack_trajetory}. These two methods attack ResNet-50 on ImageNet respectively under targeted, $\ell_{\infty}$ setting. We choose 3 samples from different attack levels as we described before (\textit{"easy", "medium", and "hard"}) for both NES and PGD to attack respectively. The blue dot represents the starting point of the attack, which is the position of the original input image, while the red star indicates the position of the adversarial example successfully generated by the attack. The attack path, shown with a color gradient, illustrates the gradual transformation from the original image to the adversarial example. We label the iteration number for PGD and query number for NES along the attack path at regular intervals in the figure to dynamically display the attack progress. NES, as a black-box attack method, updates the adversarial example incrementally through random sampling and gradient estimation, resulting in a smooth curve, indicating that the update directions are relatively stable but not entirely precise. As the attack gradually approaches the target, the path accelerates toward the adversarial example, ultimately succeeding in deceiving the model. In contrast, PGD, a white-box attack, directly optimizes using the true gradient information from the model, resulting in a more complex path with lateral loops and intersections, especially near the decision boundary, where fixed step sizes are repeatedly adjusted to approach the adversarial target. Although PGD’s path is more complex, its use of precise gradient information makes the attack more accurate near the decision boundary, while NES, through its gradual exploration, approaches the decision boundary more smoothly.}

\blue{In summary, by analyzing the query counts for each method, we observed that there are always samples that are more difficult to attack. Next, we present the relative positions of these hard-to-attack samples within the visualized decision boundary, which raises another question: why can white-box attacks succeed with relatively fewer iterations, while black-box attacks may either fail or require significantly more queries? By visualizing the attack paths of both white-box and black-box attacks, we explain why white-box attacks are more likely to succeed—because they can directly compute the attack direction using model information, whereas black-box attacks must rely on estimations to determine the attack direction. Clearly, this also partially explains why combination methods in score-based attacks perform so exceptionally well.}

\subsection{Effect of Model Architecture}
\label{sec: effect of model architecture}
In this section, we are interested in the relationship between attack performances and model architectures. It is well-known that different classification models employ distinct architectural designs as well as different decision boundaries. In query-based attacks, we explore \textit{which model architecture is most resistant to malicious queries}. We specifically compare the difficulty of crafting successful adversarial samples among various target model architectures from different families\footref{fn: family}. 
\blue{The main findings are that \textbf{complex models, such as ConvNeXt-T and ViT-B/16, demonstrate stronger robustness against adversarial attacks, with higher AQN/ASR ratios, while simpler models like ResNet-50 and VGG-19$^{\dagger}$ are more vulnerable.} The integrity and continuity of decision boundaries in complex models make them harder to attack, whereas simpler models have more fragmented boundaries, making them easier targets. Model architecture complexity and decision boundary characteristics are key factors influencing adversarial robustness.}
In transfer-based attacks, we question whether the architectural \textit{families}\footnote{A family contains variants of the same architecture, \textit{e.g.}, ResNet-34/50/152 in ResNet family, Swin-Tiny/Base/Large in Swin family. \label{fn: family}} the surrogate and target model architectures belong to determine transferability. If this is not the case, \textit{what kind of misalignment between surrogate models and target models then undermines transferability?} Specifically, we explore transferability between models within and between different families, \textit{i.e.}, intra/cross-family transferability, \bluetwo{and further extend the analysis to the higher-level \textit{mechanisms}}\footnote{Models with the same mechanism means they share a common foundational block, \textit{e.g.}, both Vit and Swin benefit from self-attention, as well as ResNet and ConvNeXt from convolution. \label{fn:mechanism}}.
\bluetwo{Our findings suggest that architectural similarity, while informative, is neither a reliable nor sufficient explanation for transferability patterns.} Instead, we propose that the \textbf{transferability is implied in the resemblance between adversarial examples generated by the surrogate and target models, and could be measured by a novel metric, Adversarial Divergence.}

\begin{figure*}[t]
\vspace{-2mm}
 \setlength{\abovecaptionskip}{-0.1cm}
\setlength{\belowcaptionskip}{-0.1cm}
 \centering
\includegraphics[width=0.85\linewidth]{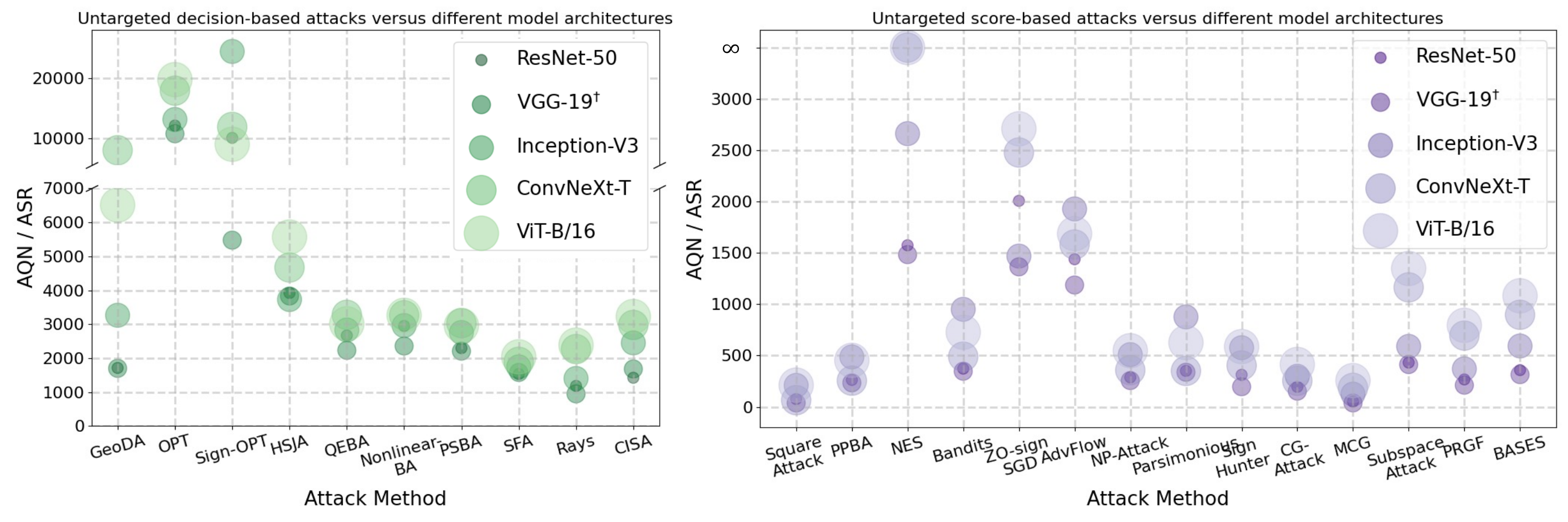}
 \caption{\textbf{Vulnerability of different model architectures}. The ratio of AQN to ASR of decision-based (\textbf{left}) and score-based attacks (\textbf{right}) versus various target model architectures. An $\infty$ ratio indicates an ASR of 0, \ie, no image is successfully perturbed, hence implying an extremely hard attacked model.}\label{query architecture}
 \vspace{-5pt}
\end{figure*}

\begin{figure*}[t]
\vspace{-2mm}
 \setlength{\abovecaptionskip}{-0.1cm}
\setlength{\belowcaptionskip}{-0.1cm}
 \centering
\includegraphics[width=0.87\linewidth]{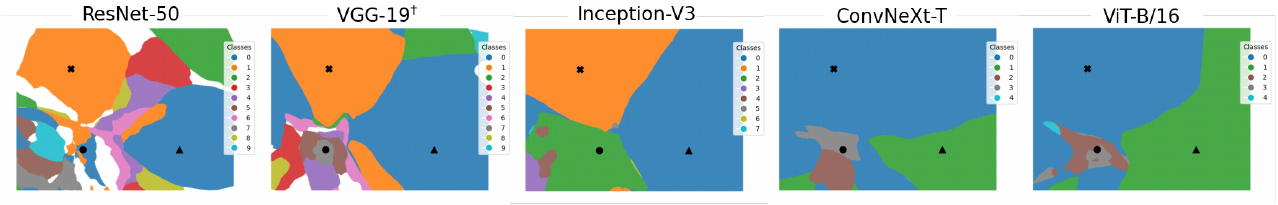}
 \caption{Visualizations of various models' decision boundaries. We also utilize the dbViz\cite{somepalli2022can} tools mentioned above to visualize the decision boundaries of these models. Following the plot settings in dbViz\cite{somepalli2022can}, we randomly sampled 3 images in ImageNet and constructed an image plane to visualize the decision boundaries. Also, we only present the top 10 categories with the highest confidence scores from target models. More details can be referred to \cite{somepalli2022can}.}\label{model decision boundaries}
  \vspace{-12pt}
\end{figure*}

\subsubsection{Vulnerability of different model architectures}
\label{sec: Vulnerability of different model architectures}
We choose to implement query-based attacks on model architectures from different families: ResNet-50 from ResNet family \cite{resnet} which introduces the identity shortcut connection, VGG-19$^{\dagger}$ from VGG family \cite{vgg} which is characterized by its multiple layers of small convolutional filters with consistent receptive fields, Inception-V3 \cite{inception} which uses "inception modules" to effectively grasp both local and broad image details, ConvNeXt-Tiny (ConvNeXt-T) from ConvNeXt family \cite{convnet} which merges depth-wise and point-wise convolutions using grouped convolutions to boost its ability to understand local and wide-ranging dependencies, ViT/B-16 from ViT family \cite{vit} which adopts self-attention and a Transformer framework to perceive global contexts and manage distant dependencies. We utilize the ratio of \blue{AQN to ASR} as a metric to measure the difficulty. The higher ratio indicates that the model is harder to be attacked. For both decision-based black-box attacks and score-based black-box attacks, we employ untargeted attacks using $\ell_{\infty}$ norm and the ImageNet dataset. 

From Fig. \ref{query architecture}, we observe that, for the majority of attack methods, attacking SOTA models (ConvNeXt-T and ViT/B-16) tends to yield higher ratio values (\ie, more queries but lower ASR), indicating they are preferable in terms of robustness. Notably, Inception-V3 also demonstrates a comparable or occasionally higher level of resilience against attacks like Sign-OPT, Bandits, AdvFlow and Parsimonious attacks. Conversely, ResNet-50 and VGG-19$^{\dagger}$ prove more susceptible to adversarial perturbations, reflected by their lower ratio values.

To investigate the possible hidden reasons, we use our analysis tool, dbViz, to visualize the decision boundaries of various target models. From the visualization results in Fig. \ref{model decision boundaries}, we can see that for models with relatively easy architectures like ResNet-50 and VGG-19$^{\dagger}$, the decision boundaries on the ImageNet dataset are quite complex and fragmented. Looking at the three samples we took for the graphing, some of these samples are located on the edge of the decision boundaries of many categories or are surrounded by the decision boundaries of several categories. In comparison, the decision boundaries of the more complex structures, Inception-V3, ConvNeXt-T and ViT-B/16, appear simpler. In other words, the decision boundaries of more complex models are relatively more intact and continuous, without fragmentation. This is bad news for adversarial attacks, as it means that often a larger attack budget is needed to push a clean sample across the decision boundary of the ground-truth category to become an adversarial sample that meets the constraints. In conclusion, our observation and analysis of decision boundary visualizations across various models suggest that the completeness and continuity of decision boundaries in target models constitute a significant factor contributing to the challenge of successful attacks.

\subsubsection{Intra/Cross-family transferability} \label{sec: Intra/Cross-family transferability}

\bluetwo{
Building on prior observations, we explore whether model architectural grouping plays a role in determining adversarial transferability. A prior study \cite{vit2cnn} reports that transferability between ViTs and non-transformer models is unexpectedly low, suggesting that structural similarity may influence transferability. Inspired by this finding, we aim to revisit this question under a finer-grained lens by first examining similarity at the \textit{family} level—across variants of the same architecture—and then extending the analysis to the higher-level notion of \textit{mechanisms}, which groups families by their underlying computational principles, as a potential factor behind transferability.
Note that to isolate the role of architectural similarity, other factors of transferability such as training data distribution and training objective are assumed to be fixed, which is exactly the commonly considered closed-set scenario.
}

\begin{figure*}[t]
\vspace{-2mm}
 \setlength{\abovecaptionskip}{-0.1cm}
\setlength{\belowcaptionskip}{-0.1cm}
 \centering
\includegraphics[width=0.95\linewidth]{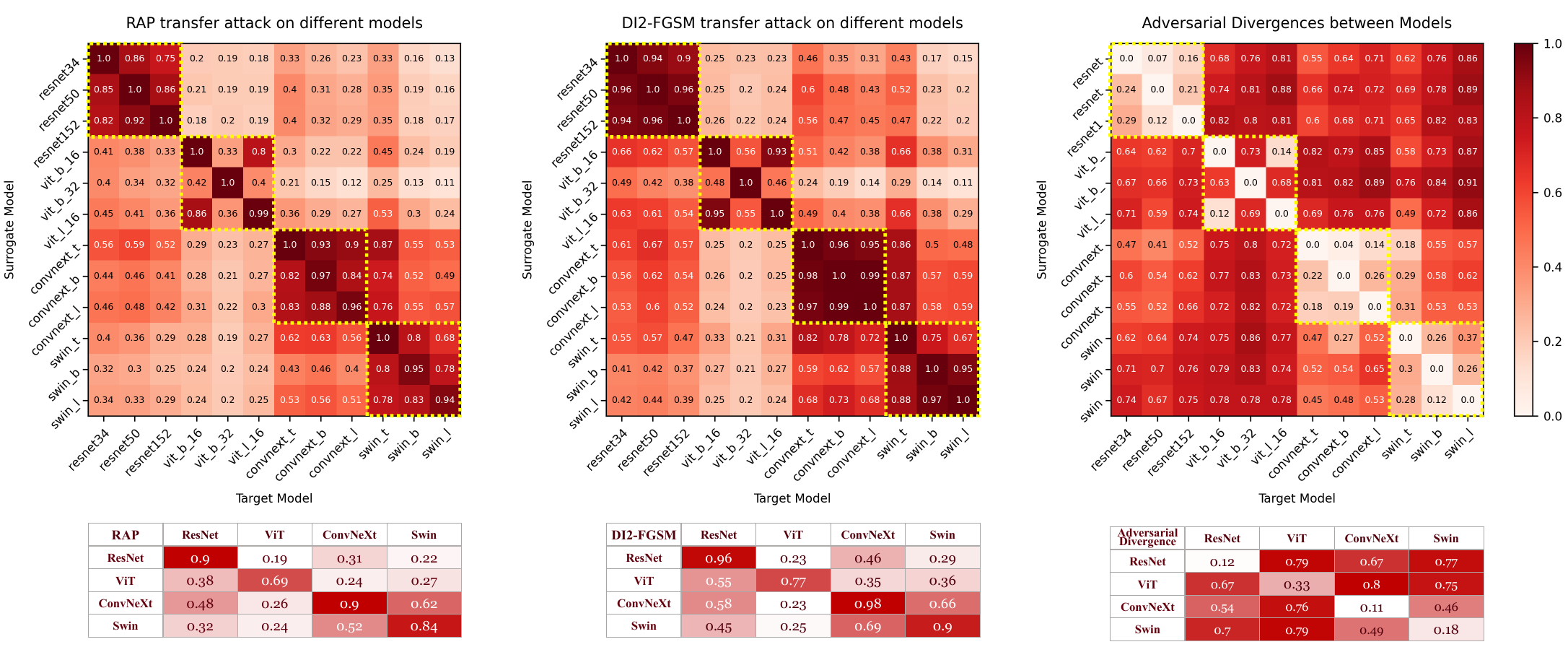}
 \caption{
 \blue{\textbf{Intra/Cross-family transferability}. 
 Transferability rate matrices under RAP attack\cite{rap} (\textbf{upper left}) and DI2-FGSM attack\cite{difgsm} (\textbf{upper middle}), where each entry represents the ASR of adversarial examples generated by the surrogate model on y-axis against the target model on x-axis. 
 Additional results of methods which display a similar phenomenon are available in Fig. \ref{add_arch} of the \textbf{Supplemental Material}. Adversarial Divergence matrix (\textbf{upper right}), where each entry represents the Adversarial Divergence between the surrogate model on y-axis and the target model on x-axis. As three variants are chosen from each of the four model families, matrices in \textbf{the second row} further summarize the first row by averaging entries \textit{w.r.t.} each family.}
 }\label{architecture}
 \vspace{-14pt}
\end{figure*}

We investigate intra/cross-family transferability via four families grouped by two underlying mechanisms: two CNNs (classic: ResNet\cite{resnet}, SOTA: ConvNeXt\cite{convnet}) and two SOTA Transformers (ViT\cite{vit} and Swin\cite{swin}). ResNet family are rather simple convolutional models with low capacities. We consider ResNet-34/50/152. ConvNeXt family are fully convolutional architectures with designs popularised by Transformers. We consider ConvNeXt-Tiny/Base/Large. ViT family modify the original Transformer to make it work with images and induce inductive biases by some training recipes and design choices. We consider ViT-B/16, ViT-B/32 and ViT-L/16. Swin families are hierarchical Transformers employing a shifting window scheme. We consider Swin-Tiny/Base/Large. We utilize all of the aforementioned 12 model architectures, leading to $12 \times 12 = 144$ surrogate-target model pairs. \blue{Fig. \ref{architecture} reports intra/cross-family transferability rate matrices, which exhibit ASRs of all pairs, under untargeted, $\ell_\infty$ attack on ImageNet.}

Rates $(i, j), |i-j| \leq 2$, highlighted by yellow bounding boxes, indicate the proportion of adversarial samples crafted on the white-box model $i$ misclassified by the black-box model $j$ within the same family, \ie, intra-family transferability. Firstly, the results on the diagonal where surrogates precisely align with targets show that SOTA Transformers and CNNs do not provide additional security than classic CNNs. None of them could resist white-box attacks. Therefore, how well an attack would perform is heavily dependent on the surrogate-target alignment rather than the target model individually. Surprisingly, we find that all models are vulnerable to intra-family transferability in a non-negligible manner, indicating that surrogates and targets shift with respect to each other but within the same family wouldn't dramatically undermine transferability.

Rates $(i, j), |i-j| > 2$ indicate the proportion of adversarial examples generated to mislead white-box model $i$ that also mislead black-box model $j$ ($i,j$ are from different families), \textit{i.e.}, cross-family transferability. The phenomenon among cross-family transferability is more heterogeneous than intra-family transferability. For the reason behind the heterogeneity, as CNNs induce convolutional inductive biases while Transformers communicate globally via self-attention modules, a natural hypothesis is that the misalignment that destroys transferability comes from performing different mechanisms, as indicated by \cite{vit2cnn}. \bluetwo{However, the second row in Fig. \ref{architecture} reveals that this pattern does not always hold.} For example, the most transferable family to ResNet is intra-mechanism ConvNeXt, but a high cross-mechanism transferability occurs between ConvNeXt and Swin. Thus, the employed mechanism is not as significant a factor in cross-family transferability as we conjectured.

\bluetwo{
These observations suggest that architectural similarity is neither a reliable nor measurable predictor of transferability.
} A more rigorous quantitative metric of transferability is desired. To this end, we hypothesize that the transferability is implied in the resemblance between adversarial examples generated by the surrogate model and target model, and could be modeled by the divergence between their conditional adversarial distributions (CADs), \ie, the distribution of perturbations conditioned on clean examples. We define this divergence as a Adversarial Divergence metric. Using our provided analysis tool (refer to \textbf{Supplementary} \ref{subsubsec: analysis module}), we capture the CADs of these models and calculate the Adversarial Divergence of each pair. The results shown in Fig. \ref{architecture} (right) align closely with the transfer success rate illustrated in Fig. \ref{architecture} (left and middle), \ie, the pair with higher ASR will report a lower Adversarial Divergence value. This alignment verifies our hypothesis that two models with high transferability share a vulnerability to a similar pattern of adversarial perturbation. \blue{
Additionally, as an effective measure of misalignment between the surrogates and targets, Adversarial Divergence could guide the selection or refinement of surrogate models to generate more transferable adversarial perturbations. 
For example, due to the unavailability of target models, one approach to improve transferability is to increase the diversity in surrogate models, which has led to the proposal of ensemble-based attacks\cite{ensemble}. Adversarial Divergence can be used to exclude similar models within the surrogate ensemble that have low Adversarial Divergence with respect to others, thereby ensuring high attack efficacy. Furthermore, if querying target models is permitted, Adversarial Divergence could be computed to capture the disparity between the surrogate and target model. Under its supervision, the surrogate model could be fine-tuned to mimic the target model in generating perturbations.
}

\begin{figure*}[t]
\vspace{-2mm}
 \setlength{\abovecaptionskip}{-0.1cm}
\setlength{\belowcaptionskip}{-0.1cm}
 \centering
\includegraphics[width=0.99\linewidth]{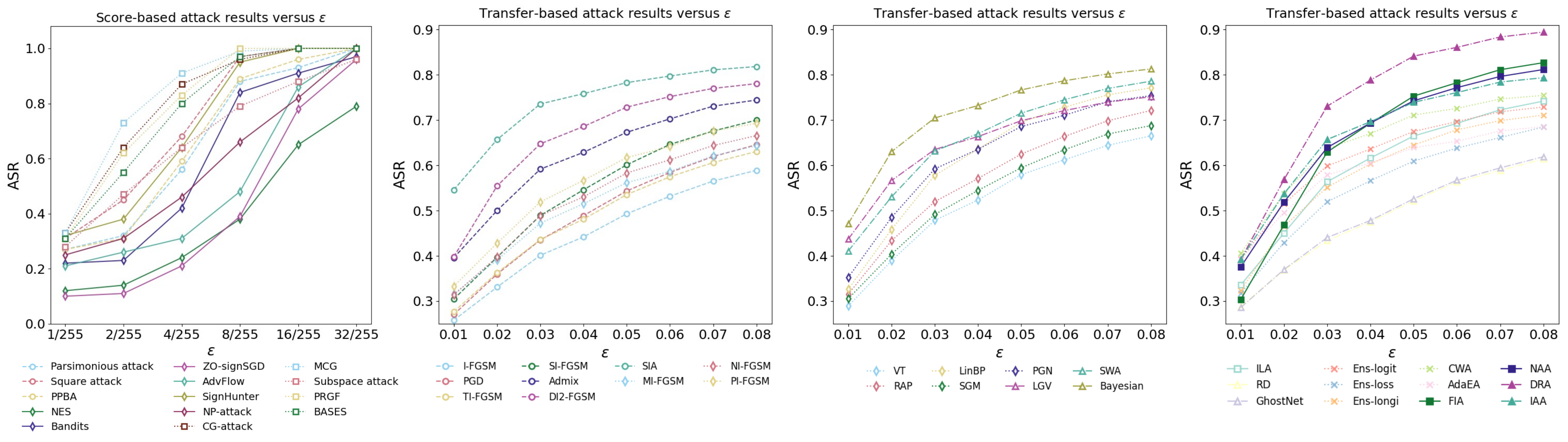}
 \caption{\textbf{Perturbation norm budget}. \blue{Attack results of score-based attacks (first subplot on the left) and transfer-based attacks (three subplots on the right) versus the perturbation norm budget $\epsilon$. Each subplot demonstrates how the attack performance varies with different values of $\epsilon$.}}\label{budget}
  \vspace{-5pt}
\end{figure*}

\begin{figure*}[t]
\vspace{-2mm}
 \setlength{\abovecaptionskip}{-0.1cm}
\setlength{\belowcaptionskip}{-0.1cm}
 \centering
\includegraphics[width=0.82\linewidth]{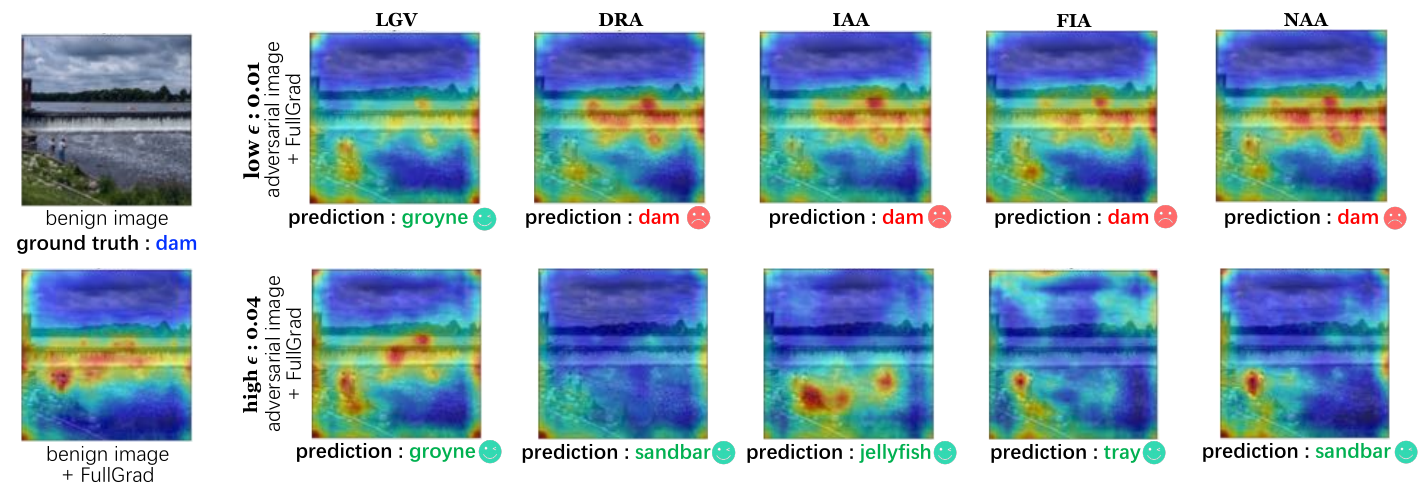}
 \caption{Comparison between adversarial examples generated by the regular methods (\textit{e.g.}, LGV) and exceptional methods (DRA, IAA, FIA, NAA). The leftmost are a benign image from "dam" category and its FullGrad. The other FullGrad images in the first and second row correspond to adversarial images generated by various methods with a low $\epsilon$ (0.01) and a high $\epsilon$ (0.04) respectively.}\label{norm_budget_grad_cam}
 \vspace{-14pt}
\end{figure*}

\subsection{Effect of Attack Budget}
\label{sec: effect of methods}

In this section, we take a closer look at the impact of the attack budgets. Fistly, we explore \textit{attacks with different perturbation norm budgets}. 
\blue{In transfer-based attacks, we claim that methods like FIA\cite{fia}, NAA\cite{naa}, DRA\cite{dra} and IAA\cite{iaa}, which aim to generate semantic adversarial perturbations rather than just increasing the classification loss with irregular noises, require a higher norm budget to be effective.}
\blue{In score-based attacks, We can observe that for combination-based methods, such as $\mathcal{CG}$-attack\cite{feng2022boosting}, BASES\cite{bases}, PRGF\cite{prgf}, and MCG\cite{yin2023generalizable}, they all achieve high ASR even with very small perturbation norms due to leveraging surrogate model information to assist in gradient estimation within their respective attack strategies.}
We also explore \textit{attacks with different query number budgets}.
\blue{In transfer-based attacks, it is find that performance of momentum-based methods like MI\cite{mifgsm}, NI\cite{nisifgsm}, PI\cite{pifgsm} and CWA\cite{cwa} deteriorates with additional rounds due to the overaccumulation of gradients.}
\blue{In decision-based attacks, CISA\cite{cisa} excels due to its adaptive sampling and dynamic step-size adjustment strategies. In score-based attacks, MCG\cite{yin2023generalizable}, PRGF\cite{prgf}, BASES\cite{bases} and $\mathcal{CG}$-attack \cite{feng2022boosting} perform outstandingly, mainly because they leverage surrogate models for gradient estimation, optimizing query efficiency.}

\subsubsection{Perturbation norm budget} \label{sec: Perturbation norm budget}
We first assess the sensitivity to perturbation norm budget $\epsilon$. The results are summarized in Fig. \ref{budget}.

\begin{figure*}[t]
\vspace{-2mm}
 \setlength{\abovecaptionskip}{-0.1cm}
\setlength{\belowcaptionskip}{-0.1cm}
 \centering
\includegraphics[width=0.78\linewidth]{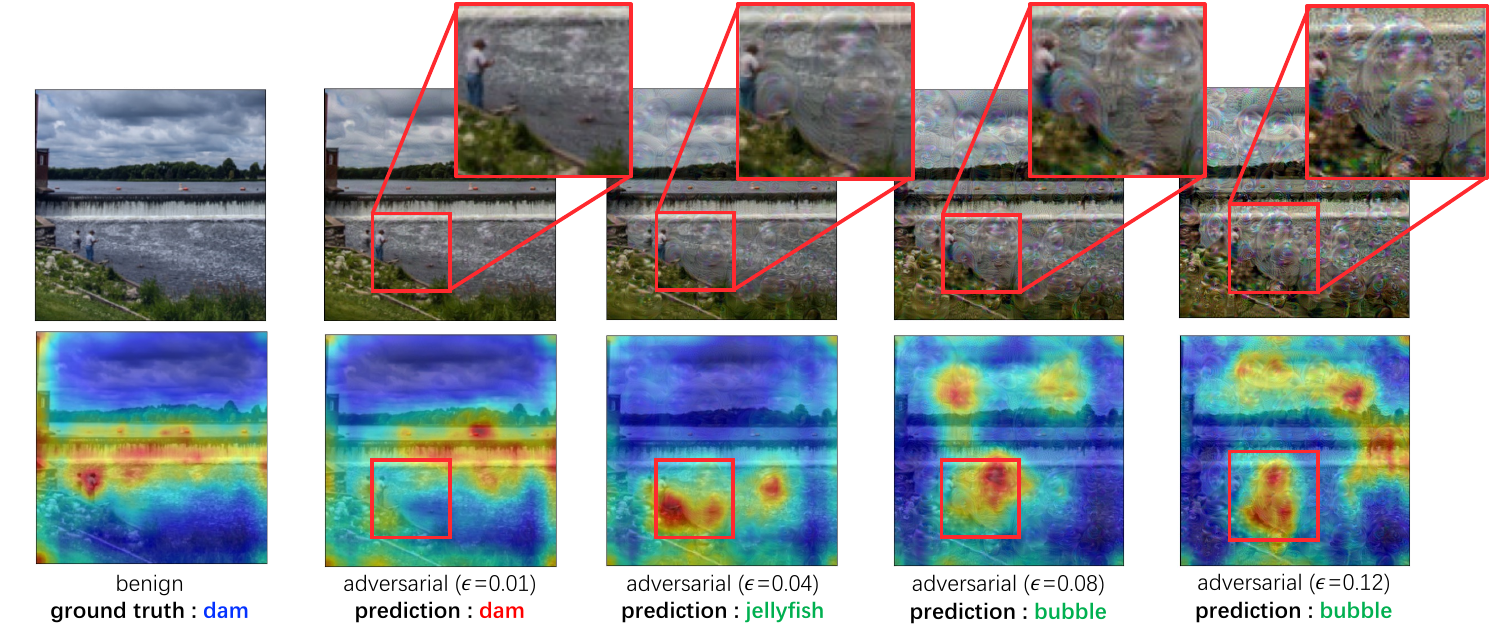}
 \caption{Adversarial images (\textbf{top row}) generated by IAA and target model attention maps (\textbf{bottom row}). Each column shows the evolution of adversarial images generated under increasing attack strengths. The red-boxed images amplify the square patches in the adversarial example. We set the maximum norm budget as 0.12 for visibility. IAA could semantically change the "dam" to "jellyfish" and finally to "bubble", but the "bubble"-like perturbations kick in only when they are obvious enough.}\label{increase_norm}
 \vspace{-5pt}
\end{figure*}

In transfer-based attacks, Fig. \ref{budget} ((three subplots on the right) shows a clear rise in ASR as the perturbation norm budget $\epsilon$ increases across all methods. Interestingly, we find the slope in ASR versus budgets remains fairly consistent across most methods, as evidenced by these parallel lines, indicating these attacks are equally sensitive to the budget. However, exceptions to this consistent pattern are observed in FIA\cite{fia}, NAA\cite{naa}, DRA\cite{dra} and IAA\cite{iaa}, as they show overwhelming attack performance with ample budgets. Unfortunately, it also drops more significantly than others in cases of low budgets. Recall that the motivations behind these methods, FIA and NAA both decrease instinct features and increase unimportant features, DRA and IAA both drag images away from the ground-truth data distribution. The common nature implied in these four attacks is to craft \textit{meaningful} noises that make the images \textit{do not "look like"} the ground-truth class. Unlike other attacks that greedily perturb the images to increase the classification loss and generate irregular noises, this nature necessitates a higher norm budget to maintain their superiority. 
This phenomenon could also be observed from a model attention perspective via our analytical tool, FullGrad, as visualized in Fig. \ref{norm_budget_grad_cam}. For instance, consider an image from the "dam" category. With a low norm budget, LGV easily, greedily misleads the target model into a really closed but different category "groyne". DRA, IAA, FIA, NAA fail to alter the features with this small budget, the target model still focuses on the dam's features and recognizes it correctly. However, when a higher norm budget is applied, these four methods do promote other trivial features. Accordingly, the target model shifts its focus away from the dam and directs its attention to other features.
We use IAA as an illustrative example to substantiate the hypothesis that \textit{the ability to generate semantic adversarial perturbations requires a relatively high norm budget} in Fig. \ref{increase_norm}. As the budget increases, the perturbation gradually contains more and more obvious features of the "bubble" category. Then this meaningful perturbation overshadows the original features and dominates the classification, as visualized in the attention maps. On the other hand, while this characteristic constrains the attack power in low $\epsilon$ cases, it could make attacks stand out when a high budget is available.

\begin{figure*}[t]
\vspace{-2mm}
 \setlength{\abovecaptionskip}{-0.1cm}
\setlength{\belowcaptionskip}{-0.1cm}
 \centering
\includegraphics[width=0.98\linewidth]{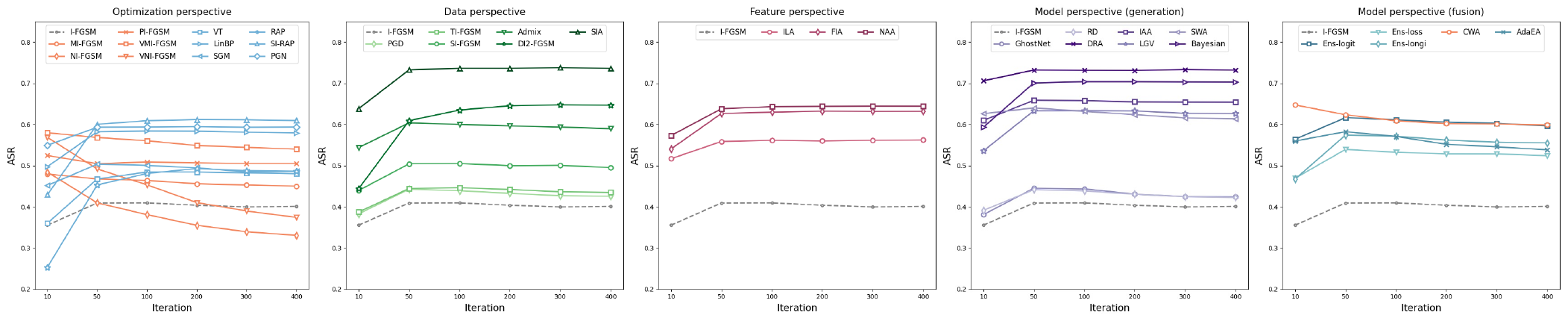}
 \caption{\textbf{Query number budget}. From left to right: attack results of transfer-based attacks from optimization, data, feature and model perspective methods with respect to the number of iteration.}\label{iter}
 \vspace{-13pt}
\end{figure*}

In query-based attacks, we only present the experimental results for score-based attacks. This is because the attack process for score-based attacks starts with a clean sample and introduces adversarial perturbations to create an adversarial sample that crosses the decision boundary of the ground-truth category within a specified $\ell_p$ norm perturbation ball. We believe that the choice of different sizes of perturbation balls will also affect the attack performance of different algorithms, which can be considered as one of the attack budgets for score-based attacks. For decision-based attacks, since the attack mechanism starts with an adversarial sample and gradually narrows the distance to the clean sample, the perturbation ball similar to the attack budget in score-based attacks is no longer applicable. For each method in Fig. \ref{budget} (first subplot on the left), ASR increases with the perturbation budget. For instance, Parsimonious Attack\cite{moon2019parsimonious} starts with an ASR of 0.27 at the smallest perturbation budget of 1/255 and reaches an ASR of 1.00 at a budget of 32/255. Similarly, the ASR for Square Attack\cite{andriushchenko2020square} increases from 0.31 to 1.00 as the perturbation budget grows. PPBA\cite{li2020projection}, NES\cite{ilyas2018black}, and Bandits\cite{ilyas2018prior} also show a trend of incrementally higher ASR with larger perturbation budgets. At the lowest budget of 1/255, the highest ASR achieved by any method is 0.31 for MCG\cite{yin2023generalizable}, and at the highest budget of 32/255, the lowest ASR is 0.79 for NES\cite{ilyas2018black}, while all other methods reach ASR of 1.00, indicating that expanding the attack budget indeed improves the performance of the attack methods, aligning with our intuitive judgment. \blue{It can be seen that $\mathcal{CG}$-attack\cite{feng2022boosting}, BASES\cite{bases}, PRGF\cite{prgf} and MCG\cite{yin2023generalizable} perform better than other methods across any attack budget, which once again proves that leveraging additional information from a surrogate model can bring significant improvements to the performance of score-based attacks. }Even with a lower attack budget, these four methods maintain a high ASR. Furthermore, methods like Square Attack\cite{andriushchenko2020square}, AdvFlow\cite{mohaghegh2020advflow}, \blue{Subspace Attack\cite{subspaceattack},} SignHunter\cite{al2020sign}, and PPBA\cite{li2020projection} also maintain good performance, with their unique attack mechanisms (for example, Square Attack\cite{andriushchenko2020square} explores the perturbation among its \(L_{\infty}\) vertices and AdvFlow\cite{mohaghegh2020advflow} introduces a flow model as the perturbation sampling model, \etc) providing references for the proposal of new methods.


\begin{figure*}
\centering
\begin{minipage}{0.45\linewidth}
\setlength{\abovecaptionskip}{-0.1cm}
\setlength{\belowcaptionskip}{-0.1cm}
\centering
\includegraphics[height=0.48\linewidth]{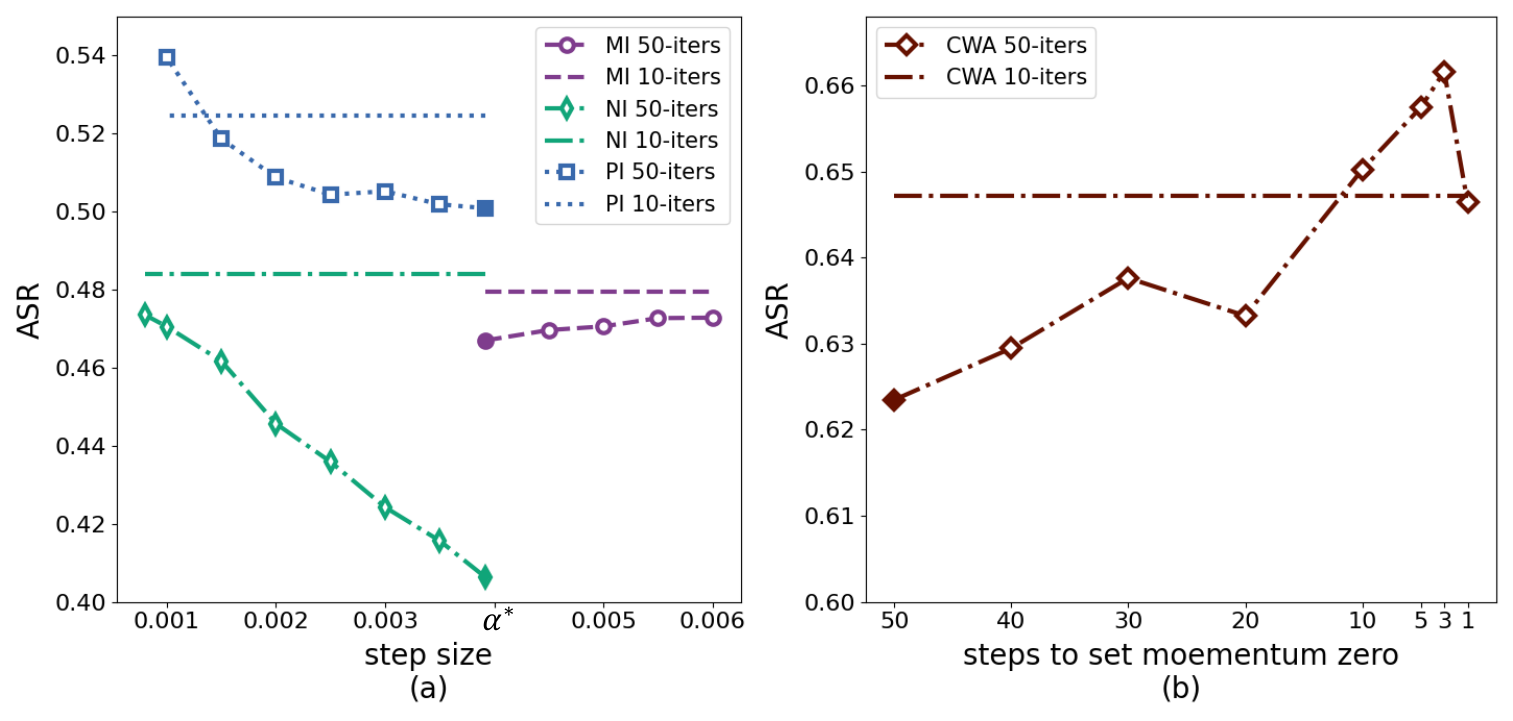}
\vspace{-4.5mm}
\caption{\blue{Momentum-based attacks with strategies to alleviate overaccumulation. \textbf{(a)} ASRs of MI, NI, PI. Lines w/ markers plot ASRs under 50 iterations (more than the optimal 10 iterations) versus step size, where the solid marker indicates ASR with default(unadapted) step size $\alpha^*$. Lines w/o markers show ASR under 10 iterations and step size $\alpha^*$.
\textbf{(b)} ASRs of CWA. Lines w/ markers plot ASRs under 50 iterations versus the frequency to set momentum zero, where the solid marker indicates ASR without setting zero. Lines w/o markers show ASR under 10 iterations without setting zero.}}\label{momentum}
\end{minipage}
\hspace{0.01\linewidth}
\begin{minipage}{0.50\linewidth}
\vspace{-1mm}
\setlength{\abovecaptionskip}{-0.1cm}
\setlength{\belowcaptionskip}{-0.1cm}
\includegraphics[height=0.58\linewidth]{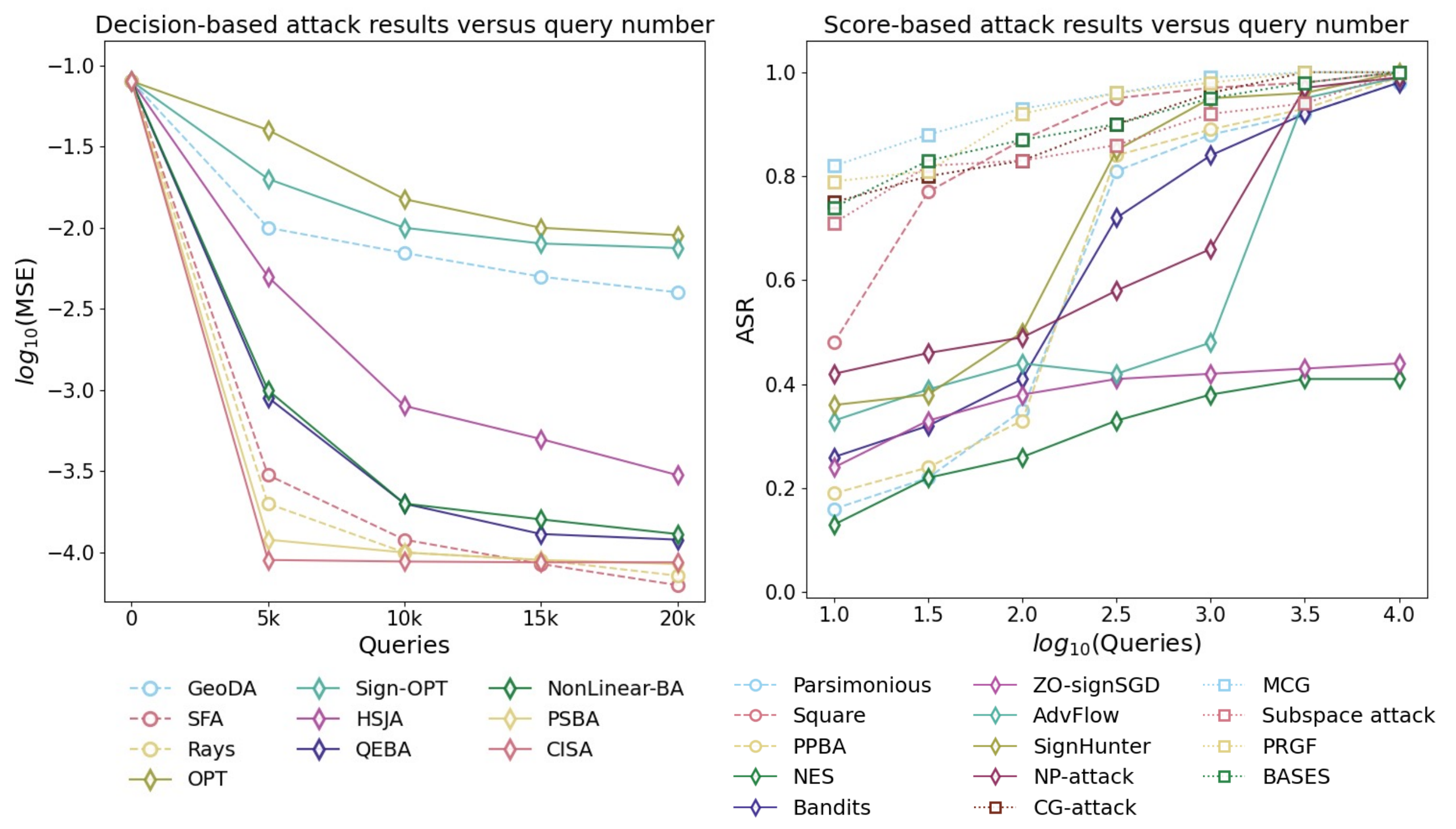}
\vspace{-3mm}
\caption{\textbf{Query number budget}. Attack results of decision-based methods (\textbf{left}) and score-based methods (\textbf{right}) versus the query number budget.}
\label{query number}
\end{minipage}
\vspace{-15pt}
\end{figure*}

\subsubsection{Query number budget}
\label{sec: Query number budget}
Then we assess methods' sensitivity to query number budget. 

In transfer-based attacks, the term \textit{query number budget} refers to the number of iterations of the white-box attack against the surrogate model. Due to the iterative nature of transfer-based attacks in crafting adversarial examples, it's therefore necessary to investigate the influence of the number of iterations on the efficacy of these methods. Fig. \ref{iter} visualizes the results of untargeted, $\ell_\infty$ attack on ImageNet, where all method shares a common maximum perturbation $\epsilon^*$ and step size $\alpha^*$. We see that basically optimal performance is attained with 50 iterations, and plateaus with further iterations for data, feature, model-perspective, and part of optimization-perspective methods. However, it is worth noting this stability does not extend to MI \cite{mifgsm}, NI \cite{nisifgsm}, PI \cite{pifgsm}, their corresponding composite attacks (VMI and VNI) \blue{and CWA\cite{cwa}}. They perform best with exactly 10 iterations as suggested in their original papers, however, deteriorate for additional rounds. 

\blue{Recall that the fundamental concept shared by MI, NI, PI, CWA is momentum-based gradient optimization. The momentum accumulates gradients, which can accelerate convergence and avoid local optima. However, excessive iterations may cause an overaccumulation of gradients, leading the update direction to deviate from the true optimal path. This can make the generation process of adversarial examples unstable, ultimately affecting the attack performance. Therefore, we propose to adapt the update step size of MI, NI and PI to alleviate the cumulative effect of the momentum term. In CWA, the update on adversarial examples depends on multiple step sizes, rendering this strategy uncontrollable. Alternatively we set the momentum term zero every several iterations. As shown in Fig. \ref{momentum}, even with a suboptimal 50 iterations, a gradual adaptation of step size or setting momentum term zero enable ASR to approach or surpass the optimal results achievable with 10 iterations. These findings validate that the overaccumulation induced by momentum might impede the convergency of attacks.}

For decision-based black-box attacks, we show the evaluation of Mean Squared Error (MSE) between adversarial examples and the clean examples. We vary the number of queries while ensuring a 100\% ASR. MSE metric represents the magnitude of adversarial perturbation. A smaller value indicates that the adversarial example is gradually approaching the clean example to satisfy the invisible constraint. Meanwhile, for score-based black-box attacks, we show how ASR changes with an increasing number of queries. We evaluate all decision-based black-box attacks and score-based black-box attacks on ResNet-50 using $\ell_{\infty}$ norm untargeted attacks, with the ImageNet dataset as the test set.

\blue{From the results in Fig. \ref{query number} (left), we can see the MSE of the CISA\cite{cisa} drops most significantly when the query number is still small. Additionally, PSBA\cite{zhang2021progressive}, NonLinear-BA\cite{li2021nonlinear} and QEBA\cite{li2020qeba}, which inherit the HSJA\cite{chen2020hopskipjumpattack} attack framework, also show promising attack performance. These three} methods exhibit similar attack behavior, with comparable rates of perturbation magnitude reduction as the number of queries increases. On the other hand, the attack performance of Sign-OPT\cite{cheng2019sign} and OPT\cite{cheng2018query} is less satisfactory, as the MSE remains high even after a significant number of queries. Furthermore, SFA\cite{chen2020boosting} and Rays\cite{chen2020rays} rapidly reduce the MSE to lower values at a relatively low query cost. 

\blue{For score-based attacks, we observe from Fig. \ref{query number} (right) that MCG\cite{yin2023generalizable}, PRGF\cite{prgf}, BASES\cite{bases} and $\mathcal{CG}$-attack\cite{feng2022boosting} significantly outperform other methods, achieving the highest ASR levels at the fastest rates.} This once again emphasizes the superiority of combining score-based attacks with surrogate models. 
Among the random search methods, the Square attack\cite{andriushchenko2020square} performs exceptionally well, achieving the highest ASR at the fastest rate.
Meanwhile, among the gradient-estimation-based methods, SignHunter\cite{al2020sign} is noteworthy as it quickly boosts the ASR to the highest level with minimal query cost. In fact, under untargeted attacks, we can see from the figure that almost all methods achieve a high ASR. It is worth noting that AdvFlow\cite{mohaghegh2020advflow}, although it eventually achieves a high ASR, incurs a significant cost as evidenced by its nearly vertical line, indicating a high number of queries for each clean example. 

\section{Conclusion}
\label{sec: conclusion}

BlackboxBench benchmark, featuring a modular, unified and extensible codebase, can contribute to validating truly effective ideas of black-box adversarial attacks. Our implementation encompasses {29} query-based attacks ({16} score-based and {13} query-based), along with {30} transfer-based attacks. These efforts resulted in {14,950} evaluations covering a wide range of attack scenarios. Our aim with BlackboxBench is to facilitate the research community, especially in rapidly comparing with existing literature during the design of new attacks, providing a grasp of the progress in the domain, and assessing defenses against adversarial threats. Furthermore, we demonstrated our benchmark is useful for analyzing various factors related to black-box adversarial vulnerability, such as data, model architecture, attack budget, and defense mechanisms. We believe the research community needs to gain a deeper understanding of how these factors impact attack effectiveness. In this endeavor, BlackboxBench can play a pivotal role.




 \vspace{-25pt}
\begin{IEEEbiography}[{\includegraphics[width=0.7in,clip,keepaspectratio]{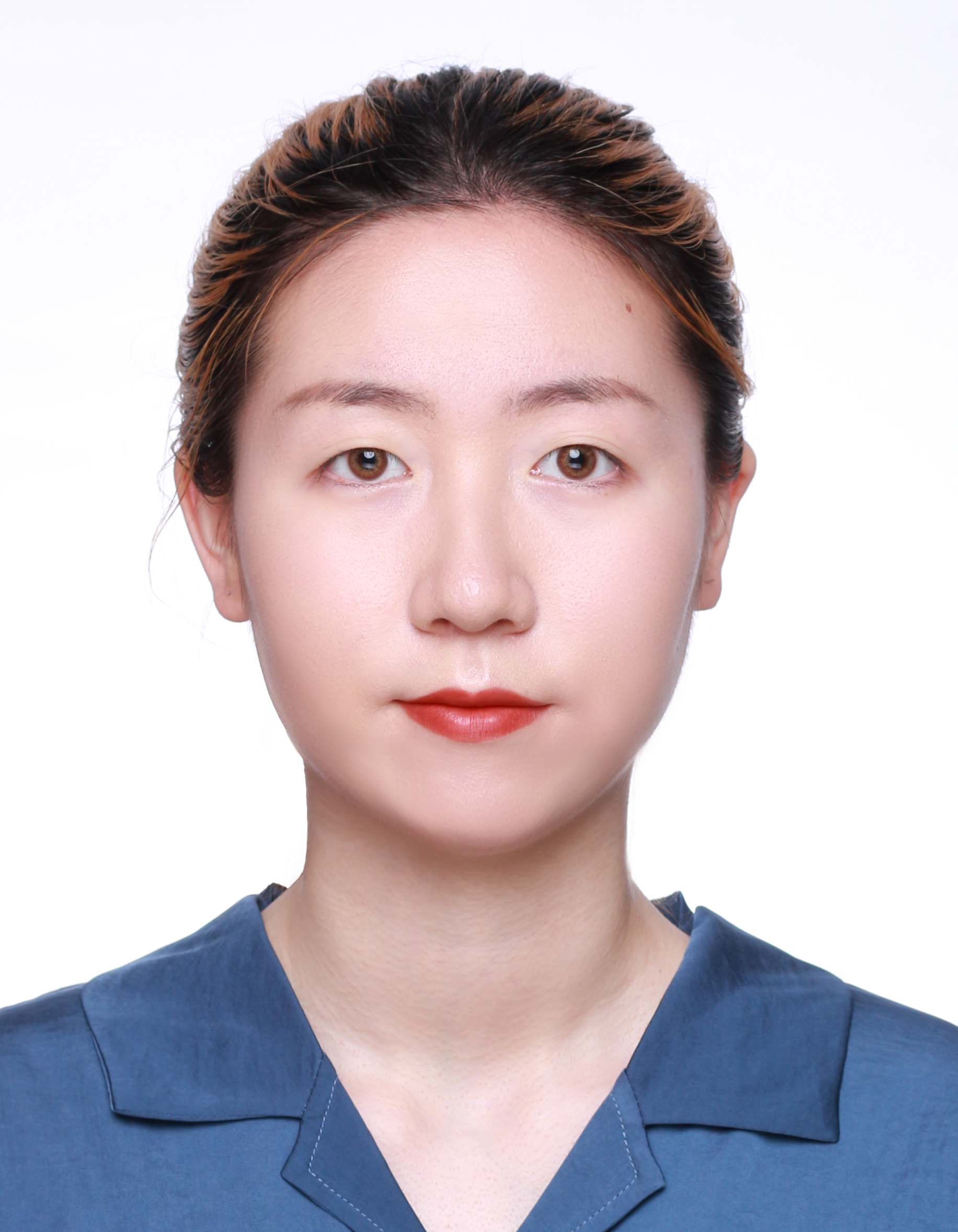}}]{Meixi Zheng}
received the bachelor’s degree and the master's degree from the Xidian University in 2019 and 2022. She is currently working toward the PhD degree with the School of Data Science, The Chinese University of Hong Kong, Shenzhen, supervised by Prof. Baoyuan Wu. Her research interests are adversarial machine learning.
\end{IEEEbiography}
\vspace{-38pt}
\begin{IEEEbiography}[{\includegraphics[width=0.7in,clip,keepaspectratio]{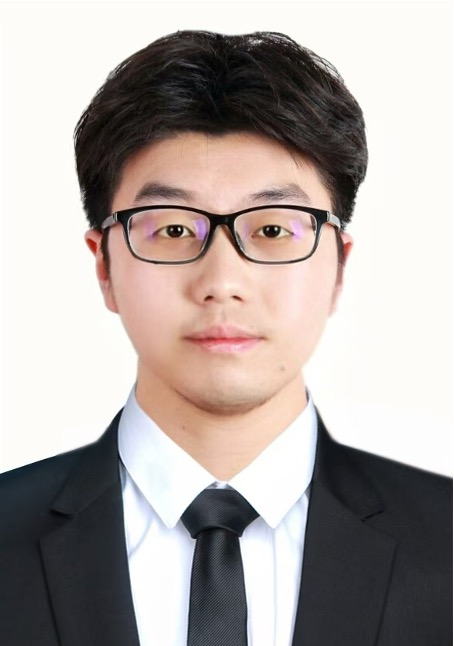}}]{Xuanchen Yan}
received his M.S. degree from The Chinese University of Hong Kong, Shenzhen, supervised by Prof. Baoyuan Wu, and his B.S. degree from Sichuan University. He is currently working at Huawei in Shanghai. His M.S. research focused on query-based black-box adversarial attacks and defenses. His research interests include the detection of copyright-protected AI-generated content (AIGC) and governance of AI data.
\end{IEEEbiography}
\vspace{-38pt}
\begin{IEEEbiography}[{\includegraphics[width=0.7in,clip,keepaspectratio]{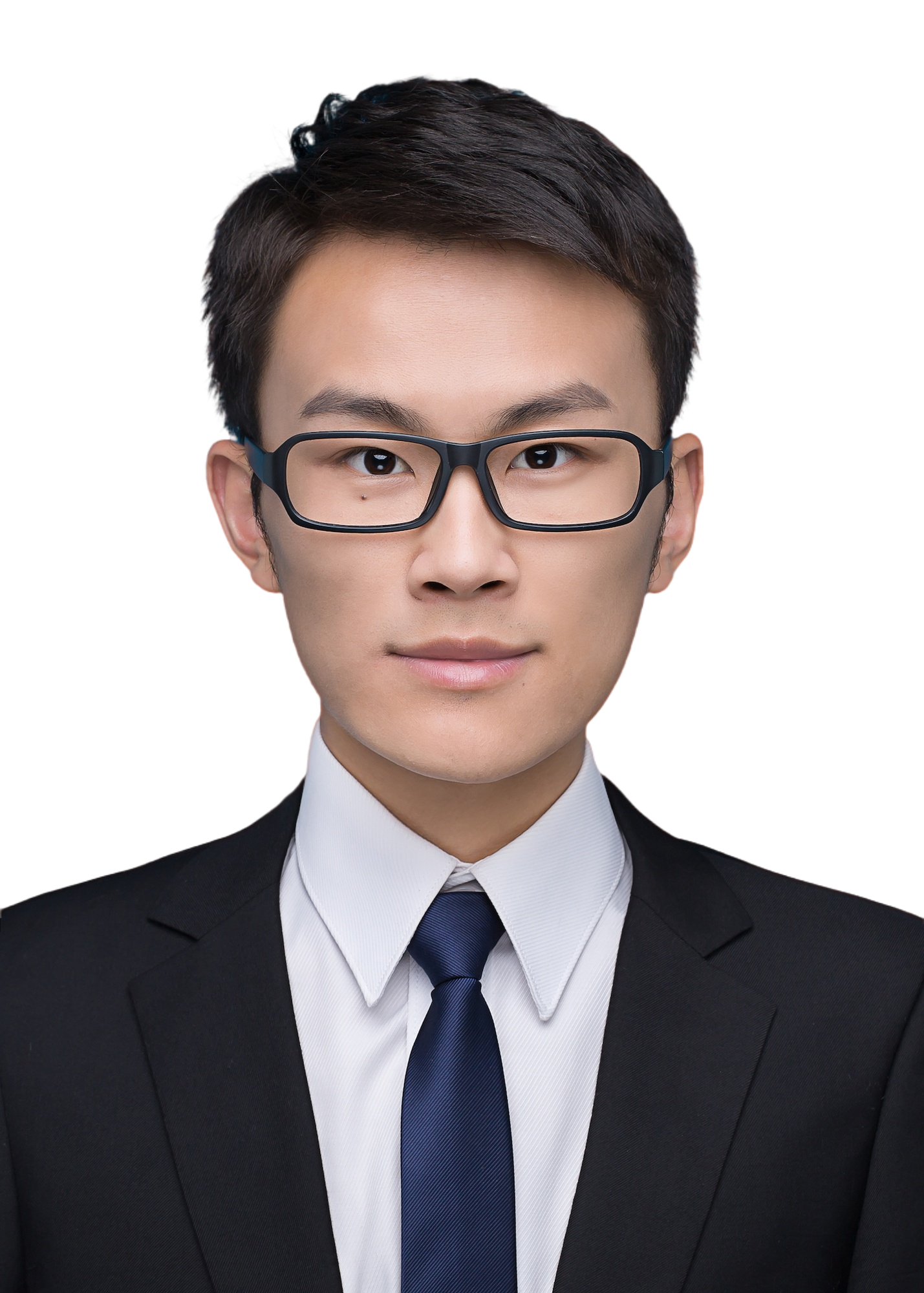}}]{Zihao Zhu}
received the bachelor’s degree from the China University of Mining and Technology in 2018 and the master's degree from the University of Chinese Academy of Sciences in 2021. He is currently working toward the PhD degree with the School of Data Science, The Chinese University of Hong Kong, Shenzhen, supervised by Prof. Baoyuan Wu. His research interests include trustworthy AI, backdoor learning.
\end{IEEEbiography}
\vspace{-38pt}
\begin{IEEEbiography}[{\includegraphics[width=0.8in,clip,keepaspectratio]{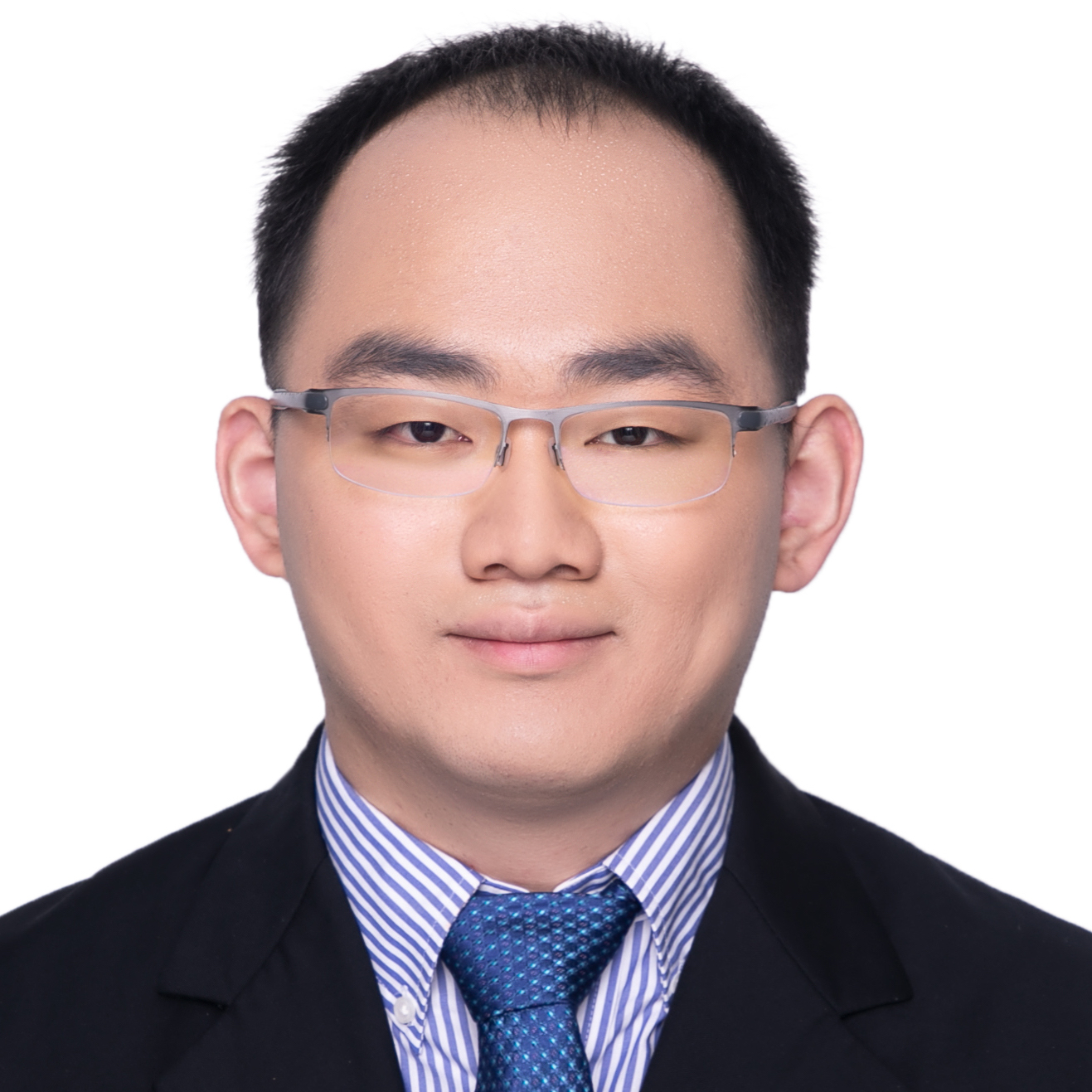}}]{Hongrui Chen}
received his master's degree from The Chinese University of Hong Kong, Shenzhen. He is a researcher at the Longgang District Key Laboratory of Intelligent Digital Economy Security (iDES) at The Chinese University of Hong Kong, Shenzhen, supervised by Prof. Baoyuan Wu. Currently, he is pursuing his PhD at The Chinese University of Hong Kong, Shenzhen. His research interests include backdoor learning and large language models.
\end{IEEEbiography}
\vspace{-38pt}

\begin{IEEEbiography}[{\includegraphics
[width=0.8in,height=1.25in,clip,
keepaspectratio]{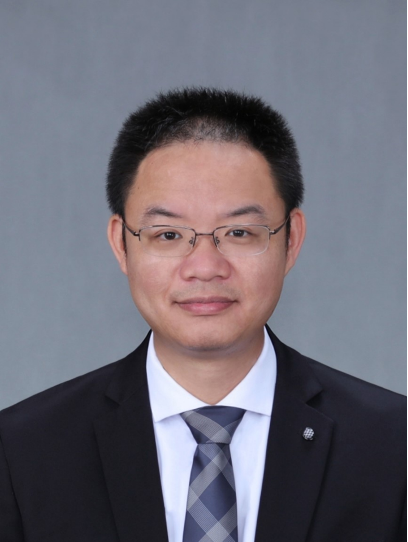}}]
{Baoyuan Wu} (Senior Member, IEEE) is a Tenured Associate Professor of the School of Data Science, Chinese University of Hong Kong, Shenzhen, Guangdong, 518172, P.R. China. His research interests are trustworthy and generative AI. He has published 100+ top-tier conference and journal papers. He is currently serving as an Associate Editor of IEEE Transactions on Information Forensics and Security and Neurocomputing, Organizing Chair of PRCV 2022, and Area Chairs of several top-tier conferences. He received the ``2023 Young Researcher Award” of The Chinese University of Hong Kong, Shenzhen.
\end{IEEEbiography}


\vfill

\clearpage

\twocolumn[
\begin{center}
    \Huge  {Supplementary Material for:\\[0.3em]BlackboxBench: A Comprehensive Benchmark of\\[0.3em] Black-box Adversarial Attacks}
\end{center}
\vspace{1cm}
]






\setcounter{section}{0}
\setcounter{equation}{0}
\numberwithin{equation}{section}
\setcounter{page}{1}
\long\def\comment#1{}
\renewcommand\thesection{\Alph{section}}

\begin{table*}[t]
  \centering
  \caption{\small{Comparison between BlackboxBench and other benchmarks.}}
    \begingroup
    \scalebox{0.8}{
    \begin{tabular}{cp{18.915em}p{12.915em}p{5.585em}p{5.585em}p{5.565em}p{11.5em}}
    \toprule
          & \multicolumn{1}{c}{\textbf{Description}} & \multicolumn{1}{c}{\textbf{Programming characteristic}} & \multicolumn{1}{c}{\textbf{|WA| / |BA|}}& \multicolumn{1}{c}{\textbf{Years}} & \multicolumn{1}{c}{\textbf{Leaderboard}} & \multicolumn{1}{c}{\textbf{Analysis \& tools}} \bigstrut\\
    \hline
    \hline
    \multirow{2}{*}{\textbf{CleverHans}\cite{papernot2016technical}} & A library provides standardized implementations of attacks and adversarial training. & All attacks are independent of each other. & \multicolumn{1}{c}{\multirow{2}{*}{12 / 1}} & \multicolumn{1}{c}{\multirow{2}{*}{2013-2020}} & \multicolumn{1}{c}{\multirow{2}{*}{\textbackslash{}}} & \multicolumn{1}{c}{\multirow{2}{*}{\textbackslash{}}} \bigstrut\\
    \hline
    \multirow{2}{*}{\textbf{FoolBox}\cite{rauber2017foolbox}} & A collection of gradient-based and decision-based adversarial attacks. & \multicolumn{1}{l}{Object-oriented programming.} & \multicolumn{1}{c}{\multirow{2}{*}{8 / 11}} & \multicolumn{1}{c}{\multirow{2}{*}{2015-2021}} & \multicolumn{1}{c}{\multirow{2}{*}{\textbackslash{}}} & \multicolumn{1}{c}{\multirow{2}{*}{\textbackslash{}}} \bigstrut\\
    \hline
    \multirow{3}{*}{\textbf{ART}\cite{nicolae2018adversarial}}   & ART provides tools to defend and evaluate ML models against the Evasion, Poisoning, Extraction, and Inference attacks. & \multicolumn{1}{l}{Object-oriented programming.} & \multicolumn{1}{c}{\multirow{3}{*}{26 / 10}} & \multicolumn{1}{c}{\multirow{3}{*}{2013-2022}}  & \multicolumn{1}{c}{\multirow{3}{*}{\textbackslash{}}} & \multicolumn{1}{c}{\multirow{3}{*}{\textbackslash{}}} \bigstrut\\
    \hline
    \multirow{3}{*}{\textbf{DEEPSEC}\cite{deepsec}} & DEEPSEC incorporates attacks, defenses, and multiple utility metrics. & All modules are independent of each other, but multiple modules can also work together. & \multicolumn{1}{c}{\multirow{3}{*}{0 / 16}} & \multicolumn{1}{c}{\multirow{3}{*}{2014-2018}}  & \multicolumn{1}{c}{\multirow{3}{*}{\textbackslash{}}} & Draw a set of findings from their systematically evaluations. \bigstrut\\
    \hline
    \multirow{2}{*}{\textbf{AdverTorch}\cite{ding2019advertorch}} 
    & AdverTorch contains modules for attacks and defenses, also scripts for robust training 
    & \multicolumn{1}{l}{Object-oriented programming.} 
    & \multicolumn{1}{c}{\multirow{2}{*}{18 / 6}}
    & \multicolumn{1}{c}{\multirow{2}{*}{2014-2020}}  
    & \multicolumn{1}{c}{\multirow{2}{*}{\textbackslash{}}}
    & \multicolumn{1}{c}{\multirow{2}{*}{\textbackslash{}}} \bigstrut\\
    \hline
    \multirow{3}{*}{\textbf{SecML}\cite{pintor2019secml}} & SecML implements: (i) evasion attacks and poisoning attacks; and (ii) explainable methods to understand attacks. & SecML has a modular architecture oriented to code reuse. Each module owns an interface. 
    & Attacks in CleverHans and Foolbox
    & \multicolumn{1}{c}{\multirow{3}{*}{\textbackslash{}}} 
    & \multicolumn{1}{c}{\multirow{3}{*}{\textbackslash{}}} & Provide four explainability methods from two perspective. \bigstrut\\
    \hline
    \multirow{3}{*}{\textbf{RealSafe}\cite{realsafe}} & RealSafe benchmarks the adversarial robustness of image classification and object detection models, and introduces robust training. & \multicolumn{1}{l}{Object-oriented programming.} & \multicolumn{1}{c}{\multirow{3}{*}{5 / 10}} & \multicolumn{1}{c}{\multirow{3}{*}{2015-2019}} & \multicolumn{1}{c}{\multirow{3}{*}{\textbackslash{}}} & Give some findings by analyzing the quantitative results. \bigstrut\\
    \hline
    \multirow{3}{*}{\textbf{AdvBox}\cite{goodman2020advbox}} & AdvBox supports black box attacks on Machine-Learning-as-a-service, as well as more attack scenarios, such as Face Recognition Attack. & \multicolumn{1}{l}{Object-oriented programming.} & \multicolumn{1}{c}{\multirow{3}{*}{8 / 2}} & \multicolumn{1}{c}{\multirow{3}{*}{2014-2017}} & \multicolumn{1}{c}{\multirow{3}{*}{\textbackslash{}}} & \multicolumn{1}{c}{\multirow{3}{*}{\textbackslash{}}} \bigstrut\\
    \hline
    \multirow{2}{*}{\textbf{DeepRobust}\cite{li2020deeprobust}} & DeepRobust contains attacks and defenses in image domain and graph domain. & \multicolumn{1}{l}{Object-oriented programming.} & \multicolumn{1}{c}{\multirow{2}{*}{8 / 2}} & \multicolumn{1}{c}{\multirow{2}{*}{2013-2019}} & \multicolumn{1}{c}{\multirow{2}{*}{\textbackslash{}}} & \multicolumn{1}{c}{\multirow{2}{*}{\textbackslash{}}} \bigstrut\\
    \hline
    \multirow{2}{*}{\textbf{TransferAttackEval}\cite{eval}} 
    & The first large-scale evaluation of transferable adversarial examples. 
    & All attacks are independent of each other. 
    & \multicolumn{1}{c}{\multirow{2}{*}{0 / 23}}
    & \multicolumn{1}{c}{\multirow{2}{*}{2018-2022}} 
    & \multicolumn{1}{c}{\multirow{2}{*}{\textbackslash{}}} & The evaluation leads to a number of new insights. \bigstrut\\
    \hline
    \multirow{2}{*}{\textbf{\blue{Torchattacks}}\cite{torchattacks}} 
    & \blue{A library of adversarial attacks that contains PyTorch-like interface and functions.} 
    & \blue{Object-oriented programming. }
    & \multicolumn{1}{c}{\multirow{2}{*}{\blue{0 / 30}}}
    & \multicolumn{1}{c}{\multirow{2}{*}{\blue{2014-2022}}} 
    & \multicolumn{1}{c}{\multirow{2}{*}{\textbackslash{}}} & \multicolumn{1}{c}{\multirow{2}{*}{\textbackslash{}}}  \bigstrut\\
    \hline
    \multirow{2}{*}{\textbf{\blue{TA-Bench}}\cite{tabench}} 
    & \blue{TA-Bench focuses on transfer-based attacks. It considers various substitute/victim models, finds a more advanced optimization back-end.}
    & \blue{Object-oriented programming}.  
    & \multicolumn{1}{c}{\multirow{2}{*}{\blue{0 / 33}}}
    & \multicolumn{1}{c}{\multirow{2}{*}{\blue{2017-2023}}} 
    & \multicolumn{1}{c}{\multirow{2}{*}{\textbackslash{}}} & \blue{Give some findings by analyzing the quantitative results.} \bigstrut\\
    \hline
    \rowcolor{mygray}
    \multirow{3}{*}{\textbf{BlackboxBench}} & A comprehensive benchmark of black-box adversarial attacks, including query-based attacks and transfer-based attacks. & Modular, unified procedural programming. & \multicolumn{1}{c}{\multirow{3}{*}{\blue{0 / 59}}} & \multicolumn{1}{c}{\multirow{3}{*}{2017-2023}} &\multirow{3}{*} {\href{https://blackboxbenchmark.github.io/}{\makecell[c]{BlackboxBench \\ LeaderBoard}}} & Analyze 15 questions and provide new insights via 10 analytical tools from 5 perspectives.\bigstrut\\
    \bottomrule
    \end{tabular}%
    }
    \endgroup
  \label{tab:difference}%
\end{table*}%

\section{Implemented Algorithms}
\label{sec:implemented algorithm}
\blue{
Despite the continuous emergence of black-box adversarial attack methods, we adhere to \textbf{two guidelines} when selecting the methods to be incorporated into BlackboxBench. \textit{Firstly}, from a chronological perspective, we select both traditional and the most recently proposed methods to illustrate the progression within this field. \textit{Secondly}, from a categorical perspective, we opt for a diversity of methods that span the numerous sub-domains of black-box adversarial attacks, facilitating our evaluation of the developmental trajectory of different technical routes.
}

\blue{For the selected methods,} we present an elaborate taxonomy of them in Fig. \ref{listmethods} in the main text. The first level of our taxonomy distinguishes three types of black-box adversarial attacks: \textit{decision-based query attacks}, \textit{score-based query attacks}, and \textit{transfer-based attacks}.
At the second level, we subdivide the existing literature within each type according to their technical routes. 
At the third level, we list the implemented attacks in our BlackboxBench.
Given the division of the first level introduced in Sec. \ref{Preliminary}, in this section, we introduce our categorization at the second level and the algorithms incorporated into BlackboxBench for each category at the third level. Regarding the method selection, we include classical techniques as baselines and advanced methods to underscore recent advancements in this field.

\subsubsection{Categorization of implemented decision-based attacks}
The primary aim of decision-based attacks is to devise effective strategies that render adversarial perturbations invisible. Existing methodologies can be categorized into two groups:
\begin{itemize}[leftmargin=15pt,itemsep=1pt,topsep=0pt]
    \item {\textbf{Random search methods}} These methods involve random direction updates and heuristic parameter adjustments. They include diverse strategies to acquire update directions, such as sampling from normal distributions, using evolutionary algorithms, gradient-free iterative techniques, and discrete optimization. (\textit{e.g.,} Boundary attack\cite{brendel2017decision}, Evolutionary attack\cite{dong2019efficient}, GeoDA\cite{rahmati2020geoda}, SFA\cite{chen2020boosting} and Rays\cite{chen2020rays}).
    \item {\textbf{Gradient-estimation-based methods}} These methods rely more on the information from the model feedback to guide the search direction. They transform decision-based attacks into optimization problems and incorporate gradient estimation, step size calculation, and binary searching as a framework. Subsequent methods enhance gradient estimation within this framework, and others introduce geometric patterns for improved adversarial detection. (\textit{e.g.,} OPT\cite{cheng2018query}, Sign-OPT\cite{cheng2019sign}, HSJA \cite{chen2020hopskipjumpattack}, QEBA\cite{li2020qeba}, NonLinear-BA\cite{li2021nonlinear}, PSBA\cite{zhang2021progressive}, Triangle attack\cite{wang2022triangle}) \blue{and CISA\cite{cisa}}.
\end{itemize}

\subsubsection{Categorization of implemented score-based attacks}
The objective of score-based attacks is to progressively move the adversarial sample closer to the decision boundary, allowing minimal input perturbations to mislead the model. Existing methods could be classified into three categories: 
\begin{itemize}[leftmargin=15pt,itemsep=1pt,topsep=0pt]
    \item {\textbf{Random search methods}} These methods use stochastic strategies (\textit{e.g.,}  sampling) to iteratively modify adversarial perturbations (\textit{e.g., } SimBA\cite{guo2019simple}, Parsimonious attack\cite{moon2019parsimonious}, Square attack\cite{andriushchenko2020square}, PPBA\cite{li2020projection} and BABIES\cite{tran2022exploiting}).
    \item \textbf{{Gradient-estimation-based methods}} These methods utilize model feedback or prior knowledge to approximate gradients for adversarial example creation (\textit{e.g.,} NES\cite{ilyas2018black}, Bandits\cite{ilyas2018prior}, ZO-signSGD\cite{liu2019signsgd}, SignHunter\cite{al2020sign}, AdvFlow\cite{mohaghegh2020advflow} and NP-attack\cite{bai2020improving}.
    \item  {\textbf{Combination-based methods}} These methods aim to enhance attack performance by integrating extra information from surrogate models with query feedback returned by the target model, to achieve fewer queries and higher attack success rate (\textit{e.g.,} \blue{Subspace attack\cite{subspaceattack}, PRGF\cite{prgf}, BASES\cite{bases}}, $\mathcal{CG}$-Attack\cite{feng2022boosting}, and MCG\cite{yin2023generalizable}).
\end{itemize}
As a result, we implemented \blue{29} query-based black-box attack methods in total, covering two types of query feedback. 
Every method is briefly described in \textbf{Supplementary} \ref{app: description of score-based attack algorithms} and \ref{app: description of decision-based attack algorithms}.

\subsubsection{Categorization of implemented transfer-based attacks}

Amongst the earliest transfer-based attacks was the Iterative Fast Gradient Sign Method (I-FGSM)\cite{ifgsm}, which optimizes Eq. (\ref{equ: white}) iteratively, serving as a stronger adversary for white-box models compared to the single-step attack, \ie, Fast Gradient Sign Method (FGSM)\cite{goodfellow2014explaining}. However, iterative attacks tend to overfit the surrogate model, causing poor transferability across models\blue{\cite{difgsm}}. \textit{So can we achieve better trade-off between attack ability and transferability?}
\blue{
Let $\mathcal{T}\left(\cdot\right)$ be the transformation function, $\mathcal{M}$ be the fusion operator, and $\textbf{I}$ be some auxiliary information (specified later). $f^{\prime\prime}(x;\theta^{\prime\prime}) \in \mathcal{F}$ indicates one refined surrogate model belonging to the surrogate ensemble $\mathcal{F}$, parameterized by $\theta^{\prime\prime}$.
}According to the general formulation:
\blue{\begin{align}\label{equ: general}
      & \arg \max _{\boldsymbol{x}^*}\mathcal{M}_{f^{\prime\prime} \in \mathcal{F}}(\mathcal{L}\left(f^{\prime\prime}\left(\mathcal{T}\left(\boldsymbol{x}^*\right);\theta^{\prime\prime}\right),y;\textbf{I}\right)), \\
      &  \text {s.t. }\left\|\boldsymbol{x}^*-\boldsymbol{x}\right\|_{p} \leq \epsilon, \nonumber
  \end{align}}
various efforts have been made to enhance the transferability of iterative attacks on the top of I-FGSM, stemming from four distinct perspectives:
\begin{itemize}[leftmargin=15pt,itemsep=1pt,topsep=0pt]
\item {\textbf{Data perspective}} Methods from this perspective use input transformation $\mathcal{T}\left(\cdot\right)$ on the benign sample to alleviate the overfitting to the surrogate model (\textit{e.g.,} PGD\cite{pgd}, TI-FGSM\cite{tifgsm},  SI-FGSM\cite{nisifgsm}, Admix\cite{admix}, DI2-FGSM\cite{difgsm}, SIA\cite{sia}).
\item {\textbf{Optimization perspective}} Methods from this perspective introduce gradient-based optimization algorithms to escape from poor local maxima or modify the gradients to facilitate transferability (\textit{e.g.,} MI-FGSM\cite{mifgsm}, NI-FGSM\cite{nisifgsm}, PI-FGSM\cite{pifgsm}, VT\cite{vt}, RAP\cite{rap}, LinBP\cite{linbp}, SGM\cite{sgm}, PGN\cite{pgn}).
\item {\textbf{Feature perspective}} Methods from this perspective distort intermediate layer features by designing a new adversarial loss function $\mathcal{L}\left(\cdot; \textbf{I}\right)$ with the aid of auxiliary information $\textbf{I}$, \blue{such as the adversarial direction guide in ILA\cite{ila}, the aggregate gradient serving as feature importance in FIA \cite{fia}, and the integrated attention serving as neuron attribution in NAA\cite{naa}.}
\blue{\item
{\textbf{Model perspective}} One main category focuses on model tuning: by refining the base surrogate model $f^{\prime}$ to a better one $f^{\prime\prime}$ (\textit{e.g.}, DRA\cite{dra}, IAA\cite{iaa}, SWA\cite{bayesian}), or generating a set $\mathcal{F}$ of diverse variants from the base surrogate model (\textit{e.g.}, RD\cite{lgv}, GhostNet\cite{ghostnet}, LGV\cite{lgv}, Bayesian attack\cite{bayesian}), this category minimizes the misalignment between surrogate and target, thereby improving transferability. Another main category focuses on the fusion strategy modeled by $\mathcal{M}$: it reconciles the different gradients of an ensemble of surrogate models to capture the intrinsic transfer information (\textit{e.g.}, Logit ensemble\cite{mifgsm}, Loss ensemble\cite{mifgsm}, Longitudinal ensemble\cite{ghostnet}, CWA\cite{cwa}, AdaEA\cite{adaea}).}
\end{itemize}
As shown in Fig. \ref{listmethods} of the main text, apart from I-FGSM, BlackboxBench incorporates \blue{29} additional transfer-based attacks, categorized into four groups. More detailed descriptions of transfer-based attacks can be found in \textbf{Supplementary} \ref{app: description of transfer-based attack algorithms}. \blue{In addition to aforementioned individual transfer attacks, BlackboxBench supports the creation of customized composite attack pipelines through the flexible combination of existing functional blocks, each representing the core of one attack. BlackboxBench officially designs 98 pairwise composite attacks as examples, which will be discussed in Sec. \ref{sec: result overview}.}

\subsubsection{Categorization of implemented defenses}
\label{sec: Categorization of implemented defenses}

Existing defenses aiming at mitigating adversarial examples can be broadly categorized into\cite{wu2023defenses}:
\begin{itemize}[leftmargin=15pt,itemsep=1pt,topsep=0pt]
\item {\textbf{Training-time approaches}} The most prominent training-time defense is Adversarial Training (AT) \cite{rebuffi2021fixing, gowal2020uncovering, wong2020fast, advwrn, advresnet, advconvswin}, which is a defense strategy designed to bolster the robustness of ML models against adversarial attacks by incorporating adversarial examples during training. While this method enhances robustness, it demands more computational resources and may compromise accuracy on clean data. AT is deployed against both query-based attacks and transfer-based attacks in BlackboxBench.
\item {\textbf{Inference-time approaches}} Compared to traditional AT, defense deployed at inference stage (\textit{e.g.}, RND \cite{qin2021random} and AAA \cite{chen2022adversarial}) are more lightweight and computationally efficient. Furthermore, the flexibility of inference-time approaches also allows them to be combined with other defense strategies, such as adversarial training, to harness the strengths of both. RND and AAA are mainly against query-based attacks in BlackboxBench. More descriptions of AT, RND, and AAA can be found in \textbf{Supplementary} \ref{app: description of black-box adversarial defenses}.

\end{itemize}


\begin{table*}[t]\color{black}
\centering 
\caption{The ten analytical tools provided in the analysis module, organized into four functional categories. Note that Attack-component effect visualization category includes five distinct plot tools. For each tool, we describe its analytical purpose and indicate the specific key questions to which it is applied in the analysis of black-box adversarial vulnerability. The tool abbreviations used here (\eg, AEV, MAV-A, DBV-A, etc.) correspond to those listed in the ``Used tool'' column of Tab. \ref{tab:questions}.}
\label{tab:tools}
    \begingroup
    \scalebox{0.79}{
\begin{tabular}{@{}cp{10em}p{22em}p{28em}@{}}
\toprule
\textbf{Functional category} & \multicolumn{1}{c}{\textbf{Analytical tool}} & \multicolumn{1}{c}{\textbf{Functionality}} & \multicolumn{1}{c}{\textbf{Application to key question in Tab. \ref{tab:questions}}} \\     
    \hline
    \hline
\multirow{4}{*}{\makecell[c]{Attack-component effect visualization\\\textbf{(AEV)}}} & Line charts, scatter charts, bar charts, bubble charts, heatmaps \centering & To systematically evaluate the effect of individual attack components (method, setting, model, dataset) by fixing the others, thus enabling visualizing underlying patterns using multi-format plots. & Applied to 14 out of 15 key questions, excluding the attack procedure analysis. \bigstrut \\
 \midrule
\multirow{8}{*}{\makecell[c]{Model attention visualization\\\textbf{(MAV)}}} & \textbf{A.} Full-Gradient Saliency Maps (FullGrad) \centering & To visualize model attention in the spatial domain by highlighting pixel-level discriminative regions, enabling revealing how adversarial examples shift the model's focus in the spatial domain. & \multirow{4}{*}{\makecell[l]{\textbf{Attack budget-Q1:} How about the methods’ reliance on the pertu-\\rbation norm budget?\\ \textbf{Attack procedure-Q1:} How does the attacker traverse from the be-\\nign example to itsperturbed counterpart?}}
\\ \cmidrule(l){2-4} 
 & \textbf{B.} Frequency Saliency Map (FSM) \centering & To visualize model attention in the frequency domain by highlighting frequency-level discriminative components, enabling revealing how adversarial examples shift the model's focus in the frequency domain. & \textbf{Defense-Q3:} Are adversarial images transferable between normally trained models and adversarially trained models? \\ \midrule
\multirow{4}{*}{\makecell[c]{Adversarial divergence measurement\\\textbf{(ADM)}}} & Conditional Adversarial Distribution (CAD) modeling \centering \& Adversarial Divergence metric & To measure the distributional divergence of perturbations between models, providing a quantitative view of adversarial transferability. & \textbf{Model architecture-Q2:} What kind of misalignment between surrogate models and target models then undermines transferability? \\ \midrule
\multirow{8}{*}{\makecell[c]{Decision boundary visualization\\\textbf{(DBV)}}} & \textbf{A.} Decision Surface \centering & To visualize the proximity of benign images to the decision boundary, thus helping analyze the individual vulnerability to adversarial attacks. & \textbf{Data-Q3:} How does the dimensionality the adversary restricted to affect its performance? \\ \cmidrule(l){2-4} 
 & \textbf{B.} dbViz \centering & To visualize the structure of decision boundaries around benign examples on the data manifold, thus helping understand the boundary complexity.\textcolor{white}{1 1 1 1 1 1 1 1 1 111111111111} & \multirow{3}{*}{\makecell[l]{\textbf{Data-Q2:} Whether there are samples harder to be attacked by all\\attack methods?\\\textbf{Model architecture-Q1:} Which model architecture is most resistant\\to malicious queries?}} \\ \bottomrule
\end{tabular}%
}
    \endgroup
\end{table*}

\section{Codebase}
\label{app: other modules}

\subsubsection{Analysis module}
\label{subsubsec: analysis module}

\bluetwo{
To facilitate our understanding of black-box adversarial vulnerability, we introduce ten analytical tools grouped into four categories: attack component effect visualization, model attention visualization, adversarial divergence measurement, and decision boundary visualization. In the following, we elaborate the types of analyses we aim to conduct, the corresponding tools employed, and the underlying principles behind these tools. Tab. \ref{tab:tools} summarizes these tools at a high level, while technical details provided in \textbf{Supplementary} \ref{app: description of analytical tools}.
}

{\textbf{Attack-component effect visualization tools}} An adversarial attack scenario comprises four fundamental components: the method, attack setting, target model, and dataset. Nevertheless, the influence of each of these components remains unclear. To unravel the effect of each component, we fix other components, expand the one's scope, and summarize the results via five plot tools (line charts, scatter charts, bar charts, bubble charts and heatmaps) to explore the underlying patterns. 

{\textbf{Model attention visualization tools}} As adversarial examples are deliberately designed to deceive classification models, comprehending how proposed methods defocus the model in the pixel domain holds substantial importance. We achieve this by leveraging Full-Gradient Saliency Maps (FullGrad)\cite{fullgrad} to visualize the discriminative regions that models capture. Additionally, as the Frequency Saliency Map (FSM)\cite{backdoorbench} allows us to visually interpret the contributions of frequency components to classification, we also adopt FSM to gain insights into model behavior from a frequency perspective.

{\textbf{Adversarial divergence measurement tools}} BlackboxBench aims to qualitatively interpret the adversarial transferability and establish a quantitative metric for evaluating transferability between white-box surrogate models and the target model. A high transferability implies that the perturbation crafted by the surrogate resembles that crafted by the target. To gain this resemblance, we adopt the Conditional Adversarial Distribution (CAD) modeling technique\cite{feng2022boosting}, which models the distribution of perturbations conditioned on clean examples. Based on the CAD, we design a transferability evaluation metric---Adversarial Divergence---to quantify the difference between models from an adversarial view by calculating the normalized Kullback-Leibler divergence between two CADs.

{\textbf{Decision boundary visualization tools}} To visualize the proximity of benign images to the decision boundary, we employ Decision Surface\cite{decisionsurface}. We also utilize dbViz\cite{somepalli2022can} to visualize the information of decision boundary around benign images, aiding in understanding the difficulty for attackers trying to misclassify a benign image into wrong categories. 

\subsection{Defense module}
While our primary focus when building the BlackboxBench lies in implementing black-box attack algorithms, we also offer defense options for users to employ attacks with defense strategies. By integrating this module into the evaluation process, we aim to create a holistic understanding of both the robustness of defensive measures and the potency of attack strategies. This module is designed to rigorously assess the effectiveness of black-box attack algorithms when targeting neural network models equipped with defensive methods. We provide both static defense and dynamic defense in BlackboxBench. As the involvement of AT defenses is accomplished through the utilization of adversarially trained target models, our defense module exclusively provides three dynamic defense methods: RND\cite{qin2021random}, AAA-Linear\cite{chen2022adversarial}, and AAA-Sine\cite{chen2022adversarial}. Our defense module offers a standardized testing environment, ensuring that evaluations are both consistent and replicable.

\subsection{Evaluation module}
The evaluation module includes three crucial metrics: average query number (AQN), median query number (MQN), and attack success rate (ASR). AQN signifies the average query number over all images made during successful query-based black-box attacks. Similarly, MQN captures the middle value among all query numbers of successfully crafted adversarial examples. ASR measures the percent of adversarial samples classified as a different class in untargeted attacks or the target class in targeted attacks. For AQN and MQN, a lower value indicates a  better attack performance, while for ASR, a higher value indicates a  better attack performance.

\section{Additional Evaluation and Analysis}



\subsection{Result overview}
\label{add: untargeted vs. targeted}
\begin{itemize}[leftmargin=10pt,itemsep=1pt,topsep=0pt]
    \item \textit{Which task is more challenging, targeted or untargeted?}
    Targeted attacks do pose a greater challenge than untargeted ones. In query-based attacks, regarding attacks with different targets, we can see that targeted query-based attacks require more query costs than untargeted attacks while exhibiting a relatively lower ASR. Similarly, achieving transferability in targeted attacks proves more challenging than in untargeted attacks. Despite our efforts to conduct evaluations with additional iterations and a logit loss following \cite{targeted}, along with a higher norm budget $\epsilon$ to boost targeted attack transferability, existing literature still shows its inability to fool models into predicting a targeted class, even though they perform well in untargeted attacks. Nevertheless, methods from the model-tuning perspective dramatically narrow the gap between targeted and untargeted tasks, illuminating a promising research direction to overcome the low transferability in targeted attack scenarios.
\end{itemize}

\subsection{Effect of Data}
\label{add: effect of data}

\subsubsection{Fluctuation among samples}
\label{add: Fluctuation among samples}
\begin{figure*}
    \centering
    \includegraphics[width=0.98\linewidth]{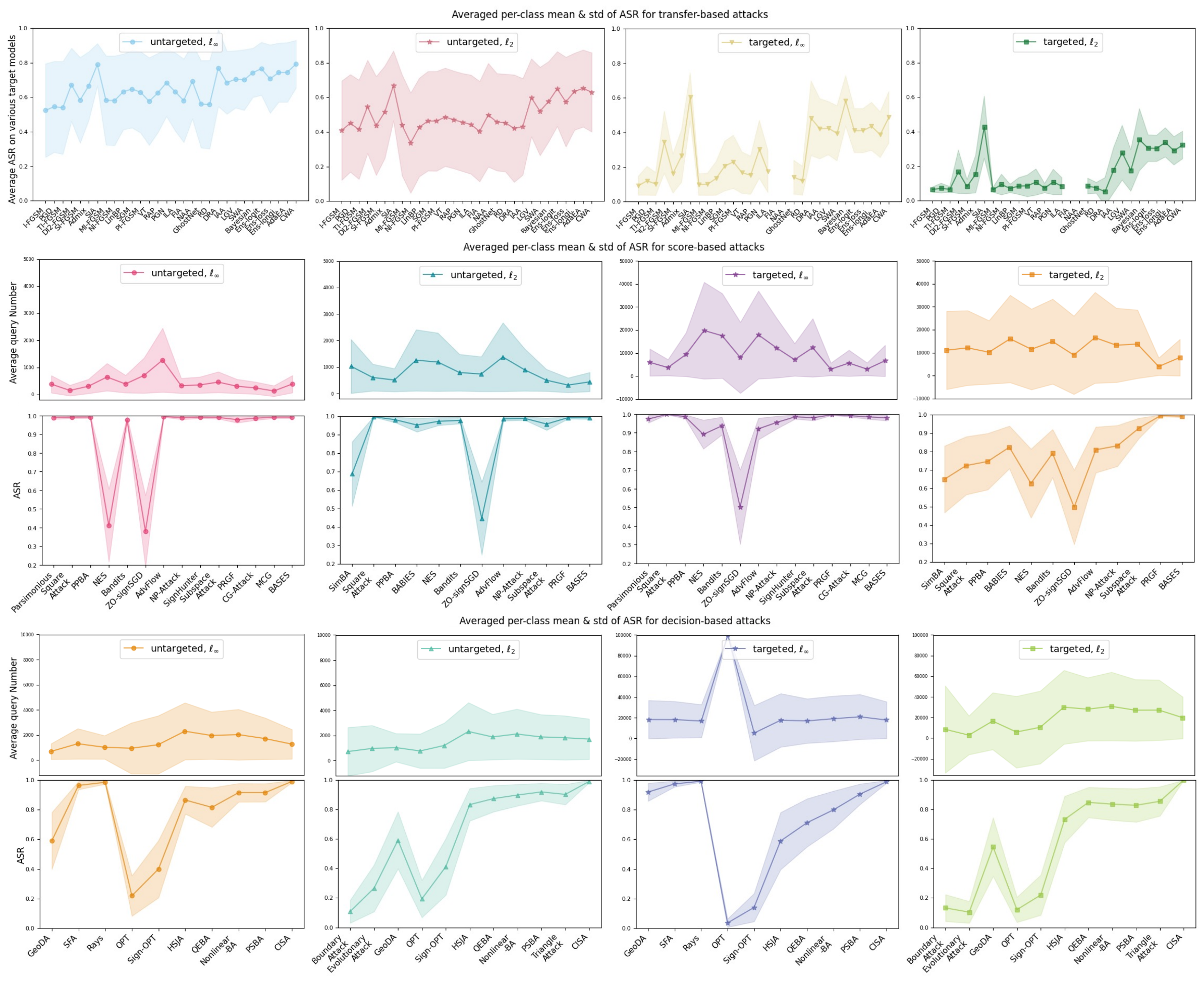}
    \caption{\blue{\textbf{Generalizability of evaluation results.}
    Let random variable $X_c \in \mathcal{X}$ be the input image from class $c$ with $\mathcal{X} \subset \mathbb{R}^d$, and $x_c^i$ the realizations. Let random variable $A_c$ be the attack result on $X_c$, and $a_c^i$ the realizations. $A_c \subset \mathbb{Z}_0^+$ takes values in $[0,1]$ and $[0, M]$ when the attack result represents ASR and AQN, respectively, where $M$ is the maximum query number.
    This figure shows the averaged mean $\pm$ std of sample-level attack results over 5 random images sampled within one class. Here, averaged mean is defined as $\bar{\mu}=\sum_{c=1}^C \hat{\mu}_c/C$, averaged std is defined as $\bar{\sigma}=\sum_{c=1}^C \hat{\sigma}_c/C$, where $\hat{\mu}_c=\sum_{i=1}^n a_c^i/n$, $\hat{\sigma}_c=\sqrt{\sum_{i=1}^n\left(a_c^i-\hat{\mu}_c\right)^2/(n-1)}$ represent the mean and std of sample-level attack results within class $c$, respectively. In our trials, $n=5, C=1000$.}}
    \label{fig:all_perclass_std}
\end{figure*}

\blue{We believe that the observed stability in attack results considered on the entire dataset does not necessarily imply a low variation in attack results of single samples within a class\footnote{The variance of dataset-level attack result is $Var(\sum_{c=1}^C A_c/C)=\sum_{c=1}^C Var(A_c)/C^2$. Thus, the variance of dataset-level attack result is approximately $C$ times smaller than that of sample-level. Given that $A_c \in [0,N]$ (e.g., ASR of $X_c$ $\in \{0,1\}$), the maximum of $Var(A_c)$ is $N^2/4$. Thus the maximum variance of dataset-level attack result is $\sqrt{N^2/4C}$.}. In contrast, Fig. \ref{fig:all_perclass_std}, which plots the mean $\pm$ std of sample-level attack results averaged across classes, indicates a large fluctuation in attack results during the switch of samples. }

\begin{figure*}[t]
\vspace{-2mm}
 \setlength{\abovecaptionskip}{-0.1cm}
\setlength{\belowcaptionskip}{-0.1cm}
 \centering
\includegraphics[width=0.9\linewidth]{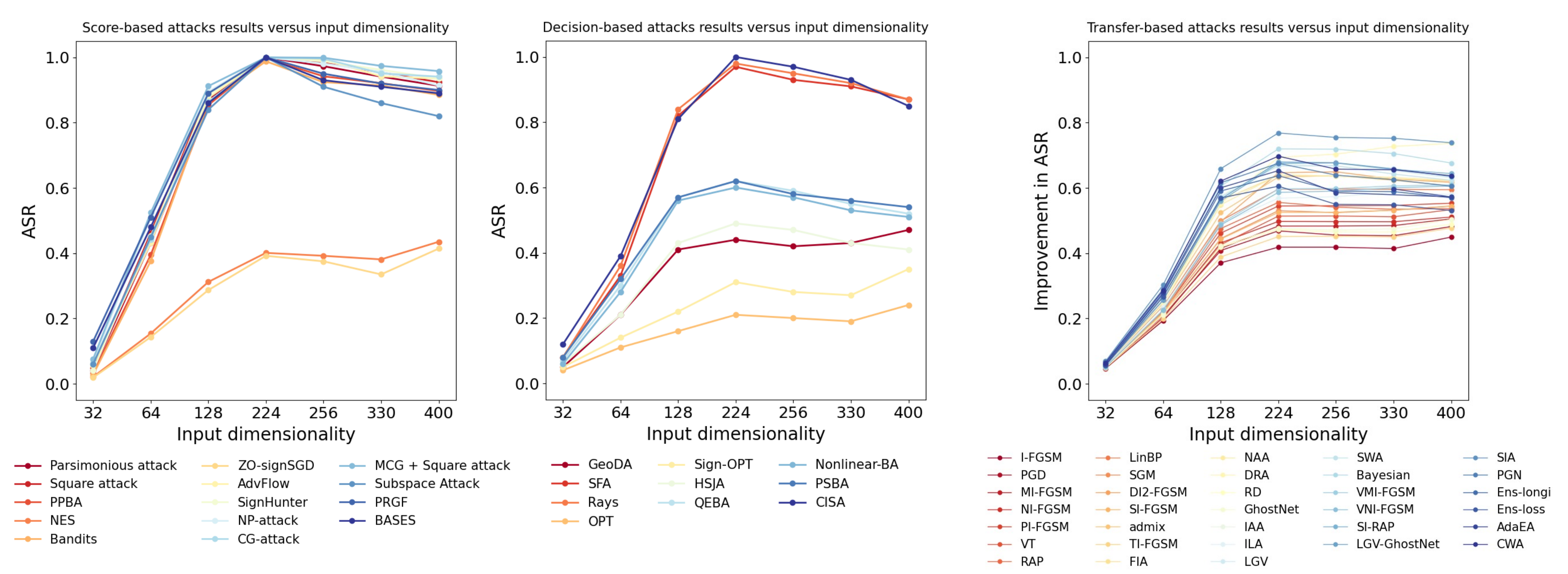}
 \caption{\textbf{Input dimensionality}. The ASR versus input dimensionality of score-based (\textbf{left}) and decision-based (\textbf{middle}). The improvement in ASR versus input dimensionality of  transfer-based attacks (\textbf{right}).}\label{imagesize}
\end{figure*}

\begin{figure*}[t]
\vspace{-2mm}
 \setlength{\abovecaptionskip}{-0.1cm}
\setlength{\belowcaptionskip}{-0.1cm}
 \centering
\includegraphics[width=0.8\linewidth]{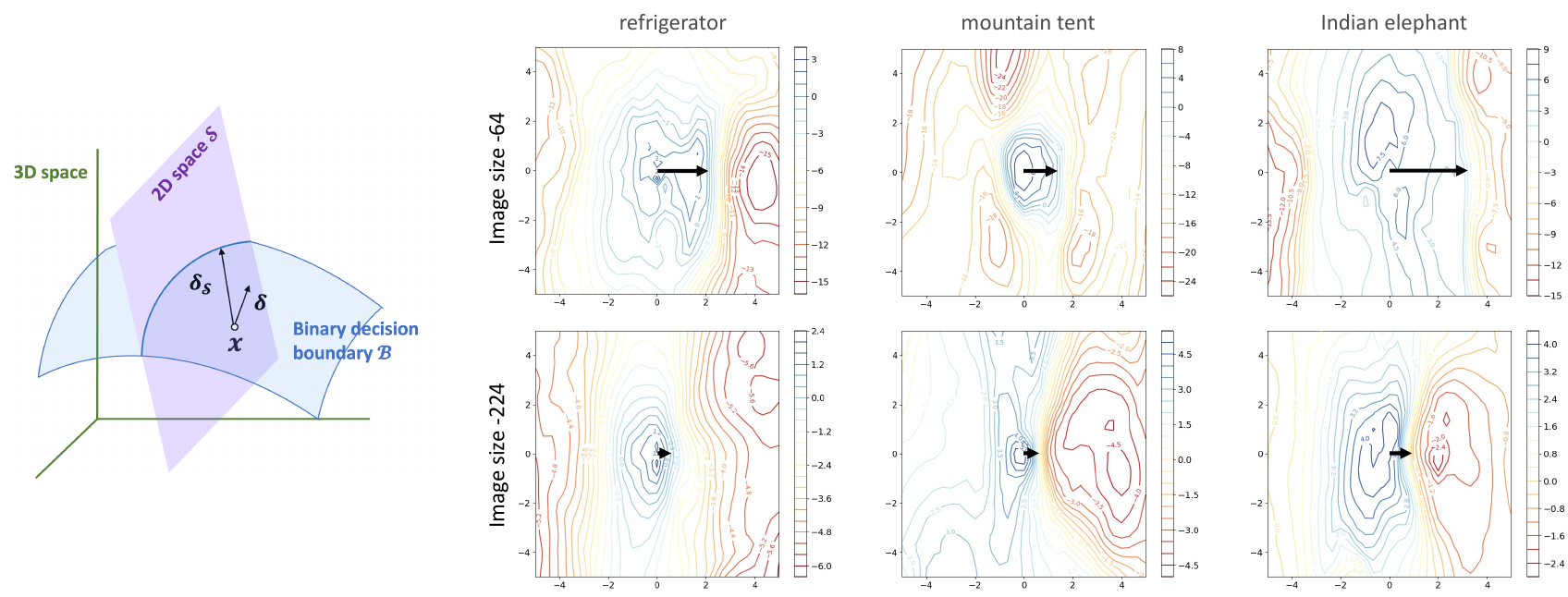}
 \caption{Pictorial representation (\textbf{left}) of how dimensionality impacts black-box adversarial attacks. Decision boundary visualization (\textbf{right}) with different images of dimensionality 64 and 224.}\label{decision_boundary}
\end{figure*}

\subsubsection{Input dimensionality}
\label{sec: Input dimensionality}
We also investigate \textit{how the dimensionality the adversary restricted to affects its performance?} We observe that \textbf{adversarial attacks on images of lower dimensions are less successful}, and suggest that the constraints of the dimensionality results in a longer distance needed to cross the decision boundary, thus requires larger perturbations to be effective.
We now quantify the sensitivity to input dimensionality via conducting untargeted, $\ell_{\infty}$ attack on ImageNet\cite{imagenet} dataset with varying image size. 
It's important to note that when working with a clean dataset with downsampled images, the clean accuracy naturally decreases, potentially confusing our analysis. Therefore, we use the \textit{improvement in ASR} after adding adversarial perturbations on benign examples resized to different dimensionalities. This metric only serves as the measure of the transfer-based attack performance. Query-based attacks define the percentage of finally adversarial images in initially non-adversarial images as ASR and thus are not influenced by the dropped clean accuracy. 
Fig. \ref{imagesize} shows that all adversaries suffer from a considerable failure rate when facing the robustness existing in low dimensional classification tasks. 

This intuition behind this phenomenon is clearly illustrated in Fig. \ref{decision_boundary} (left). Consider a toy example: $\mathcal{B}$ is the decision boundary of binary classification problem, example $\boldsymbol{x}$ lies in a 2-dimensional subspace $\mathcal{S}$ of a 3-dimensional space $\mathbb{R}^3$. $\delta$ and $\delta_{\mathcal{S}}$ points in the distance to the nearest decision boundary in 3D space and 2D space respectively. Due to the dimensionality constraint, the attack in lower dimensionality requires a higher perturbation norm budget, as a result, a lower (improvement in) ASR. This intuition could be verified through our analytical tool, Decision Surface, as depicted in Fig. \ref{decision_boundary} (right) where black arrows indicate the distance to cross the decision boundary along the perturbation direction. Notably, inputs in lower dimensionalities require longer distances to traverse the boundary. 


\subsection{Effect of Defense}
\label{sec: effect of defense}

\begin{figure*}[t]
\vspace{-2mm}
 \setlength{\abovecaptionskip}{-0.1cm}
\setlength{\belowcaptionskip}{-0.1cm}
 \centering
\includegraphics[width=0.97\linewidth]{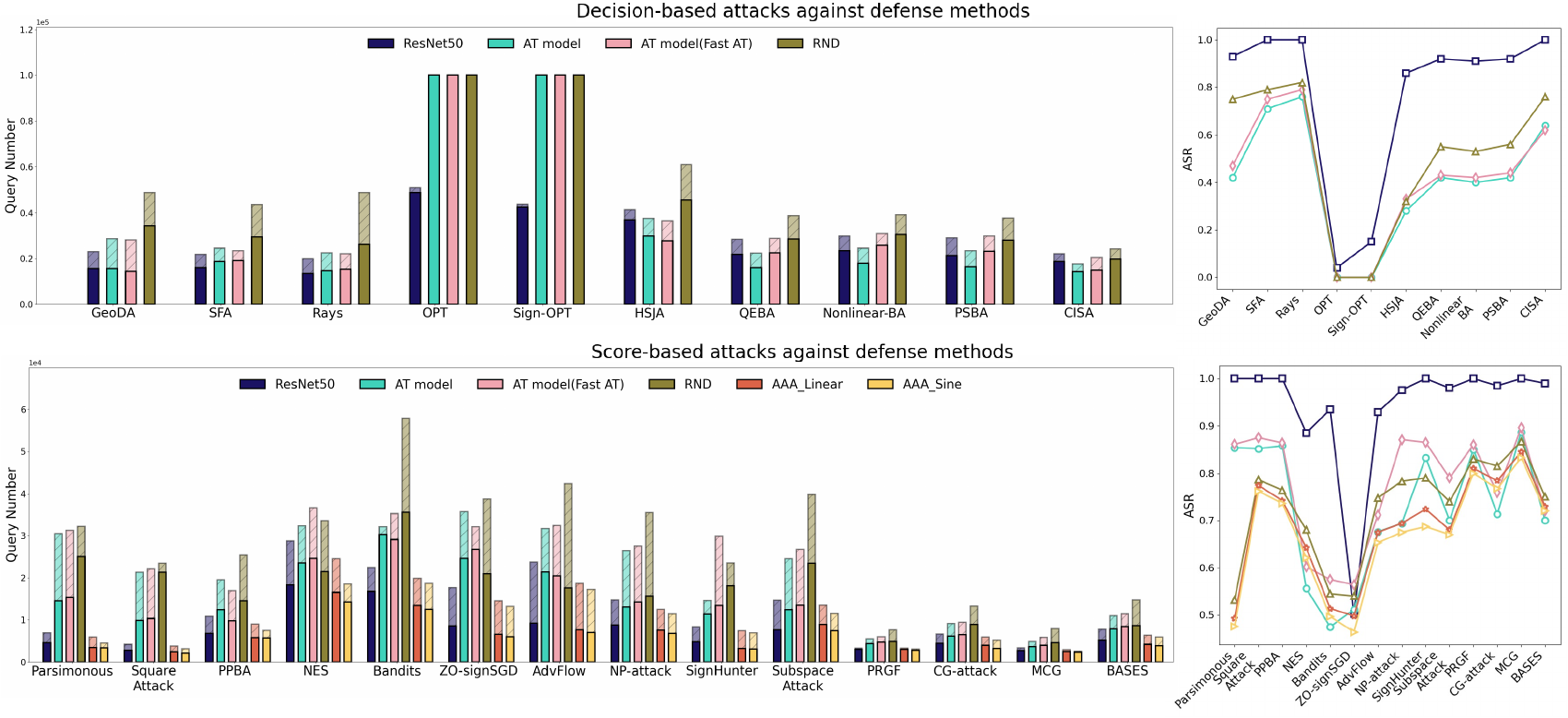}
 \caption{\blue{\textbf{Different query-based attacks against defenses}. The bar chart and line chart represent the query number and ASR, respectively, of decision-based attacks (\textbf{top}) and score-based attacks (\textbf{bottom}) against different defenses. We use solid bars to represent the MQNs and shaded bars to represents AQNs. For the two chosen AT models, one is from \cite{advresnet}, the other is trained using fast AT.}}\label{defense}
\end{figure*}

\begin{figure*}[t]
\vspace{-2mm}
 \setlength{\abovecaptionskip}{-0.1cm}
\setlength{\belowcaptionskip}{-0.1cm}
 \centering
\includegraphics[width=0.85\linewidth]{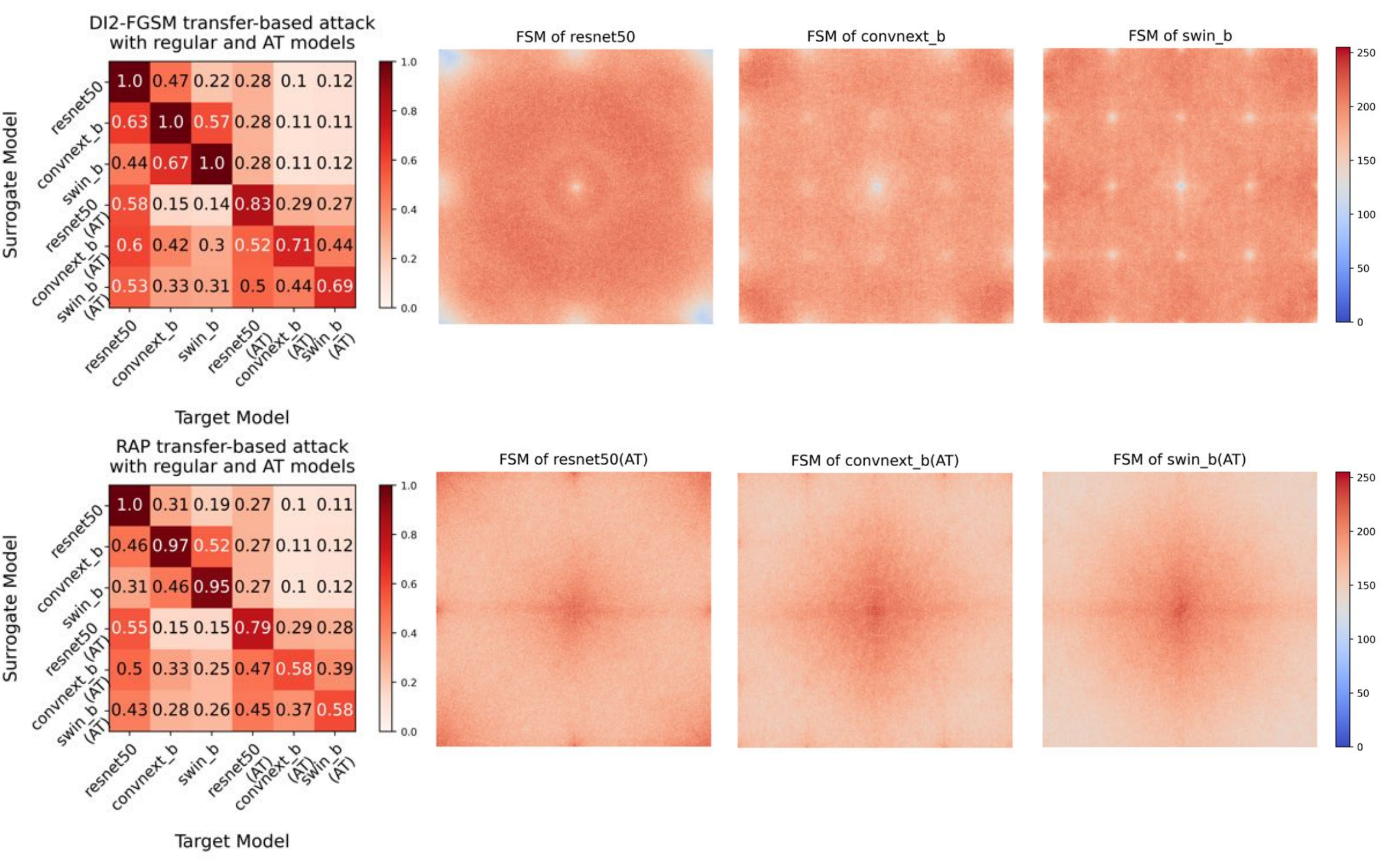}
 \caption{\textbf{Intra/Cross-training-strategy transferability}. The matrices in the first column represent the ASRs for DI2-FGSM (\textbf{top}) and RAP (\textbf{bottom}) attacks, with each entry corresponding to a surrogate-target pair. The y-axis and x-axis represent surrogate models and target models, respectively. We employ the regular and AT versions of three model architectures for our evaluations. Additional results of more attacks exhibiting similar trends can be found In Fig. \ref{add_defense}. The other columns show the magnitude of the FSM for six models involved in this analysis. Low frequencies are located at the center, and high frequencies at the corners.}\label{defense_transfer}
\end{figure*}

In our previous experimental analyses, we focused on clean models. Now, we shift our focus to attacks on target models involved of defense mechanisms. Here we consider defenses mentioned in \textbf{Supplementary} \ref{sec: Categorization of implemented defenses}.
In the context of query-based attacks, we aim to investigate \textit{the evasion capabilities of different attack methods against training-time defense and inference-time defense.} 
We also conduct an in-depth discussion on the \textit{advantages and disadvantages of these two types of defenses}. 
\blue{In decision-based black-box attacks, random search methods like \textbf{SFA \cite{chen2020boosting} and Rays\cite{chen2020rays} perform the best}, achieving ASRs over 70\%, while traditional methods such as OPT \cite{cheng2018query} and Sign-OPT \cite{cheng2019sign} struggle under defenses. In score-based attacks, combination-based methods like \textbf{MCG \cite{yin2023generalizable} stand out with the highest ASR and fewest queries}, maintaining strong performance even when defenses are applied.
Besides, we can also observe that training-time defense methods (e.g., adversarially trained models) effectively suppress the attack success rates of various methods, while inference-time defense methods primarily focus on increasing the number of queries required for each attack.}
As for transfer-based attacks, in Sec. \ref{sec: Intra/Cross-family transferability}, we have observed the strong intra-family transferability and heterogeneous cross-family transferability among clean models, we further expand on the previous analyses and now examine \textit{the transferability between normally trained models and adversarially trained models}, \textit{i.e.}, intra/cross-training-strategy transferability. 
\blue{\textbf{It is observed that stronger transferability occurs within models that share the same training strategies, rather than between different strategies.} This is attributed to the different spatial frequency features used by the models for predictions.}

\subsubsection{Different query-based attacks against defenses} In this study, we aim to investigate the evasion capabilities of different attack methods against defenses. For our evaluation, we have chosen the ImageNet dataset, which includes two AT models following the ResNet architecture. We specifically evaluate decision-based black-box attacks and score-based black-box attacks using targeted attacks with $\ell_{\infty}$ norm. We will present the AQN, MQN and ASR achieved by different attack methods against the defenses in Fig. \ref{defense}. It is important to note that some defense methods may increase the query numbers of attack methods, but the corresponding increase in ASR indicates that the defense primarily raises the cost of the attack. 
In decision-based black-box attacks, methods like OPT \cite{cheng2018query} and Sign-OPT \cite{cheng2019sign} face challenges under both training-time and inference-time defenses, often failing to attack even one clean example. However, attacks like QEBA \cite{li2020qeba}, NonLinear-BA \cite{li2021nonlinear}, and PSBA \cite{zhang2021progressive} can bypass defenses to some extent, achieving ASRs of around 40\%. Remarkably, random search methods such as SFA \cite{chen2020boosting} and Rays \cite{chen2020rays} consistently outperform, with ASRs exceeding 70\% across all defense scenarios, indicating that attacks effective on clean models also perform well against defended models. In score-based black-box attacks, there's a notable variation in attack performance across different methods, with query numbers ranging from a few thousand to 30,000-40,000. Among all defenses, the combination-based method MCG \cite{yin2023generalizable} exhibits the highest ASR and the fewest queries, demonstrating only a 10\%-15\% reduction in ASR compared to attacking ResNet-50.

\subsubsection{Training-time defense vs. inference-time defense}
\label{add: training-time defense vs. inference-time defense}
Defense methods against adversarial attacks fall into two categories: training-time and inference-time defenses. Assessing their respective pros and cons is essential. Referring to Fig. \ref{defense}(top), we compare AT and RND \cite{qin2021random}, as AAA \cite{chen2022adversarial} exclusively defends against score-based attacks.
The results reveal that RND primarily increases the query cost for attackers to achieve performance comparable to, or slightly lower than, those without defense strategies, leading to a trade-off between performance and query cost. 
Conversely, the query numbers for attacking AT models are similar to those on the clean model (ResNet-50), while the achieved ASR significantly diminishes, indicating a different kind of resistance. In summary, AT could effectively reduce ASR but at the cost of considerable model training overhead. RND provides lightweight and real-time defense while not significantly reducing ASR. 

For score-based attacks (see Fig. \ref{defense} (bottom)), the comparison between AT models and RND echoes the earlier discussion and is omitted. Particularly, our focus shifts to comparing AAA with AT and RND. Remarkably, AAA markedly reduces ASR within a smaller number of queries, suggesting a more efficient defense strategy by misleading attackers to upadte adversarial examples in the wrong direction. Compared to AT and RND, AAA represents a more proactive approach, positioning it as an ideal lightweight defense method.

\subsubsection{Intra/Cross-training-strategy transferability} To comprehensively assess the impact of different training strategies, in addition to normally trained ResNet50, ConvNeXt-B, Swin-B the same as in Sec. \ref{sec: Intra/Cross-family transferability}, we also exploit their adversarially trained counterparts, denoted as ResNet50(AT)\cite{advresnet}, ConvNeXt-B(AT)\cite{advconvswin} and Swin-B(AT)\cite{advconvswin}. This study considers all possible combinations of these models as surrogate-target pairs. For each pair, we subject them to the untargeted, $\ell_\infty$ attack on the ImageNet and present the transferability between them in Fig. \ref{defense_transfer}. We can see that, in general, the transferability between regular models and AT models is extremely low. For example, in DI2-FGSM, the model with highest transferability to ResNet50(AT) is ConvNeXt-B(AT) instead of ResNet50, the model with highest transferability to ConvNeXt-B is Swin-B instead of ConvNeXt-B(AT). That is to say, strong transferability mostly occurs within the same training strategies, but not between different strategies. 

We gain intuition on the cause of this phenomenon by analyzing the spatial frequency features utilized by models to make predictions, via our analysis tool---FSM. The FSMs of these six models are shown in Fig. \ref{defense_transfer}. 
Firstly, we find that FSMs of normal models consistently display uniform attention distribution across the frequency spectrum, with certain frequency components (e.g., very low frequency) universally disregarded.  
We also observe the FSM similarity among AT models: their attention spreads over the whole frequency spectrum and predominantly focuses on the frequencies ignored by the normal counterparts. 
As per Eq. (\ref{equ: fsm}), FSM is the Discrete Fourier Transform (DFT) of model gradient \textit{w.r.t} input. Given that adversarial perturbations are largely parallel to the gradient, the magnitude of the Fourier spectrum of adversarial perturbation crafted on a model aligns with the magnitude of the model's FSM, \ie, the perturbation span frequency ranges which the model is sensitive to. Due to the similarity in frequency attention maps among models with the same training strategy, the crafted perturbations also maintain similarity.
The transferability of adversarial perturbations between models arises from shared non-robust features. Consequently, perturbations transfer effectively between models with the same training strategy, yet exhibit reduced transferability between models employing different training strategies. 
Note that all models transfer well to ResNet50 might be attributed to their coverage of a broader frequency spectrum compared to ResNet50.

\begin{figure*}[h]
\vspace{-2mm}
 \setlength{\abovecaptionskip}{-0.1cm}
\setlength{\belowcaptionskip}{-0.1cm}
 \centering
\includegraphics[height=5.2cm]{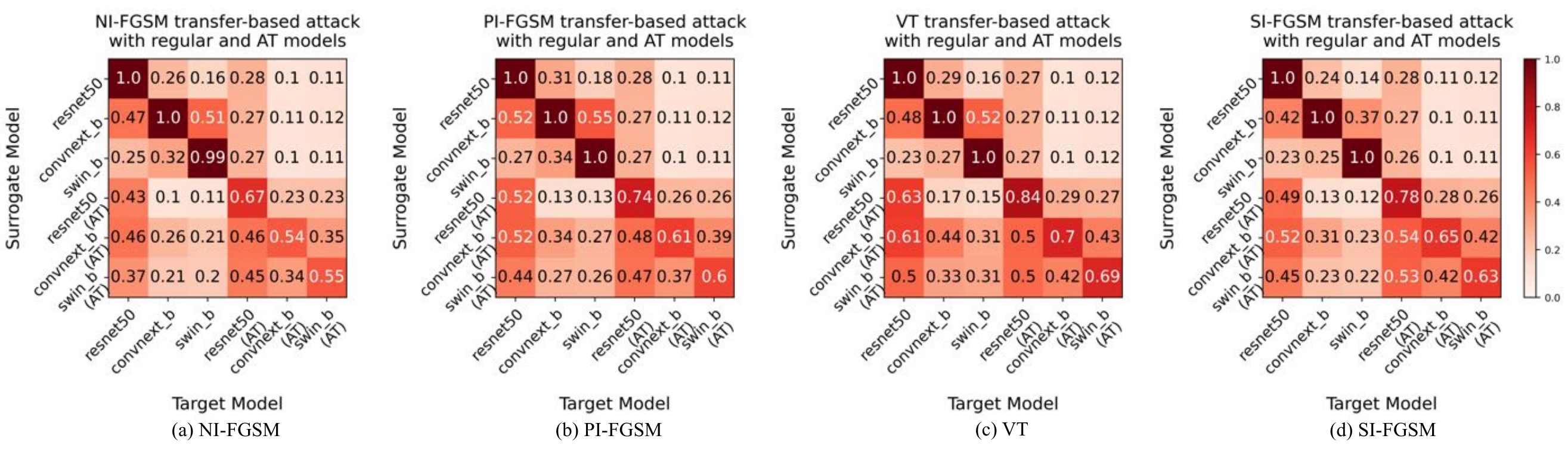}
 \caption{\small{The matrices representing the ASR for \textbf{(a)} NI-FGSM\cite{nisifgsm}, \textbf{(b)} PI-FGSM\cite{pifgsm}, \textbf{(c)} VT\cite{vt}, \textbf{(d)} SI-FGSM\cite{nisifgsm} attacks, with each entry corresponding to a surrogate-target pair. The y-axis and x-axis represent surrogate models and target models, respectively. We employ the regular and AT versions of three model architectures for our evaluations.}}\label{add_defense}
\end{figure*}


\subsection{Attack Procedure Analysis}
\label{sec: visualization analysis}

\begin{figure*}
\vspace{-2mm}
 \setlength{\abovecaptionskip}{-0.1cm}
\setlength{\belowcaptionskip}{-0.1cm}
 \centering
\includegraphics[height=12.6cm]{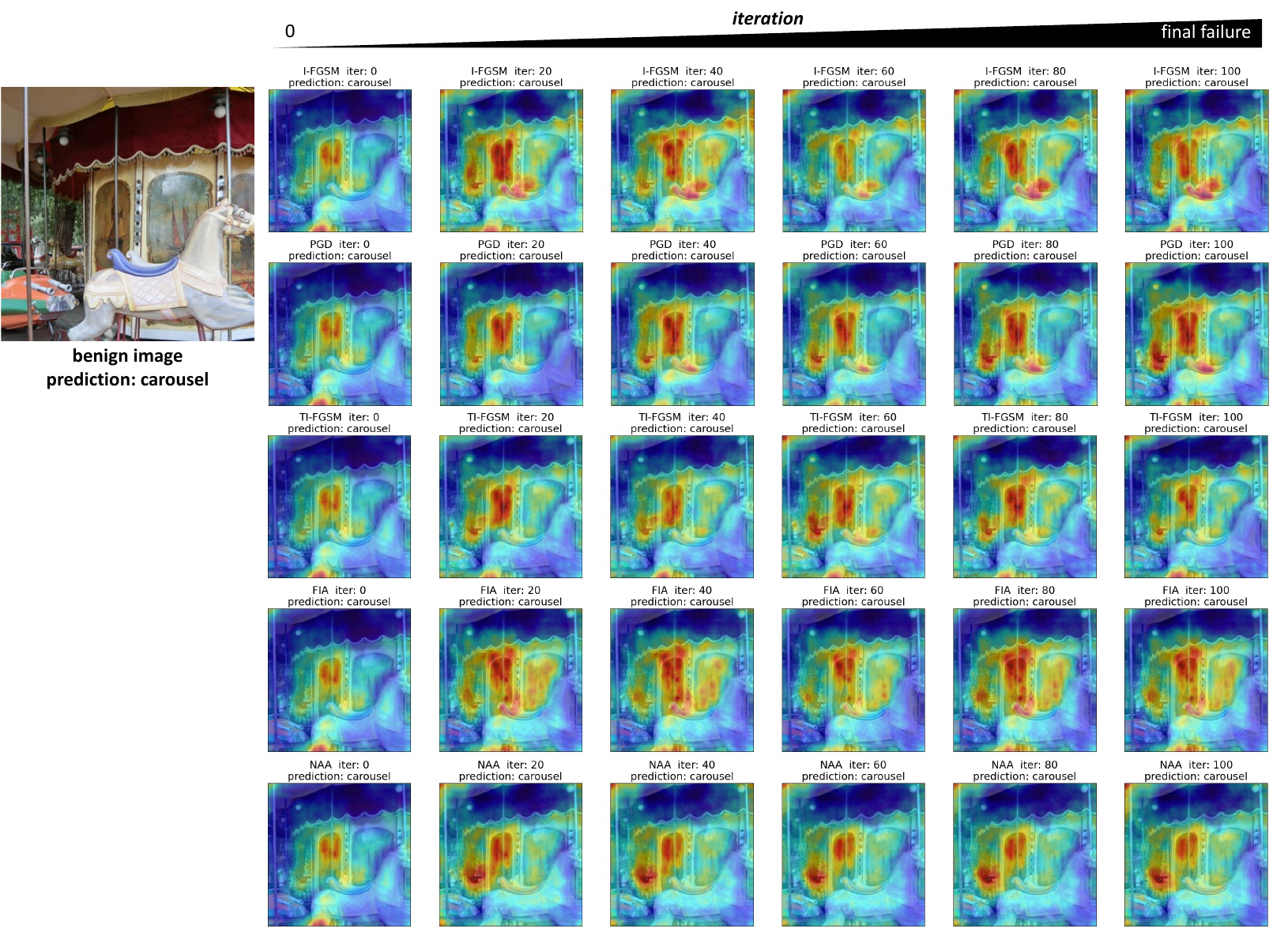}
 \caption{\small{The benign image from the category of carousel (\textbf{first up}), and the attention maps on its perturbed counterparts generated by failed transfer-based attacks (\textbf{others}). Each row plots the evolution of adversarial images crafted by a certain attack. For these failed attacks, we plot the adversarial images at regular intervals between the first iteration and the final iteration. The title of each image indicates the method, current iteration number and target model prediction. It's observed that when attacks fail, the discriminative regions remain unchanged.}}
 \label{add_vis_grad_cam_fail}
\end{figure*}

\begin{figure*}
\vspace{-2mm}
 \setlength{\abovecaptionskip}{-0.1cm}
\setlength{\belowcaptionskip}{-0.1cm}
 \centering
\includegraphics[height=22cm]{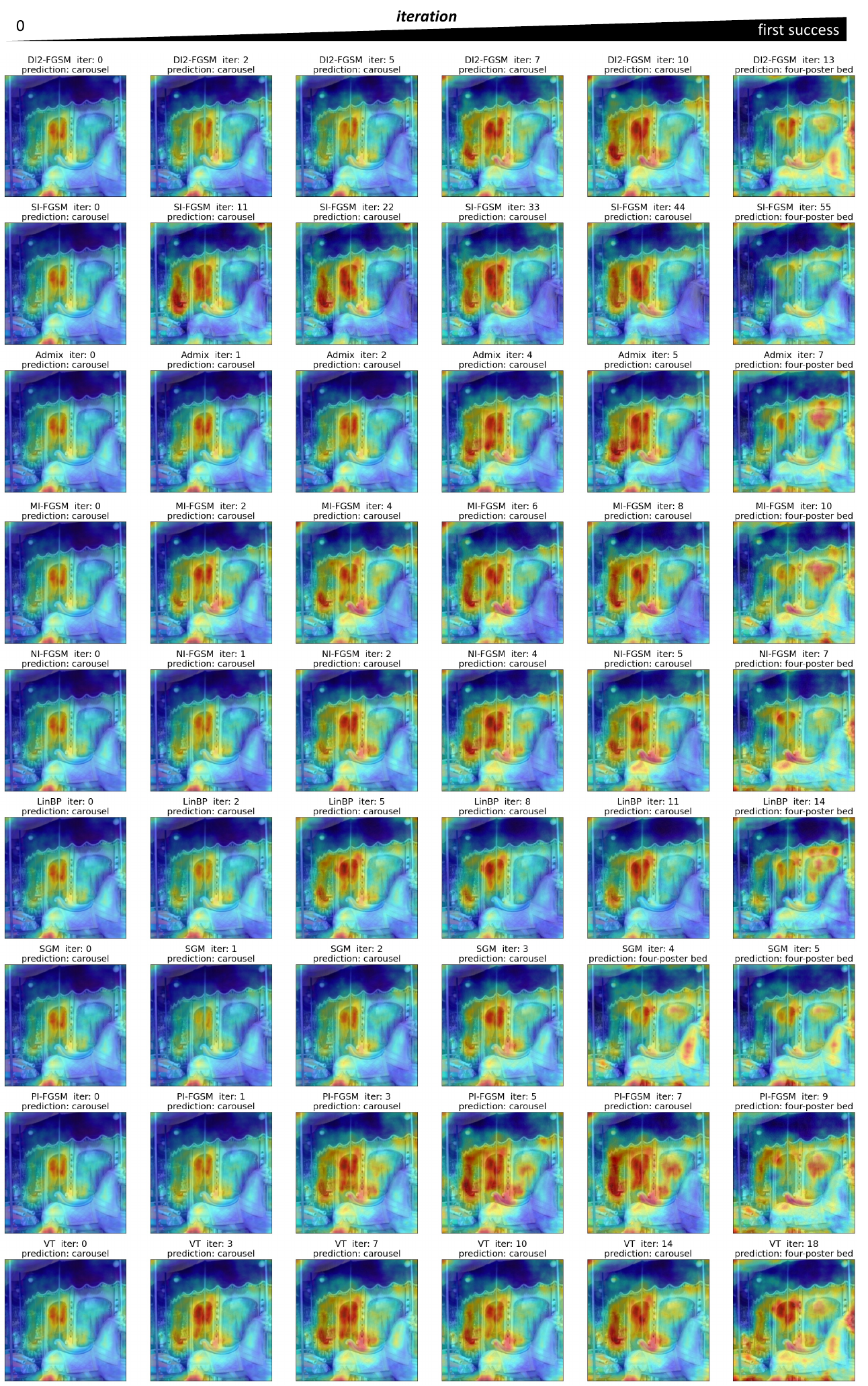}
 \caption{\small{Given a benign image from the category of \textit{carousel} (same as the benign image in Fig. \ref{add_vis_grad_cam_fail}), the attention maps on its perturbed counterparts generated by successful transfer-based attacks. Each row plots the evolution of adversarial images crafted by a certain attack. For these successful attacks, we plot the adversarial images at regular intervals between the first iteration and the first successfully attacked iteration. The title of each image indicates the method, current iteration number and target model prediction. It's observed that when attacks succeed, the discriminative regions change to trivial features, and the important features are not captured.}}
 \label{add_vis_grad_cam_suc1}
\end{figure*}

\begin{figure*}
\vspace{-2mm}
 \setlength{\abovecaptionskip}{-0.1cm}
\setlength{\belowcaptionskip}{-0.1cm}
 \centering
\includegraphics[height=22cm]{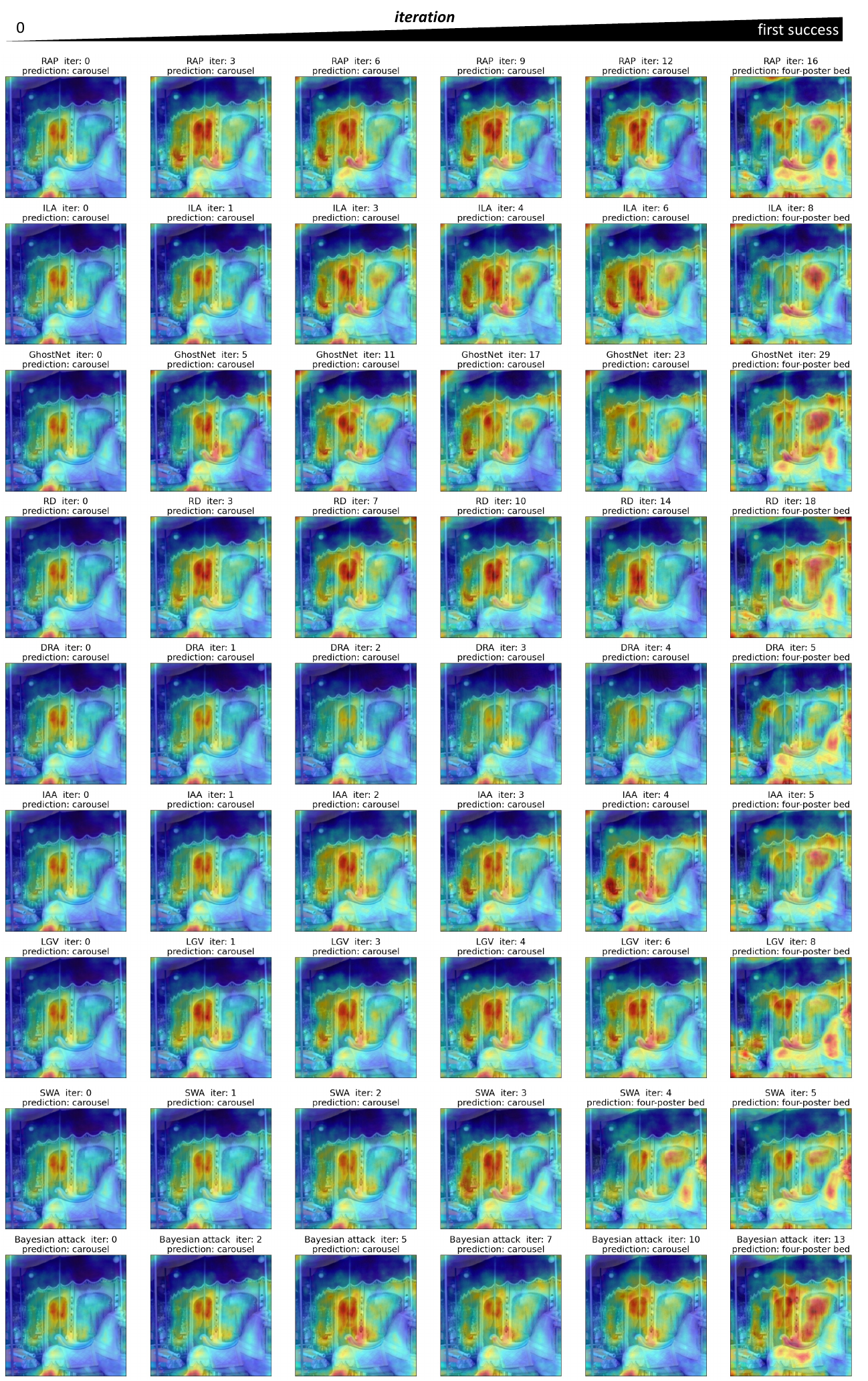}
 \caption{\small{The rest part of Fig. \ref{add_vis_grad_cam_suc1}}}
 \label{add_vis_grad_cam_suc2}
\end{figure*}

\begin{figure*}
\vspace{-2mm}
 \setlength{\abovecaptionskip}{-0.1cm}
\setlength{\belowcaptionskip}{-0.1cm}
 \centering
\includegraphics[height=12cm]{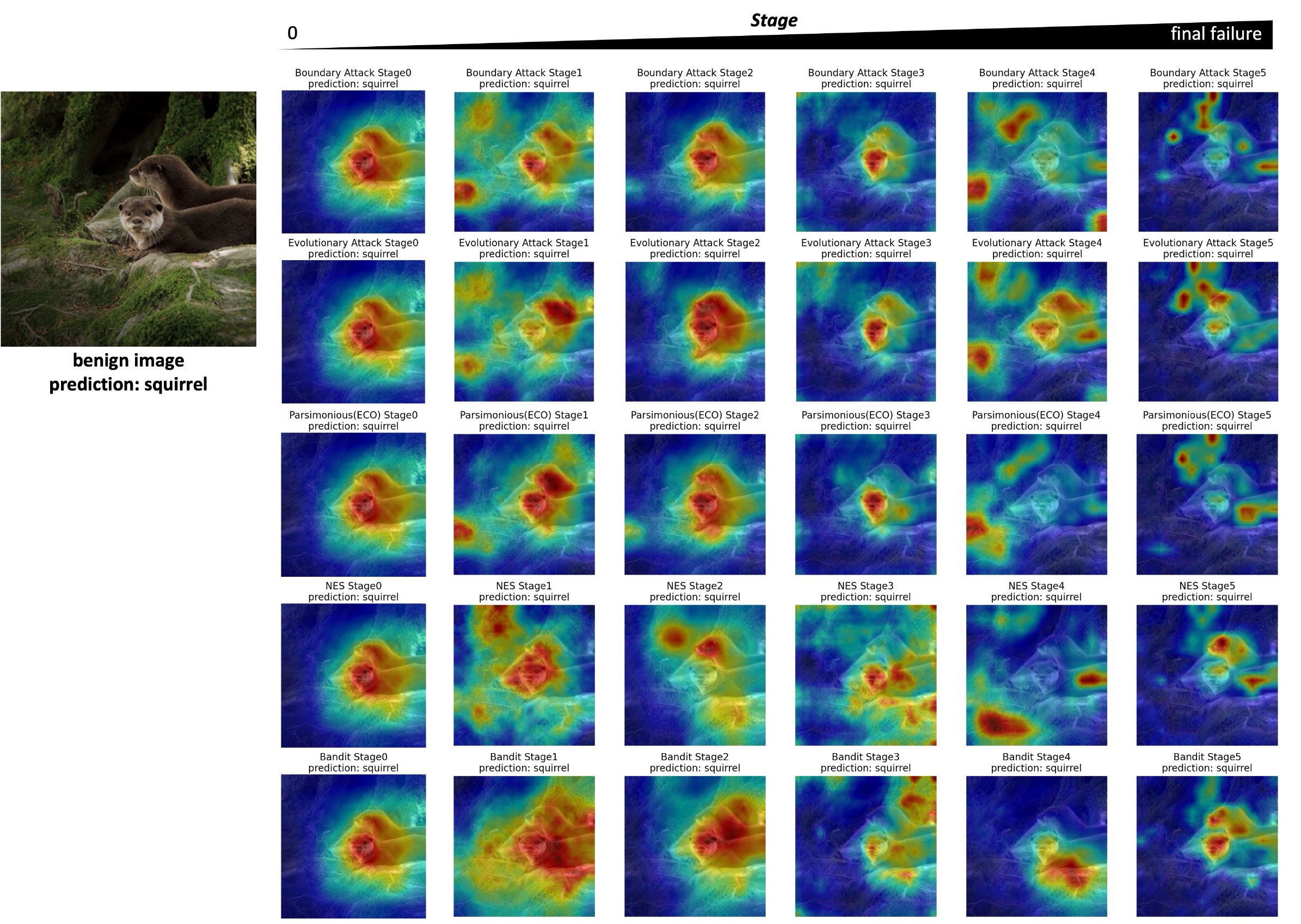}
 \caption{\small{The benign image from the category of squirrel (\textbf{first up}), and the attention maps on its perturbed counterparts generated by failed decision-based attacks(\textbf{the first 2 rows}) and failed score-based attacks(\textbf{the following 3 rows}). Each row plots the evolution of adversarial images crafted by a certain attack. For these failed attacks, we plot the adversarial images at each stage, where we divide the entire attack process into 6 stages(from the beginning to the end, no matter fails or succeeds). The title of each image indicates the method, current stage period, and target model prediction. It's observed that when attacks fail, the prediction label provided by the target model remains unchanged.}}
 \label{query_grad_cam_fail}
\end{figure*}


\begin{figure*}
\vspace{-2mm}
 \setlength{\abovecaptionskip}{-0.1cm}
\setlength{\belowcaptionskip}{-0.1cm}
 \centering
\includegraphics[height=22cm]{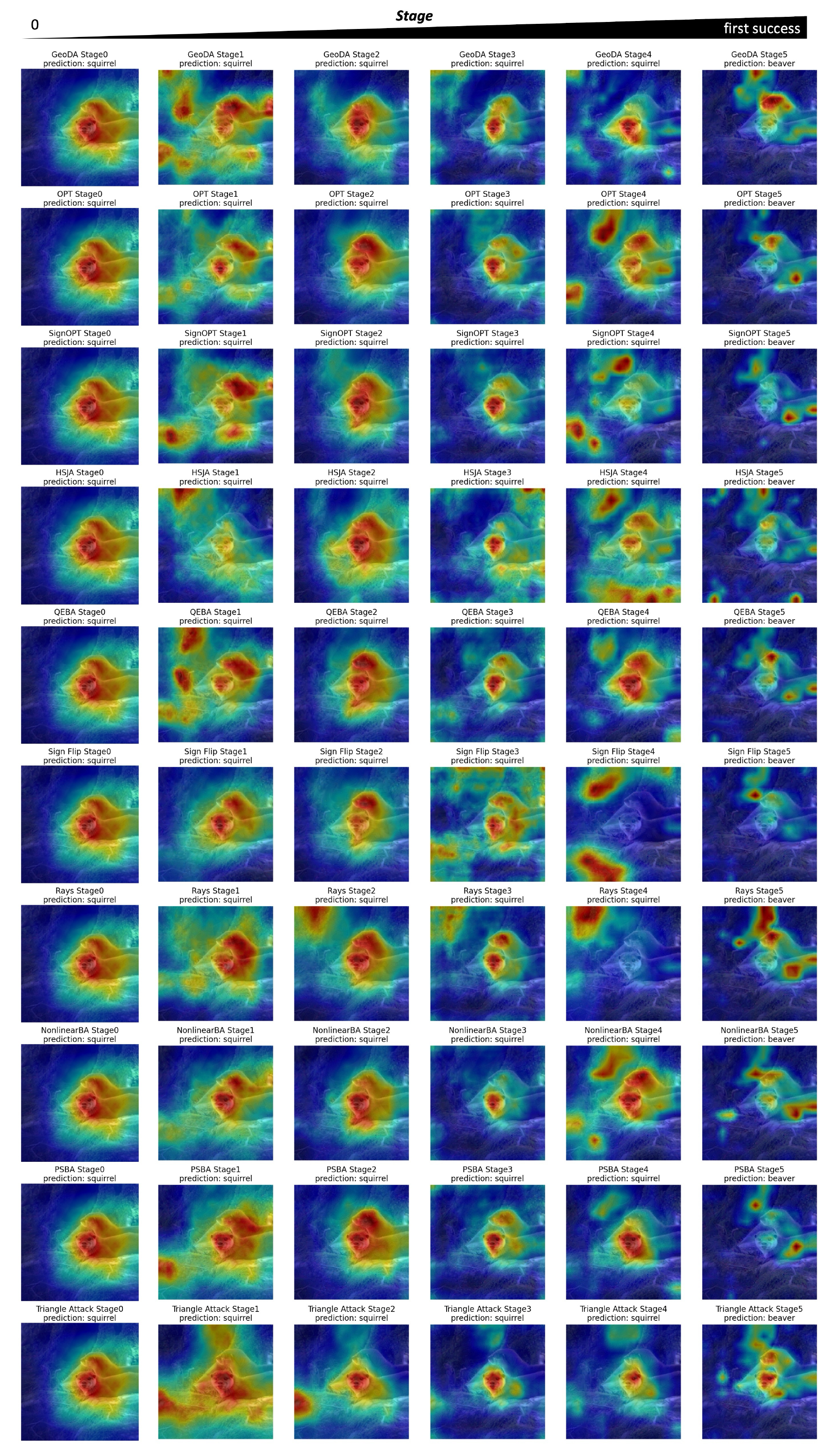}
 \caption{\small{Given a benign image from the category of \textit{squirrel} (same as the benign image in Fig. \ref{query_grad_cam_fail}), the attention maps on its perturbed counterparts generated by successful decision-based attacks. Each row plots the evolution of adversarial images crafted by a certain attack. For these successful attacks, we also plot the adversarial images at each stage, where we divide the entire attack process into 6 stages(from the beginning to the end, no matter fails or succeeds). The title of each image indicates the method, current stage period, and target model prediction. It's observed that when attacks succeed, the prediction label provided by the target model will be changed to \textit{beaver}.}}
 \label{query_grad_cam_decision_suc}
\end{figure*}

\begin{figure*}
\vspace{-2mm}
 \setlength{\abovecaptionskip}{-0.1cm}
\setlength{\belowcaptionskip}{-0.1cm}
 \centering
\includegraphics[height=23cm]{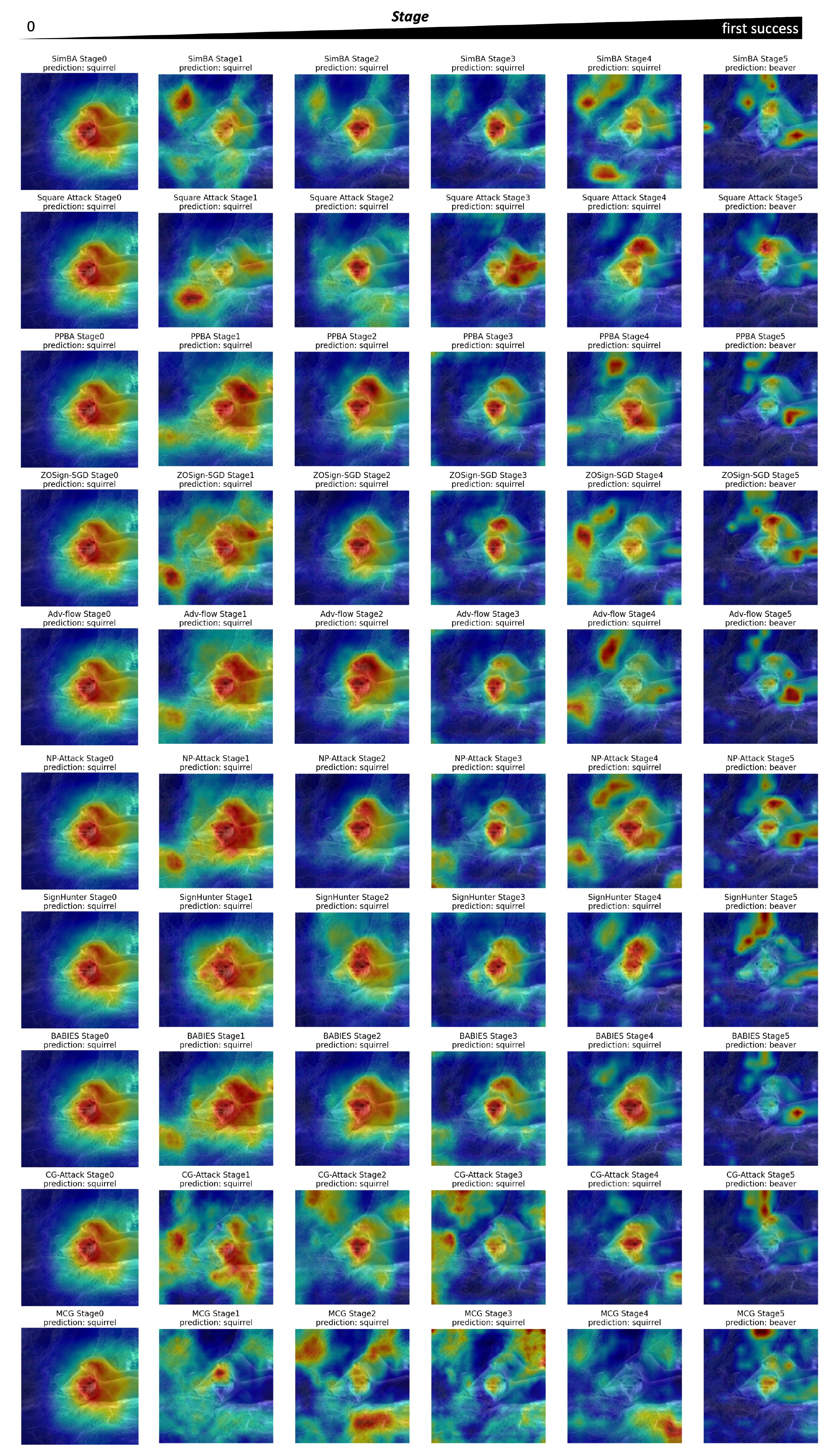}
 \caption{\small{The attention maps of successful score-based attacks which follows the setting as Fig. \ref{query_grad_cam_decision_suc}}}
 \label{query_grad_cam_score_suc}
\end{figure*}

For a more qualitative analysis, we explore what is most salient to the target model when an image is under some attack. Using our analytical tools---FullGrad, we visualize the evolution of model attention maps on the perturbed images along the iteration in transfer-based attacks, or the path traversed by query-based attackers. We are interested in both \textit{how successful attacks succeed} and \textit{how failed attacks fail}. The main observation is that, in both transfer-based and query-based attacks, \textbf{successful adversarial attacks manage to shift the model's attention from critical, discriminative regions to trivial areas, leading to misclassification}, whereas failed attacks cannot divert the model's focus from the key regions, allowing the model to correctly classify the images despite the attack.

In transfer-based attacks, given an image from the category of \textit{carousel}, which is selected to be successfully perturbed by about $80\%$ of all implemented methods, the attention maps on its unsuccessfully perturbed counterparts along the iteration are visualized in Fig. \ref{add_vis_grad_cam_fail}, the successful cases are visualized in Fig. \ref{add_vis_grad_cam_suc1} and Fig. \ref{add_vis_grad_cam_suc2}. 
Among the failed attacks, the target model consistently focuses on the original discriminative regions and recognizes them correctly as the attack runs. While in successful attacks, the adversarial images gradually defocus the models from these regions. Finally, the models are misled to focus on those trivial areas and recognize the adversarial images as the category of \textit{four-poster bed}.

In query-based attacks, we choose an image from the category of \textit{squirrel}, which is also selected to be successfully attacked by about $80\%$ of all methods. Following the similar presentation of FullGrad in transfer-based attacks, we show the attention maps through the attack process of unsuccessful attacks in Fig. \ref{query_grad_cam_fail} and successful attacks of both decision-based attacks and score-based attacks in Fig. \ref{query_grad_cam_decision_suc} and Fig. \ref{query_grad_cam_score_suc}. We can also observe that in samples where attacks failed, the target model continuously concentrated on the critical regions or nearby regions, which failed to mislead the target model and thus the target model can correctly identify them throughout the attack process. Conversely, during successful attacks, the adversarial images incrementally shift the model's attention away from these critical regions (from the central part to another part). Eventually, the models are deceived into concentrating on insignificant areas, leading to the misclassification of the adversarial images as the category of a \textit{beaver}.

\begin{figure*}[h]
\vspace{-2mm}
 \setlength{\abovecaptionskip}{-0.1cm}
\setlength{\belowcaptionskip}{-0.1cm}
 \centering
\includegraphics[height=6.7cm]{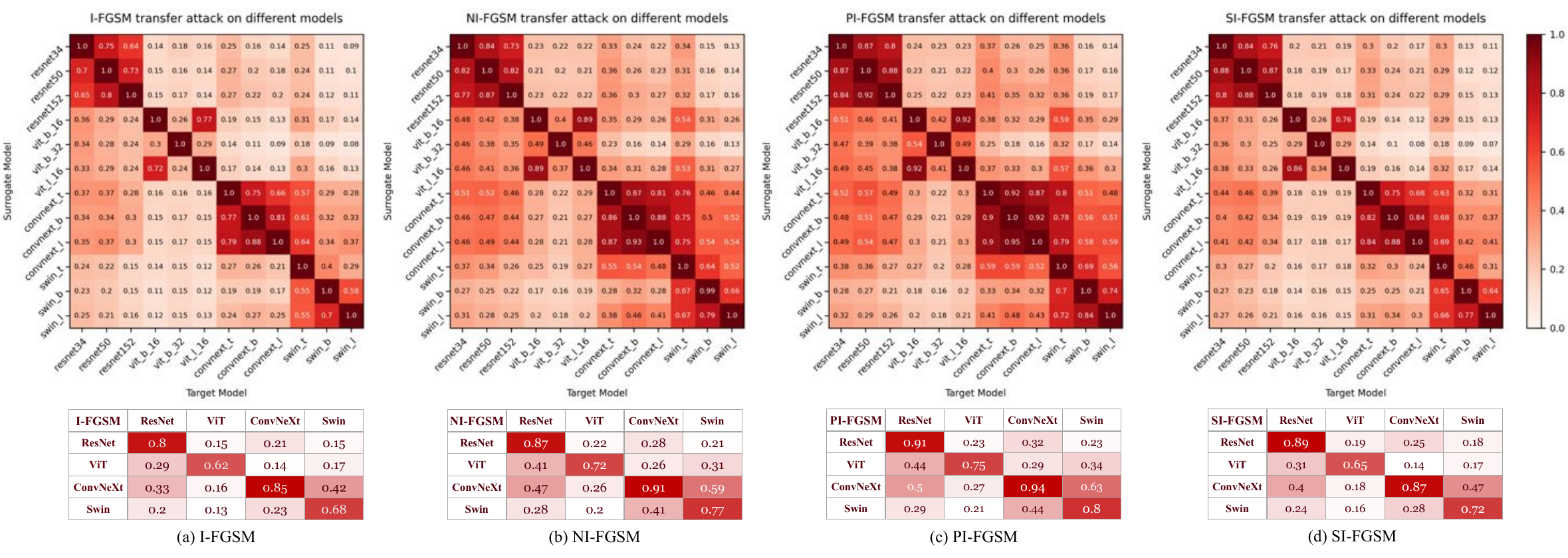}
 \caption{\small{Matrices of ASR in \textbf{(a)} I-FGSM\cite{ifgsm}, \textbf{(b)} NI-FGSM\cite{nisifgsm}, \textbf{(c)} PI-FGSM\cite{pifgsm} and \textbf{(d)} SI-FGSM\cite{nisifgsm} attacks, where each entry correspond to a surrogate-target pair. The y-axis represents surrogate models and the x-axis represents target models. We choose three variants in every model family for our evaluations. Matrices in the second row summarizes the first row by averaging ASR \textit{w.r.t.} family.}}\label{add_arch}
\end{figure*}

\section{Implementation details on experiments}\label{app: implementation details on experiments}

\subsection{Pretrained model download links}
\label{app:pretrained model download links}
For evalutions on CIFAR-10 models, we use the following implementations(embedded links):
\begin{itemize}[leftmargin=15pt,itemsep=1pt,topsep=0pt]
\item 
\textbf{regular models} \\
pretrained for query-based black-box attacks:
\href{https://drive.google.com/file/d/19pBDfd3I_veQdCyAtUZgIyByvzdnLEfK/view?usp=drive_link}{ResNet-50},
\href{https://drive.google.com/file/d/1Nag-ujkemtvaG6_US-lnbJLO9oCrFwX7/view?usp=drive_link}{VGG-19$^{\dagger}$},
\href{https://drive.google.com/file/d/1zAOZnluOfUe-HX81nxMdnqxVzO8j233d/view?usp=drive_link}{Inception-V3},
\href{https://drive.google.com/file/d/1L1yQgmJ1mzkTz5USGo-KseqQuXEBMEf2/view?usp=drive_link}{DenseNet-121}
\\
pretrained for transfer-based black-box attacks:
\href{https://www.kaggle.com/datasets/firuzjuraev/trained-models-for-cifar10-dataset/download?datasetVersionNumber=1}{ResNet-50}, \href{https://github.com/bearpaw/pytorch-classification/tree/master}{VGG-19$^{\dagger}$}, \href{https://www.kaggle.com/datasets/firuzjuraev/trained-models-for-cifar10-dataset/download?datasetVersionNumber=1}{Inception-V3}, \href{https://github.com/bearpaw/pytorch-classification/tree/master}{DenseNet-BC}, \href{https://github.com/bearpaw/pytorch-classification/tree/master}{ResNeXt-29}, \href{https://github.com/bearpaw/pytorch-classification/tree/master}{WRN-28-10}, \href{https://drive.google.com/file/d/16GibXwzzMYsgivfz5tKdK0UAoXiIsHCW/view?usp=drive_link}{PyramidNet$^{\dagger}$}, 
\href{https://drive.google.com/file/d/16GibXwzzMYsgivfz5tKdK0UAoXiIsHCW/view?usp=drive_link}{GDAS}.
\item 
\textbf{AT models} \\
\blue{pretrained for query-based black-box attacks:
\href{https://storage.googleapis.com/dm-adversarial-robustness/cifar10_linf_wrn28-10_with.pt}{WRN-28-10 (\textbf{AT}\cite{gowal2020uncovering}, optimization trick, $\ell_{\infty}$, 8/255)}, 
\href{https://storage.googleapis.com/dm-adversarial-robustness/cifar10_linf_wrn28-10_cutmix_ddpm_v2.pt}{WRN-28-10 (\textbf{AT}\cite{rebuffi2021fixing}, ddpm, $\ell_{\infty}$, 8/255)}
\\
pretrained for transfer-based black-box attacks:
\href{https://github.com/RobustBench/robustbench}{WRN-28-10 (\textbf{AT}\cite{advwrn}, $\ell_{\infty}$, 8/255)}.}
\end{itemize}
Models for ImageNet dataset are from following implementations:
\begin{itemize}[leftmargin=15pt,itemsep=1pt,topsep=0pt]
\item 
\textbf{regular models}
\href{https://download.pytorch.org/models/resnet50-19c8e357.pth}{ResNet-50}, 
\href{https://download.pytorch.org/models/vgg19_bn-c79401a0.pth}{VGG-19$^{\dagger}$}, 
\href{https://download.pytorch.org/models/inception_v3_google-1a9a5a14.pth}{Inception-V3}, 
\href{https://download.pytorch.org/models/densenet121-a639ec97.pth}{DenseNet-121}, 
\href{https://download.pytorch.org/models/vit_b_16-c867db91.pth}{ViT-B/16}, 
\href{https://download.pytorch.org/models/resnet152-b121ed2d.pth}{ResNet-152}, 
\href{https://download.pytorch.org/models/mobilenet_v2-b0353104.pth}{MobileNet-V2}, 
\href{http://data.lip6.fr/cadene/pretrainedmodels/senet154-c7b49a05.pth}{SENet-154}, 
\href{https://download.pytorch.org/models/resnext101_32x8d-8ba56ff5.pth}{ResNeXt-101}, 
\href{https://download.pytorch.org/models/wide_resnet101_2-32ee1156.pth}{WRN-101}, 
\href{http://data.lip6.fr/cadene/pretrainedmodels/pnasnet5large-bf079911.pth}{PNASNet}, 
\href{https://download.pytorch.org/models/mnasnet1.0_top1_73.512-f206786ef8.pth}{MNASNet}, 
\href{https://github.com/SwinTransformer/storage/releases/download/v1.0.0/swin_base_patch4_window7_224_22kto1k.pth}{Swin-B}, 
\href{https://download.pytorch.org/models/convnext_small-0c510722.pth}{ConvNeXt-S}
\href{https://dl.fbaipublicfiles.com/convnext/convnext_base_1k_224_ema.pth}{ConvNeXt-B}.
\item
\textbf{AT models} \\
\blue{pretrained for query-based black-box attacks: 
\href{https://huggingface.co/madrylab/robust-imagenet-models/resolve/main/resnet50_l2_eps3.ckpt}{ResNet-50 (\textbf{AT}\cite{advresnet}, $\ell_{2}$, 3)},
\href{https://huggingface.co/madrylab/robust-imagenet-models/resolve/main/resnet50_linf_eps8.0.ckpt}{ResNet-50 (\textbf{AT}\cite{advresnet}, $\ell_{\infty}$, 8/255)},
\href{https://drive.google.com/file/d/1yvP1KTlQuxU9UTLdnBCFCYavdex3A-KL/view?usp=drive_link}{ResNet-50 (\textbf{FastAT}, $\ell_{\infty}$, 1/255)}\\
pretrained for transfer-based black-box attacks: \href{https://github.com/RobustBench/robustbench}{ResNet-50 (\textbf{AT}\cite{advresnet}, $\ell_{\infty}$, 4/255)}, 
\href{https://github.com/RobustBench/robustbench}{Swin-B (\textbf{AT}\cite{advconvswin}, $\ell_{\infty}$, 4/255)}, 
\href{https://github.com/RobustBench/robustbench}{ConvNeXt-B (\textbf{AT}\cite{advconvswin}, $\ell_{\infty}$, 4/255)}.}
\end{itemize}

The clean accuracies of above models on benign test sets are shown in Tab. \ref{tab:models}. On CIFAR-10, they are evaluated on 10,000 test set. On ImageNet, they are evaluated on the 50,000 validation set.

\begin{table*}[t]
  \centering
  \caption{\small{Clean accuracies of the models adopted in BlackboxBench. Models in \textcolor{blue}{blue} are treated as surrogate models. \textbf{AT} indicates adversarially trained models. PyramidNet$^{\dagger}$ indicates PyramidNet\cite{pyramidal} + ShakeDrop\cite{shakedrop} + AutoAugment\cite{autoaugment}. VGG-19 $^{\dagger}$ indicates VGG-19 with batch normalization.}}
  \begingroup
  \setlength{\tabcolsep}{4pt} 
    \scalebox{0.66}{\begin{tabular}{cccccc>{\centering\arraybackslash}p{2.5cm}>{\centering\arraybackslash}p{3.3cm}>{\centering\arraybackslash}p{2.7cm}cc}
    \toprule
     \multicolumn{1}{c}{\multirow{6}[5]{*}{\textbf{Query}}} & \multirow{2}[2]{*}{\textbf{CIFAR-10}} & {ResNet-50\cite{resnet}} & {VGG-19$^{\dagger}$\cite{vgg}} & {Inception-V3\cite{inception}} & {DenseNet-121\cite{densenet}} & WRN-28-10 (\textbf{AT} w/ DDPM) \cite{rebuffi2021fixing} & WRN-28-10 (\textbf{AT} w/ optimization trick) \cite{gowal2020uncovering}\\
          &       & 93.65\% & 93.95\% & 93.74\% & 94.06\% & 60.75\% & 62.80\% \\
          \cmidrule{2-11}          & \multirow{2}[2]{*}{\textbf{ImageNet}} & {ResNet-50\cite{resnet}} & {VGG-19$^{\dagger}$\cite{vgg}} & Inception-V3\cite{inception} & ConvNeXt-S\cite{convnet} & ViT-B/16\cite{vit} & ResNet-50 (\textbf{AT}-$\ell_2$) \cite{advresnet} & ResNet-50 (\textbf{AT}-$\ell_\infty$)\cite{advresnet}&  ResNet-50 (Fast\textbf{AT})\cite{wong2020fast} \\
          &       & 76.13\% & 74.22\% & 77.29\% & 83.62\% & 81.07\% & 62.83\% & 54.53\% &43.46\%  \\
    \midrule
    \midrule
    \multicolumn{1}{c}{\multirow{6}[5]{*}{\textbf{Transfer}}} & \multirow{2}[2]{*}{\textbf{CIFAR-10}} & \textcolor[rgb]{ .016,  .196,  1}{ResNet-50\cite{resnet}} & \textcolor[rgb]{ .016,  .196,  1}{VGG-19$^{\dagger}$\cite{vgg}} & \textcolor[rgb]{ .016,  .196,  1}{Inception-V3\cite{inception}} & \textcolor[rgb]{ .016,  .196,  1}{DenseNet-BC\cite{densenet}} & ResNeXt-29\cite{resnext} & WRN-28-10\cite{wrn} & PyramidNet$^{\dagger}$\cite{linbp} & GDAS\cite{gdas}  & WRN-28-10 (\textbf{AT})\cite{advwrn} \\
          &       & 95.30\% & 93.34\% & 94.77\% & 96.68\% & 96.31\% & 96.21\% & 98.44\% & 97.19\% & 89.69\% \\
\cmidrule{2-11}          & \multirow{4}[3]{*}{\textbf{ImageNet}} & \textcolor[rgb]{ .016,  .196,  1}{ResNet-50\cite{resnet}} & \textcolor[rgb]{ .016,  .196,  1}{VGG-19$^{\dagger}$\cite{vgg}} & \textcolor[rgb]{ .016,  .196,  1}{Inception-V3\cite{inception}} & \textcolor[rgb]{ .016,  .196,  1}{DenseNet-121\cite{densenet}} & \textcolor[rgb]{ .016,  .196,  1}{ViT-B/16\cite{vit}} & ResNet-152\cite{resnet} & MobileNet-V2\cite{mobilenet} & SENet-154\cite{senet} & ResNeXt-101\cite{resnext} \\
          &       & 76.15\% & 74.24\% & 77.29\% & 74.65\% & 81.07\% & 78.31\% & 71.88\% & 81.32\% & 79.31\% \\
\cmidrule{3-11}          &       & WRN-101\cite{wrn} & PNASNet\cite{pnasnet} & MNASNet\cite{mnasnet} & Swin-B\cite{swin} & ConvNeXt-B\cite{convnet} & ResNet-50 (\textbf{AT})\cite{advresnet} & Swin-B (\textbf{AT})\cite{advconvswin} & ConvNeXt-B (\textbf{AT})\cite{advconvswin} &  \\
          &       & 78.84\% & 82.90\% & 73.46\% & 85.20\% & 83.80\% & 64.02\% & 76.16\% & 76.02\% &  \\
    \bottomrule
    \end{tabular}}
    \endgroup
  \label{tab:models}%
\end{table*}%

\subsection{Experimental setting for Result Overview} 
In the experiments of query-based  attacks, we implemented both untargeted and targeted attacks. For untargeted attacks, we assign the groud-truth label provided by the validation set for each clean example. In this case, one attack is successful if the predicted class of the adversarial example is different from the ground-truth label. The maximum number of queries is set to 10,000 for all untargeted attacks. For targeted attacks, we also assign the groud-truth label as untargeted attacks for each clean example. Besides, we choose the target label for each clean example according to the logit returned by the target model. There are 5 target types of target labels, which are "random", "least likely", "most likely", "median" and "specified label", for users to choose from. In our evaluation, we choose to assign a target label for each clean example by "median", which is the median label of the logit given by the target model.  For all targeted attacks, the maximum number of queries is set to 100,000 since if we still set the number to 10,000, which is the same as untargeted attacks, most attacks will fail to  craft even one adversarial example. We only select the attack setting with ImageNet as our dataset and ResNet-50 as our target model. We present the performance of both $\ell_2$ norm and $\ell_{\infty}$ norm attacks. The adversarial perturbation is bounded by $\ell_\infty=0.03$ and $\ell_2=3$ for all attacks.

In the experiments of transfer-based attacks, we test all four attack settings on ImageNet. We employ ResNet50 as the surrogate model and encompass all 17 models listed in Tab. \ref{tab:models} as target models. \blue{For attacks in model-fusion category, the surrogate models are ensembled by ResNet50, VGG-19$^{\dagger}$, Inception-V3, DenseNet-121 and ViT-B/16.} The reported performances are averaged ASR over all target models. For untargeted attacks, we utilize Cross Entropy loss. The adversarial perturbation is bounded by $\ell_\infty=0.03$ with step size $1/255$ and $\ell_2=3$ with step size $0.3$ for all attacks. The number of iteration is method-specific. For targeted attacks, we utilize the logit loss. The adversarial perturbation is bounded by $\ell_2=0.06$ with step size $0.006$ and $\ell_2=6$ with step size $0.6$ for all attacks. The number of iteration varies by method, generally exceeding that of untargeted attacks. \blue{The composite attacks are conducted under $\{$untargeted, $\ell_{\infty}\}$ setting. When SGM is incorporated with model-fusion-based attacks, the surrogate models is only ensembled by ResNet50, DenseNet-121 and ViT-B/16, as SGM relies on residual modules in surrogate models.}

\subsection{Experimental setting for Effect of Data} 
With regard to character of clean samples, we randomly extracted one image from each class of the ImageNet Large Scale Visual Recognition Challenge 2012 (ILSVRC2012) validation set, totaling 1,000 images, to form the test dataset.  Based on this, we perform $\{$untargeted, $\ell_{\infty}\}$ attacks with $\ell_{\infty}$ bound to be 0.03 using all attack algorithms on ResNet-50. 

Regarding input dimension analysis, to craft adversarial examples of specific sizes, for query-based and transfer-based attacks, we both replace the official pre-processing pipeline of each neural network with resizing transformation. All images are resized to $n \times n$, $n \in\{32,64,128,224,256,$ $330,400\}$. In the experiments of transfer-based attacks, we adopt the same $\{$untargeted, $\ell_{\infty}\}$ attack configuration as discussed in the result overview analysis. Note that the pre-processing for adversarial examples in evaluation module is omitted here due to the potential impact of its resize operation on our analysis.  In the experiments of query-based attacks, we also adopt the same $\{$untargeted, $\ell_{\infty}\}$ attack configuration as discussed in the section of result overview.

\subsection{Experimental setting for Effect of Model Architecture} 
 For query-based attacks, different model architectures imply different decision boundaries. We specifically compare the difficulty of crafting successful adversarial samples among various target model architectures. We utilize the ratio of AQN to ASR as metrics to measure the difficulty of attacks. For both decision-based black-box attacks and score-based black-box attacks, we employ untargeted attacks using $\ell_{\infty}$ norm bounded to be 0.03 and the ImageNet dataset. We specifically set the maximum query number of each evaluation to be 10,000.
 
 For intra/cross-family transferability analysis, we conduct $\{$untargeted, $\ell_{\infty}\}$ attacks on ImageNet dataset. The surrogate model and target model are chosen from twelve model architectures, encompassing four variants within each of three model families. The reported performances are ASR of each surrogate-target pair. We utilize Cross Entropy loss. The adversarial perburbation is bounded by $\ell_\infty=0.03$ and the step size is set as $1/255$ for all attacks. The number of iterations is method-specific.

\subsection{Implementation details for Effect of Attack budget} 
In the context of perturbation norm budget analysis, for transfer-based experiments, we adopt the same $\{$untargeted, $\ell_{\infty}\}$ attack configuration as discussed in the result overview analysis. However, the maximum perturbation is defined as $\{0.01, 0.02, 0.03, 0.04, 0.05, 0.06, 0.07, 0.08\}$ with a corresponding step size set at one-tenth of the budget.

In query number budget analysis, for transfer-based attacks, we evaluate the $\{$untargeted, $\ell_{\infty}\}$ attack using the same configuration discussed in the result overview analysis but the iteration number, common across all methods, is drawn from  the set $\{10,50,100,200,300,400\}$. For decision-based attacks, we vary the query numbers while keeping the sample being successfully attacked to evaluate the distance between original example and adversarial example. A smaller value indicates that the adversarial example is gradually approaching the clean example to satisfy the adversarial constraint. Meanwhile, for score-based black-box attacks, we will show how the attack success rate (ASR) changes with an increasing number of queries. We evaluate all decision-based black-box attacks and score-based black-box attacks on ResNet-50 using $\ell_{\infty}$ norm untargeted attacks, with the ImageNet dataset as the test set.

\subsection{Experimental setting for Effect of Defense} 
Regarding different attacks against defense in query-based attacks, we specifically evaluate decision-based black-box attacks and score-based black-box attacks using targeted attacks with $\ell_{\infty}$ norm bounded to be 0.03 and the maximum query number is set to be 100,000.

Regarding training-time-based defense versus inference-time-based defense, we directly add random Gaussian noise on each input to implement RND\cite{qin2021random} while deploying the AAA model class to implement AAA-Linear\cite{chen2022adversarial} and AAA-Sine\cite{chen2022adversarial} respectively according to our configuration in our code. 

Regarding intra/cross-training-strategy transferability analysis, we adhere to the configuration specified in the intra/cross-family transferability analysis. However, the models to compose surrogate-target pairs are chosen from the regular and Adversarial Training versions of three distinct model architectures.

\section{Hyper-parameters}
\label{app: hyperparameters}
The hyper-parameters for score-based attacks, decision-based attacks and transfer-based attacks are specified in Tab. \ref{tab:hyperparameter_score}, Tab. \ref{tab:hyperparameter_decision} and Tab. \ref{tab:hyperparameter}. Due to potential variation in effectiveness across the different surrogate model architectures and datasets, hyper-parameter configurations are meticulously detailed for each scenario. One can reproduce all the evaluation results with these configurations. Additionally, we provide our rules to select the hyper-parameters, as follows:
\begin{itemize}[leftmargin=15pt,itemsep=1pt,topsep=0pt]
    \item \par \textbf{Whenever available, we adopt the hyper-parameter configuration recommended in the original paper.}
\par \par Generally, the original paper will provide the the values of hyper-parameters specific to one or some scenarios to help reader reproduce their experimental results. For example, the Bayesian Attack\cite{bayesian} outlined their experimentation details: "We tested untargeted $\ell_{\infty}$ attacks in the black-box setting, ... On CIFAR-10, ... and used ResNet-18 as source model. While on ImageNet, ... and used ResNet-50 as the source model. ... In possible finetuning, we set $\gamma=0.1 /\left\|\Delta \mathbf{w}^*\right\|_2$ and a finetuning learning rate of 0.05 if SWAG was incorporated. ... We use an SGD optimizer with a momentum of 0.9 and a weight decay of 0.0005 and finetune models for 10 epochs on both CIFAR-10 and ImageNet. We set the batch size of 128 and 1024 on CIFAR-10 and ImageNet, respectively. ...". It provides the configuration for $\{untargeted, L_\infty\}$ white-box attack against ResNet-18 and ResNet-50 on CIFAR-10 and ImageNet datasets. In these instances, we adhere to the prescribed values provided by the authors in the corresponding scenarios.
    \item \par \textbf{For the scenarios without suggestions, we engage in the search for reasonable, effective hyper-parameter values when datasets and surrogate models architectures vary, then fix the values for different attack types.}
\par \par Taking the above Bayesian Attack as an example, maintaining hyper-parameter consistency across different attack settings could guarantee the fairness of our evaluations. However, this set of hyper-parameters for finetuning surrogate models needs to adapt to different surrogate model architectures. Similar considerations apply to methods like DI2-FGSM \cite{difgsm}, where image resizing hyper-parameters are irrelevant to attack settings but contingent upon distinct datasets. 
\end{itemize}

\section{Unified algorithms}
\label{Unified algorithms}
\begin{figure}[h]
 \setlength{\abovecaptionskip}{-0.1cm}
\setlength{\belowcaptionskip}{-0.1cm}
 \centering
\includegraphics[width=9cm]{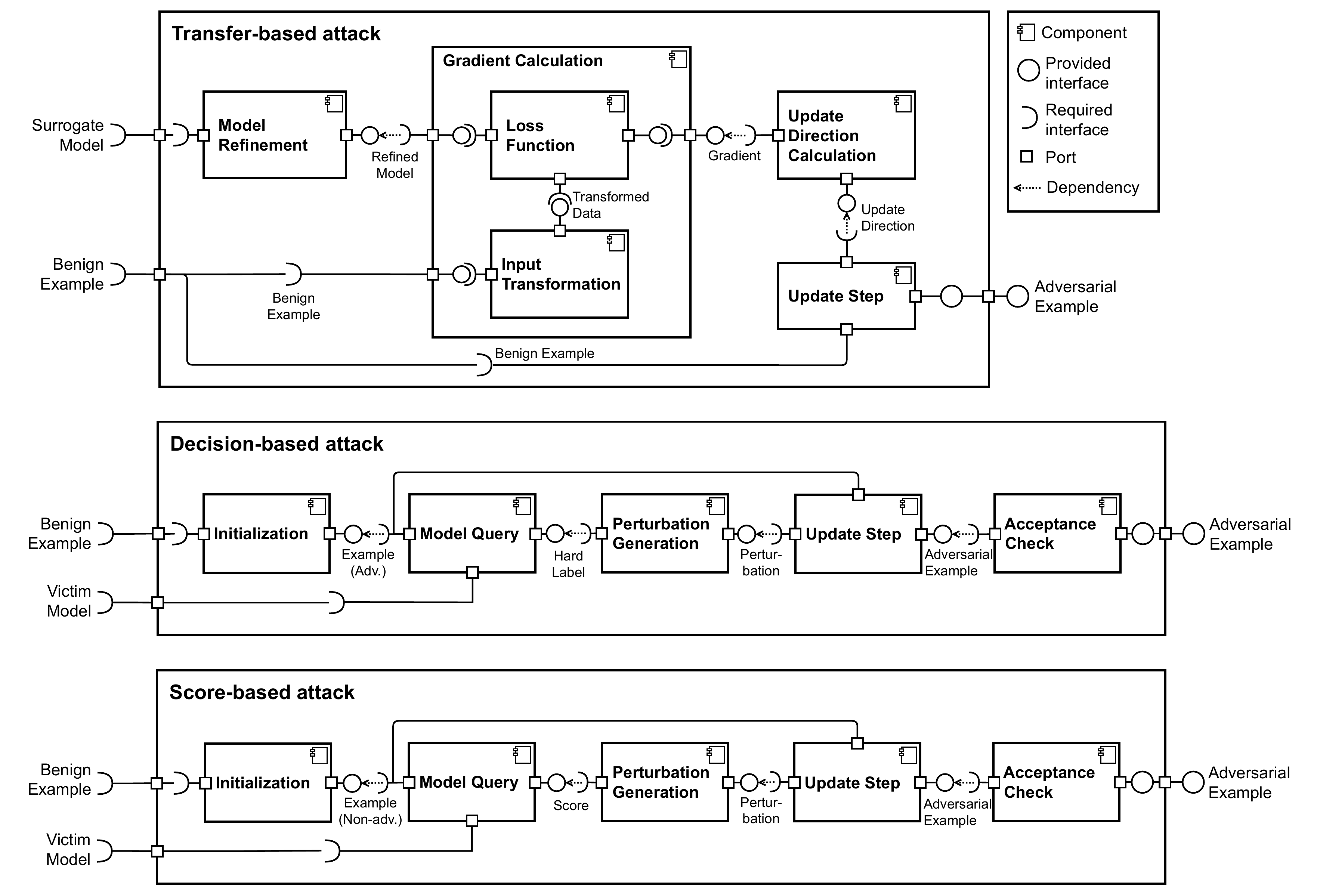}
 \caption{UML component diagrams of attack modules for transfer-based attack methods (\textbf{top}), decision-based attack methods (\textbf{middle}), and score-based attack methods (\textbf{bottom}).}\label{uml}
\end{figure}

BlackboxBench unifies decision-based attacks, score-based attacks and transfer-based attacks in Alg.\ref{alg: decision}, Alg.\ref{alg: score} and Alg.\ref{alg: transfer} respectively. Any attack process in BlackboxBench could be composed of the functional blocks in this algorithm. \blue{Based on the unified pipelines, we build our attack module, as illustrated by UML component diagrams in Fig. \ref{uml}.}

\subsection{Decision-based attack algorithm}\label{app: decision unified algorithm}
\begin{algorithm}[ht]\label{alg: decision}
\scriptsize
\caption{Unified Decision-based Black-box Adversarial Attacks}
\hspace*{0.02in} {\bf Input:} Target model $f$, Benign example $\boldsymbol{x}$, Label $y$: The true label of the input example $\boldsymbol{x}$; \\
\hspace*{0.02in} {\bf Hyper-parameter:} Maximum perturbation $\epsilon$, Step size $\alpha$, Max query numbers\\
\hspace*{0.02in} {\bf Attack-specific Functions:} Adversarial example initialization function $\mathcal{I}$, Perturbation generation function $\mathcal{G}$\\
\begin{algorithmic}[1]
\STATE \textbf{Initialize:} Initialize the adversarial example as $\boldsymbol{x}_0^* \leftarrow \mathcal{I}(\boldsymbol{x})$;
\WHILE{$t < N$ \textbf{and} ${\left\|\boldsymbol{g}_{t}\right\|_p} < \epsilon$} 
\STATE \textcolor{brown}{\# Perturbation generation step}
\STATE Generate the perturbation $\boldsymbol{g}_{t+1}$ using the perturbation generation function $\mathcal{G}\left(\boldsymbol{x}_t^*\right)$;
\STATE \textcolor{brown}{\# Update adversarial example}
\STATE Update the adversarial example $\boldsymbol{x}_{t+1}^*$ by applying the perturbation $\boldsymbol{g}_{t+1}$ with the step size $\alpha$: 
$$ \boldsymbol{x}_{t+1}^* \leftarrow \boldsymbol{x}_t^* + \alpha \cdot \boldsymbol{g}_{t+1} $$
\STATE \textcolor{brown}{\# Acceptance check on the updated adversarial example}
\IF{$\boldsymbol{x}_{t+1}^*$ satisfies the acceptance criteria (e.g., the adversarial example leads to a misclassification by $f$)}
\STATE \textbf{Break the loop}
\ENDIF
\ENDWHILE
\RETURN The final adversarial example $\boldsymbol{x}^* = \boldsymbol{x}_N^*$.
\end{algorithmic}
\end{algorithm}


\subsection{Score-based attack algorithm}\label{app: score unified algorithm}
\begin{algorithm}[ht]\label{alg: score}
\scriptsize
\caption{Unified Score-based Black-box Adversarial Attacks}
\hspace*{0.02in} {\bf Input:} 
Target model $f$, Benign example $\boldsymbol{x}$, Label $y$\\
\hspace*{0.02in} {\bf Hyper-parameter:} 
Maximum perturbation $\epsilon$, Step size $\alpha$, Max query numbers $N$\\
\hspace*{0.02in} {\bf Attack-specific Functions:}
Adversarial example initialization function $\mathcal{I}$, Perturbation generation function $\mathcal{G}$\\
\begin{algorithmic}[1]
\STATE \textbf{Initialize:} Initialize the adversarial example as $\boldsymbol{x}_0^* \leftarrow \mathcal{I}(\boldsymbol{x})$;
\WHILE{$t < N$ \textbf{and} $\boldsymbol{x}_t^*$ is not adversarial (i.e., $f(\boldsymbol{x}_t^*)$ predicts $y$ with high confidence)}
\STATE \textcolor{brown}{\# Perturbation generation step}
\STATE Generate the perturbation $\boldsymbol{g}_{t+1}$ using the perturbation generation function $\mathcal{G}\left(\boldsymbol{x}_t^*\right)$;
\STATE \textcolor{brown}{\# Update adversarial example}
\STATE Update the adversarial example $\boldsymbol{x}_{t+1}^*$ by applying the perturbation $\boldsymbol{g}_{t+1}$ with the step size $\alpha$: 
$$ \boldsymbol{x}_{t+1}^* \leftarrow \boldsymbol{x}_t^* + \alpha \cdot \boldsymbol{g}_{t+1} $$
\STATE \textcolor{brown}{\# Check if updated adversarial example is successful}
\IF{$f(\boldsymbol{x}_{t+1}^*) \neq y$ \textbf{or} the confidence of $y$ is sufficiently reduced}
\STATE \textbf{Break the loop}
\ENDIF
\ENDWHILE
\RETURN The final adversarial example $\boldsymbol{x}^* = \boldsymbol{x}_t^*$.
\end{algorithmic}
\end{algorithm}

\subsection{Transfer-based attack algorithm}\label{app: transfer unified algorithm}
\begin{algorithm}[ht]\label{alg: transfer}
\scriptsize
\caption{Unified Transfer-based Black-box Adversarial Attacks}
\hspace*{0.02in} {\bf Input:}
A surrogate model $f^{\prime}$ with parameter $\theta^{\prime}$, a benign example $\boldsymbol{x}$, label $y$;\\
\hspace*{0.02in} {\bf Hyper-parameters:}
Maximum perturbation $\epsilon$, step size $\alpha$, iterations $T$, hyper-parameter set specific to attacks;\\
\hspace*{0.02in} {\bf Functions specific to attacks:} Input transformation function $\mathcal{T}$, loss function $\mathcal{L}$, gradient calculation function $\widetilde{\nabla}$, update direction calculation function $\mathcal{U}$, surrogate model refiner $\mathcal{R}$; \\
\begin{algorithmic}[1]
\label{alg1}
\STATE $\boldsymbol{g}_0=0 ; \boldsymbol{x}_0^*=\boldsymbol{x}$;
\STATE Obtain auxiliary information $\textbf{I}$; obtain the refined surrogate model $f^{\prime\prime}=\mathcal{R}(f^{\prime})$, parameterized by $\theta^{\prime\prime}$;
\FOR{$t=0$ to $T-1$}
\STATE \textcolor{brown}{\# Input transformation block}
\STATE Get the transformed images $\mathcal{T}\left(\boldsymbol{x}_t^*\right)$;
\STATE \textcolor{brown}{\# Gradient calculation block}
\STATE Input $\mathcal{T}\left(\boldsymbol{x}_t^*\right)$ to $f^{\prime\prime}$ and obtain the loss $\ell=\mathcal{L}\left(f^{\prime\prime}\left(\mathcal{T}\left(\boldsymbol{x}_t^*\right);\theta^{\prime\prime}\right),y;\textbf{I}\right)$;
\STATE Input the loss $\ell$ to $\widetilde{\nabla}$ and calculate the gradient $\widetilde{\nabla}_x(\ell)$;
\STATE\textcolor{brown}{\# Update direction calculation block}
\STATE Update perturbation $\boldsymbol{g}_{t+1}$ as $\mathcal{U}(\widetilde{\nabla}_x(\ell))$;
\STATE\textcolor{brown}{\# Update step}
\STATE Update $\boldsymbol{x}_{t+1}^*$ by one update step specific to different norms: \\
\STATE \quad \quad \quad
$\ell_\infty:\boldsymbol{x}_{t+1}^*=\operatorname{Clip}_x^\epsilon\left\{\boldsymbol{x}_t^*+\alpha \cdot \operatorname{sign}\left(\boldsymbol{g}_{t+1}\right)\right\} \quad
L_2:\boldsymbol{x}_{t+1}^*=\operatorname{Clip}_x^\epsilon\left\{\boldsymbol{x}_t^*+\alpha \cdot \frac{\boldsymbol{g}_{t+1}}{\left\|\boldsymbol{g}_{t+1}\right\|_2}\right\}
$
\ENDFOR
\RETURN $\boldsymbol{x}^*=\boldsymbol{x}_T^*$.
\end{algorithmic}
\end{algorithm}

\section{Description of black-box adversarial attacks}
Here we provide a detailed description of all implemented attacks in BlackboxBench.
\subsection{Description of decision-based attacks}\label{app: description of decision-based attack algorithms}
\textbf{Random search methods}
\begin{itemize}[leftmargin=15pt,itemsep=1pt,topsep=0pt]
    \item \textbf{Boundary Attack} \cite{brendel2017decision} Boundary attack initializes with an already adversarial example and then gradually reduces the distance to the target image using rejection sampling with a suitable proposal distribution, often a normal distribution. By following a given adversarial criterion, the attack aims to find smaller adversarial perturbations.
    \item \textbf{Evolutionary Attack} \cite{dong2019efficient} Evolutionary attack utilizes the refined covariance matrix adaptation evolution strategy (CMA-ES) \cite{hansen2001completely} to generate adversarial examples. This evolutionary approach improves the efficiency of the attack by employing the Gaussian distribution to example noises, thereby capturing the local geometries of the search directions.
    \item \textbf{Geometric Decision-based Attack (GeoDA)} \cite{rahmati2020geoda} GeoDA operates under the intuition that small adversarial perturbations should be sought in directions where the decision boundary of the classifier closely approaches the data examples. By leveraging the low mean curvature of the decision boundary in the vicinity of the data examples, GeoDA effectively estimates the normal vector to the decision boundary, aiding in the search for perturbations.
    \item \textbf{Sign Flip Attack (SFA)} \cite{chen2020boosting} SFA takes a gradient-free iterative approach to craft adversarial perturbations. In each iteration, SFA reduces the distance (usually measured using the $\ell_{\infty}$ norm) through projection and randomly flips the signs of partial entries in the adversarial perturbation, progressively moving closer to an optimal perturbation.
    \item \textbf{Rays} \cite{chen2020rays} Rays tackles the problem of finding the nearest decision boundary by reformulating it as a discrete optimization problem. It directly searches among a set of ray directions generated from the solutions of the optimization problem, employing the $\ell_{\infty}$ norm as the attack criterion.
\end{itemize}
\textbf{Gradient-estimation-based methods}
\begin{itemize}[leftmargin=15pt,itemsep=1pt,topsep=0pt]
    \item \textbf{OPT} \cite{cheng2018query} OPT transforms the decision-based black-box attack into a real-valued continuous optimization problem and employs the Random Gradient-Free (RGF) method to locate the stationary point. Notably, OPT guarantees the convergence of decision-based black-box attacks, marking a significant advancement in this field.
    \item \textbf{Sign-OPT} \cite{cheng2019sign} Sign-OPT turns to estimate the sign of the gradient instead of the gradient itself. This approach offers the advantage of requiring only a single query in each iteration. 
    \item \textbf{HopSkipJumpAttack (HSJA)} \cite{chen2020hopskipjumpattack} HSJA introduces a novel framework for decision-based black-box attacks. There are mainly 3 modules for attackers to implement in each iteration. Attackers need to estimate the gradient for the next update first, search for the step size through geometric progression, and finally perform a binary search to project the point onto the decision boundary. HSJA utilizes Monte Carlo estimation to approximate the gradient, and subsequent works have proposed improved strategies for gradient estimation. 
    \item \textbf{Query-Efficient Boundary-based black-box Attack (QEBA)} \cite{li2020qeba} QEBA proposes gradient estimation within a small representative subspace, employing techniques such as spatial transformation, discrete cosine transformation (DCT).
    \item \textbf{NonLinear-BA} \cite{li2021nonlinear} NonLinear-BA follows the overall framework of HSJA as well, and leverages popular deep generative models such as AEs, VAEs, and GANs to estimate gradients, which corresponds to the gradient estimation by Monte Carlo method in HSJA.
    \item \textbf{Progressive-Scale based projective Boundary Attack (PSBA)} \cite{zhang2021progressive} Based on the research of HSJA, QEBA, and NonLinear-BA, PSBA is proposed. PSBA also follows the attack framework of HSJA and incorporates gradient estimation strategies similar to QEBA and NonLinear-BA. It progressively searches for the optimal scale in a self-adaptive manner across spatial, frequency, and spectrum scales.
    \item \textbf{Triangle Attack} \cite{wang2022triangle} Triangle attack breaks away from the requirement of adversarial points being restricted to the decision boundary or gradient estimation at each iteration. Triangle attack builds triangles iteratively based on the current adversarial example and learned or searched angles. It aims to find an adversarial point by utilizing geometric information in the low-frequency space generated by DCT\cite{ahmed1974discrete}. 
    \item \blue{\textbf{Customized Iteration and Sampling Attack (CISA)}\cite{cisa} CISA) enhances query efficiency by integrating surrogate model information and iterative sampling strategies. CISA addresses the challenges of low query efficiency in decision-based attacks by incorporating a transfer-based phase using a surrogate model. This phase generates intermediate adversarial examples, which are then refined using customized sampling based on noise sensitivity. By utilizing the surrogate model's gradients to guide dual-direction iterative updates, CISA improves the likelihood of crossing the decision boundary with minimal noise. The approach also adapts step sizes and sampling variance based on previous queries, leading to more efficient noise compression and fewer queries overall.}
\end{itemize}
\subsection{Description of score-based attacks}\label{app: description of score-based attack algorithms}
\textbf{Random search methods}
\begin{itemize}[leftmargin=15pt,itemsep=1pt,topsep=0pt]
    \item \textbf{Simple Black-box Attack (SimBA)} \cite{guo2019simple} SimBA employs a stochastic approach by randomly selecting a direction, which will be sampled from a set of orthonormal basis vectors, to perform either addition or subtraction for the perturbation at each step.
    \item \textbf{Parsimonious Attack} \cite{moon2019parsimonious} Parsimonious attack introduces a novel methodology to determine the perturbation among the vertices of the $\ell_{\infty}$ ball denoted as $\mathbb{B}_{\boldsymbol{x}, \epsilon}$. Consequently, this formulation gives rise to a discrete set maximization problem.
    \item \textbf{Square Attack} \cite{andriushchenko2020square} Square attack also operates within the $\ell_{\infty}$ ball $\mathbb{B}_{\boldsymbol{x}, \epsilon}$, exploring the perturbation among its vertices. However, in contrast to SimBA and Parsimonious attacks, the Square attack restricts its search to a randomly exampled local patch at each step.
    \item \textbf{Projection \& Probability-driven Black-box Attack (PPBA)} \cite{li2020projection} PPBA pursues perturbation exploration within a low-dimensional and low-frequency subspace, which is constructed through the application of the discrete cosine transform (DCT) \cite{ahmed1974discrete}  and its inverse transform (IDCT). 
    \item \textbf{BABIES} \cite{tran2022exploiting} BABIES implements $\ell_2$ black-box adversarial attack in the frequency domain. BABIES leverages quadratic interpolation to approximate adversarial losses and guide perturbation updates, resulting in improved query efficiency for constraint optimization on an $\ell_2$ sphere.
\end{itemize}
\textbf{Gradient-estimation-based methods}
\begin{itemize}[leftmargin=15pt,itemsep=1pt,topsep=0pt]
    \item \textbf{Natural Evolution Strategy (NES)} \cite{ilyas2018black} NES employs the natural evolution strategy \cite{wierstra2014natural} to estimate gradients by sampling perturbations from Gaussian distribution and adjust the attack direction according to the feedback from the targeted model.
    \item \textbf{Bandits} \cite{ilyas2018prior} Babdits formulates the score-based attack into a bandit optimization problem. By considering different priors, such as time-dependent and data-dependent priors, Bandits can carry over latent vectors to improve gradient estimation.
    \item \textbf{AdvFlow} \cite{mohaghegh2020advflow} AdvFlow further expands upon the NES attack by replacing the Gaussian distribution with a complex distribution modeled by a pre-trained normalizing flow model \cite{tabak2013family}.
    \item \textbf{Neural Process based black-box attack (NP-attack)} \cite{bai2020improving} NP-attack incorporates the structure information of the image using Neural Process(NP)\cite{garnelo2018neural}, an efficient autoencoder method, to model a distribution over regression functions, allowing for reduced query counts and more efficient reconstruction of adversarial examples.
    \item \textbf{Zeroth-Order sign-based Stochastic Gradient Descent (ZO-signSGD)} \cite{liu2019signsgd} ZO-signSGD proposes updating the perturbations based on the idea of signSGD by incoporating zeroth-order optimization. The sign operation in ZO-signSGD can mitigate the negative effect of gradient noise.
    \item \textbf{SignHunter} \cite{al2020sign} SignHunter introduces a divide-and-conquer strategy to expedite gradient sign direction estimation. Instead of sequentially flipping the gradient sign of each pixel, SignHunter performs group-wise flipping operations on specific regions of the image. 
\end{itemize}
\textbf{Combination-based methods}
\begin{itemize}[leftmargin=15pt,itemsep=1pt,topsep=0pt]
    \item \blue{\textbf{Subspace Attack}\cite{subspaceattack} Subspace Attack leverages information from a surrogate model by combining it with low-dimensional subspace exploration to enhance the attack's efficiency. The surrogate model provides an approximate gradient or important directions that guide the selection of the subspace in which the search for adversarial perturbations occurs. By using the surrogate model's gradient estimates, Subspace Attack refines its search within this reduced space, ensuring that the perturbations are aligned with the target model's vulnerabilities. This combination of surrogate model information and subspace optimization allows the attack to achieve high success rates with fewer queries.}
    \item \textbf{$\mathcal{CG}$-Attack} \cite{feng2022boosting} $\mathcal{CG}$-Attack leverages the evolution strategy (ES)\cite{beyer2002evolution} and a Conditional Glow (c-Glow, $\mathcal{CG}$) model as the search distribution. By utilizing the information of a surrogate model, $\mathcal{CG}$-Attack learned the parameters of the c-Glow model. When attacking the target model, $\mathcal{CG}$-Attack refers to parital parameters learned from the c-Glow model while updating the other parameters based on feedback from the target model in the attacking process. This partial transfer mechanism addresses the issue of surrogate biases and allows for flexible adjustment of the conditional adversarial distribution (CAD) for the target model. 
    \item \blue{\textbf{Blackbox Attacks via Surrogate Ensemble Search (BASES)}\cite{bases} BASES combines surrogate model information with a novel query-efficient optimization technique. The key idea is to use a perturbation machine (PM) that generates adversarial examples by minimizing a weighted loss function over a set of surrogate models. Each surrogate model in the ensemble is assigned a weight, and the weights are iteratively adjusted using feedback from the target model's output. By leveraging surrogate models, BASES efficiently explores the adversarial space in a low-dimensional search space, significantly reducing the number of queries required while maintaining a high success rate in fooling the target model.}
    \item \blue{\textbf{Prior-guided Random Gradient-Free (PRGF)}\cite{prgf} PRGF also combines score-based black-box attack techniques with surrogate model information to efficiently estimate the gradient of the target model. The surrogate model provides gradient estimates, which are used to guide the search for adversarial perturbations. PRGF projects random perturbations onto the gradient directions suggested by the surrogate model, refining the estimation of the target model's gradient based on its output scores. This projection strategy allows PRGF to reduce the number of queries required while maintaining high attack success rates, as the surrogate model helps steer the attack toward the most vulnerable directions.}
    \item {\textbf{Meta Conditional Generator (MCG)} \cite{yin2023generalizable} MCG utilizes the information from surrogate models to boost the attack performance. MCG  involves three steps: (a) fine-tuning the meta generator using inner optimization with adversarial loss function, (b) fine-tuning the surrogate model using historical information and the current task to mimic the behavior of the target model, and (c) attacking the target black-box model using the adapted generator combined with off-the-shelf attack methods, where the generated adversarial examples are used to refine the surrogate model.}
\end{itemize}

\subsection{Description of transfer-based attacks}\label{app: description of transfer-based attack algorithms}
\begin{itemize}[leftmargin=15pt,itemsep=1pt,topsep=0pt]
    \item \textbf{Iterative Fast Gradient Sign Method (I-FGSM) \cite{ifgsm}} I-FGSM is the first work to optimize the general function of transfer-base attacks multiple times with a small step size and iteratively update adversarial examples. Compared with the one-step attack FGSM\cite{goodfellow2014explaining}, I-FGSM performs better in the white-box setting and worse in the black-box setting.
\end{itemize}
\textbf{Optimization perspective}
\begin{itemize}[leftmargin=15pt,itemsep=1pt,topsep=0pt]
    \item \textbf{Projected Gradient Descent (PGD)}\cite{pgd} PGD is an iterative version of FGSM\cite{goodfellow2014explaining} starting from a random perturbation in the $\ell_p$-ball. 
    \item \textbf{Momentum Iterative Fast Gradient Sign Method (MI-FGSM)}\cite{mifgsm} To mitigate the overfitting to white-box models, MI-FGSM integrates the idea of momentum to the gradient ascent of I-FGSM to escape from poor local maxima.
    \item \textbf{Nesterov Iterative Fast Gradient Sign Method (NI-FGSM)}\cite{nisifgsm} Beyond the stability from MI-FGSM, NI-FGSM further leverages the looking ahead property of Nesterov Accelerated Gradient to escape from poor local maxima easier and faster.
    \item \textbf{Pre-gradient guided momentum Iterative Fast Gradient Sign Method (PI-FGSM)}\cite{pifgsm} Given that the accumulated momentum in NI-FGSM does not serve as an optimal look ahead direction, PI-FGSM instead looks ahead by relying on the gradient of last iteration.
    \item \textbf{Variance Tuning (VT)}\cite{vt} VT tunes the gradient of the adversarial example at each iteration with the gradient variance from last iteration, thereby enhancing the stability of the update direction.
    \item \textbf{Reverse Adversarial Perturbation (RAP)}\cite{rap} When adversarial examples are located at a sharply local maximum, even a minor misalignment between the surrogate and the target results in a large decrease in adversarial loss. Thus, RAP encourages adversarial examples not only of high adversarial loss but also within local flat regions.
    \item \textbf{Linear BackPropagation (LinBP)}\cite{linbp} Motivated by Goodfellow et al.'s interpretation of adversarial examples as a result of linearity of DNNs \cite{goodfellow2014explaining}, LinBP enhances linearity by backpropagating the loss as if no non-linear activation function is encountered in the forward pass.
    \item \textbf{Skip Gradient Method (SGM)}\cite{sgm} Observed that gradients flowing through skip connections are more transferable, SGM crafts adversarial examples by utilizing more gradients from skip connections rather than residual modules.
\blue{\item \textbf{Penalizing Gradient Norm (PGN)}\cite{pgn} PGN enhances adversarial example transferability by targeting flat local regions of the loss landscape through an efficient approximation of the gradient regularization norm.}
\end{itemize}
\textbf{Data perspective}
\begin{itemize}[leftmargin=15pt,itemsep=1pt,topsep=0pt]
    \item \textbf{Diverse Inputs Iterative Fast Gradient Sign Method (DI2-FGSM)\cite{difgsm}} Inspired by applying data augmentation to prevent networks from overfitting, DI2-FGSM adds inputs transformed by random resizing and random padding with probability $p$ into the attack process for the sake of transferability.
    \item \textbf{Scale-invariant Iterative Fast Gradient Sign Method (SI-FGSM)}\cite{nisifgsm} SI-FGSM augments models via the loss-preserving input transformation, scaling, to alleviate overfitting on white-box models.
    \item \textbf{Admix\cite{admix}} To improve the attack transferability, admix adopts information from images in other classes by admixing two images in a master and slave manner for the gradient calculation.
    \item \textbf{Translation-invariant Fast Gradient Sign Method (TI-FGSM)}\cite{tifgsm} As adversarial examples generated against a white-box model are highly correlated with the discriminative region of the white-box model at the given input point, the tranferability to models with different discriminative regions is reduced. TI-FGSM shifts images by translation operation to generate adversarial examples less sensitive to the discriminative regions of the white-box model.
\blue{    \item \textbf{Structure Invariant Attack (SIA)}\cite{sia} Existing input-transformation-based attacks, which apply a single global transformation globally, produce limited diversity in transformed images, thereby restricting their transferability. SIA enhances the diversity of adversarial examples by applying different transformations locally across various parts of an input image without altering the image's global structure to improve its transferability.}

\end{itemize}
\textbf{Feature perspective}
\begin{itemize}[leftmargin=15pt,itemsep=1pt,topsep=0pt]
    \item \textbf{Feature Importance-aware Attack (FIA)}\cite{fia} FIA generates adversarial examples by only distorting “important” (object-related) features, where the feature importance is defined as how the features contribute to the final decision, intuitively, gradient. FIA aggregates gradients from the randomly masked input to highlight object-aware gradients and neutralize model-specific gradients.
    \item \textbf{Neuron Attribution-based Attack (NAA)}\cite{naa} Beyond FIA, NAA provides more accurate neuron importance measures that not only completely attribute a model’s output to each neuron but also reflect the polarity and magnitude of neuron importance.
    \item \textbf{Intermediate Level Attack (ILA)}\cite{ila} ILA expects to use the suboptimal perturbation found by the basic attack as a proxy and stray from it in exchange for increasing the norm. Considering increasing the norm of perturbation in image space is perceptible, FIA increases the norm in feature space.
\end{itemize}
\blue{\textbf{Model perspective: model generation}}
\begin{itemize}[leftmargin=15pt,itemsep=1pt,topsep=0pt]
    \item \textbf{Random Directions (RD)}\cite{lgv} RD adds random directions in the weight space to a regularly trained DNN to increase its transferability. The success is sensitive to the magnitude of noise but demonstrates the feasibility of exploiting variations in the weight space.
    \item \textbf{Ghost Networks (GhostNet)}\cite{ghostnet} GhostNet generates a vast number of ghost networks on the fly by imposing erosion on certain intermediate structures of the base network and performs longitudinal ensemble attack on this a huge candidate pool.
    \item \textbf{Distribution-Relevant Attack (DRA)}\cite{dra} The main intention is dragging the adversarial image into the target distribution to mislead classifiers to classify the image as the target class. DRA achieves this by approximating the gradient of the target class data distribution through score matching and subsequently moving the initial image towards it using Langevin Dynamics.
    \item \textbf{Intrinsic Adversarial Attack (IAA)}\cite{iaa} Similiar to DRA, IAA also boosts the alignment between adversarial attack and intrinsic attack but further replaces activation function ReLU with Softplus to make the classifier smoother and decrease the weight for certain residual modules to reduce those impact.
    \item \textbf{Large Geometric Vicinity (LGV)}\cite{lgv} LGV collects weights in a single run along the SGD trajectory with a high constant learning rate and iteratively attacks the collected models. The high constant learning rate is key to sample much wider models with better generalization and consequently produce flatter adversarial examples same as RAP.
    \item \textbf{Stochastic Weight Averaging (SWA)}\cite{bayesian} SWA samples models in a sufficiently large vicinity via SGD with a high constant learning rate and averages the weights to move into a broader space. SWA needs to update the batch-normalization statistics of the averaged model before attacks against it.
    \item \textbf{Bayesian Attack\cite{bayesian}} To consider model diversity as well as low time complexity, Bayesian Attack models the Bayesian posterior of the surrogate model, obtaining an ensemble of infinitely many DNNs from a single run of training. Adversarial examples are subsequently generated on this Bayesian model.
\end{itemize}
\blue{\textbf{Model perspective: model fusion}
\begin{itemize}[leftmargin=15pt,itemsep=1pt,topsep=0pt]
    \item \textbf{Logit ensemble}\cite{mifgsm} Based on the fact that an adversarial example effective across multiple models may reveal an intrinsic transferable weakness, logit ensemble method attacks multiple models simultaneously by fusing their logit activations.
    \item \textbf{Loss ensemble}\cite{mifgsm} As an alternative ensemble scheme, loss ensemble method aggregates multiple surrogate models in loss.
    \item \textbf{Longitudinal ensemble}\cite{ghostnet} By attack only one model from the set of surrogate models rather than using all in each iteration, longitudinal ensemble method significantly reduces computational overhead compared to traditional ensemble approaches, facilitating efficient generation of adversarial examples.
    \item \textbf{Common Weakness Attack (CWA)}\cite{cwa} CWA enhances transferability by exploiting two key properties of model ensembles: the flatness of the loss landscape and proximity to models' local optima. By leveraging Sharpness Aware Minimization (SAM) and Cosine Similarity Encourager (CSE) to optimize these properties, CWA generates more transferable adversarial examples across different models.
    \item \textbf{Adaptive ensemble attack (AdaEA)}\cite{adaea} To address the limitation of traditional ensemble adversarial attacks that equally fuse outputs from surrogate models, AdaEA adaptively modulates the fusion of model outputs based on their contribution to the adversarial goal and introduces a disparity-reduced filter to align update directions. (It is important to note that the hyperparameters $\eta$ and $\beta$ of AdaEA have a high correlation with the surrogate models used. Care must be taken when setting their values.)
\end{itemize}}

\section{Description of black-box adversarial defenses}
\label{app: description of black-box adversarial defenses}

\subsection{Description of static defenses}
\label{app: description of static defenses}
Static defenses refer to defense strategies that are established during or before the training phase of the model and do not change in response to varying attacks. The most widely-used static defense method is \textbf{Adversarial Training (AT)}
\begin{itemize}[leftmargin=15pt,itemsep=1pt,topsep=0pt]
    \item \textbf{Adversarial Training (AT)}  AT is a method in machine learning designed to improve the robustness of models against adversarial examples—inputs crafted to deceive models into making mistakes. This method is particularly relevant in the field of deep learning where neural networks can be surprisingly vulnerable to slight, often imperceptible, perturbations of their inputs. During the training phase, the model is exposed not only to the original training data but also to adversarial examples. These adversarial examples are typically generated by applying small but strategically chosen perturbations to the training data, intended to mislead the model. By training on both regular and adversarial examples, the model learns to generalize better and become more resistant to adversarial manipulation.
\end{itemize}

\subsection{Description of dynamic defenses}
\label{app: description of dynamic defenses}
Dynamic defenses refer to defense strategies that can dynamically adjust in response to changes in adversarial attacks. Dynamic defense is particularly important in this field because attackers often change their attack strategies to circumvent existing defense mechanisms. Besides, dynamic defenses are often lightweight since there is no cost for training. 
\begin{itemize}[leftmargin=15pt,itemsep=1pt,topsep=0pt]
    \item \textbf{Random Noise Defense (RND)\cite{qin2021random}} RND is to add random noise to the responses of machine learning models during the inference stage for each query, which adds uncertainty for attackers attempting to make queries. This uncertainty makes it difficult for attackers to accurately discern the model's behavior or extract sensitive information from the model. RND also aims to find a balance between maintaining model accuracy and enhancing defense performance, which is achieved by adjusting the magnitude of the noise to ensure effective defense without overly compromising accuracy.
    \item \textbf{Adversarial Attack on Attackers (AAA)\cite{chen2022adversarial}}  AAA introduces a defense mechanism  to particularly counteract score-based query attacks (SQAs) on target models. The AAA approach aims to mitigate SQAs by manipulating the output logits of the model. This manipulation is designed to confuse the SQAs, leading them towards incorrect attack directions. Specifically, AAA-sine utilizes a sine function or a similar periodic function to adjust or perturb the model's outputs (logits). This perturbation aims to confuse and mislead score-based query attacks (SQAs), making it difficult for attackers to accurately estimate how to modify inputs to create adversarial samples. AAA-line adopts linear functions or adjustments in the model's output logits. This could apply a straightforward, possibly constant or proportional change to the logits to disrupt the attack patterns.
\end{itemize}

\section{Description of analytical tools}
\label{app: description of analytical tools}
In this section, we provide implementation details of our analytical tools.

\textbf{Attack-component effect visualization tool} In BlackboxBench, we expand the scopes of the three components of the attack scenario to include values under-explored in existing literature. This strategic expansion serves a twofold purpose: from the perspective of methods, they could be tested more rigorously. From the perspective of components, their effects on attack performance could be thoroughly investigated. During the assessment of each component across its extended range, the other two components are fixed. The outcomes are graphically summarized through line charts, scatter charts, bar charts, bubble charts, and heatmaps.

\textbf{Model attention visualization tool} BlackboxBench visualizes model attention by means of saliency map-based interpretability methods. Saliency methods are useful tools to comprehend neural networks by assigning each input feature a score to measure its distribution. Specifically, we employ \textit{Full-Gradient Saliency Maps} (FullGrad)\cite{fullgrad} and \textit{Frequency Saliency Map} (FSM)\cite{backdoorbench} to achieve interpretability in pixel and frequency domains respectively.

FullGrad is a variant of a saliency map that not only captures the importance of each pixel but also provides that of groups of pixels, which correspond to structures. Consider a neural network function $f: \mathbb{R}^D \rightarrow \mathbb{R}$ with inputs $\boldsymbol{x} \in \mathbb{R}^D$. Let $c_l$ be the channel of layer $l$ in a neural network. Let $\psi\left(\cdot\right)$ be the post-processing operations. The FullGrad saliency map is given by
\begin{equation}
    S_f(\boldsymbol{x})=\psi\left(\nabla_{\boldsymbol{x}} f(\boldsymbol{x}) \odot \boldsymbol{x}\right)+\sum_{l \in L} \sum_{c \in c_l} \psi\left(f^b(\boldsymbol{x})_c\right),
\end{equation}
where $\nabla_{\boldsymbol{x}} f(\boldsymbol{x})$ is the input-gradient, $f^b(\boldsymbol{x})$, short-hand notation for $\nabla_b f(\boldsymbol{x}, \mathbf{b}) \odot \mathbf{b}$, is the bias-gradient. Together, they constitute full-gradients.

FSM further measures the importance of each frequency feature. Consider a neural network function $f: \mathbb{R}^{H \times W \times C} \rightarrow \mathbb{R}$ with inputs $\boldsymbol{x} \in \mathbb{R}^{H \times W \times C}$. Let $\mathcal{F}$ be the channel-wise Discrete Fourier Transform (DFT) and $\mathcal{F}^{-1}$ be the Inverse Discrete Fourier Transform (IDFT). We define the frequency counterpart of $f$, which is the classifier for spectrum $\tilde{\boldsymbol{x}}=\mathcal{F}(\boldsymbol{x})$, as follows:
\begin{equation}
F(\tilde{\boldsymbol{x}})=f(\boldsymbol{x})=f\left(\mathcal{F}^{-1}(\tilde{\boldsymbol{x}})\right).
\end{equation}
Inspired by the saliency map in the pixel domain, FSM is given by
\begin{equation}
    \begin{aligned}
S_F(\tilde{\boldsymbol{x}}) & =\frac{\partial F(\tilde{\boldsymbol{x}})}{\partial \tilde{\boldsymbol{x}}(u, v, c)} \\
& =\sum_{c=1}^C \sum_{h=0}^{H-1} \sum_{w=0}^{W-1} \sum_{c^{\prime}=0}^C \frac{\partial f(\boldsymbol{x})}{\partial \boldsymbol{x}\left(h, w, c^{\prime}\right)} \cdot \frac{\partial \boldsymbol{x}\left(h, w, c^{\prime}\right)}{\partial \tilde{\boldsymbol{x}}(u, v, c)}
\\
& =\sum_{c=1}^C \sum_{h=0}^{H-1} \sum_{w=0}^{W-1} \frac{\partial f_s(\boldsymbol{x})}{\partial \boldsymbol{x}(h, w, c)} e^{2 \pi i\left(\frac{u h}{H}+\frac{v w}{W}\right)}.
\end{aligned}\label{equ: fsm}
\end{equation}
As we can see, FSM is exactly the DFT of the model gradient.

\textbf{Adversarial divergence measurement tool} When harnessing adversarial transferability from transfer-based attacks, the performance is notably destroyed by the misalignment between a pair of surrogate model $f_s$ with parameter $\theta_s$ and target models $f_t$ with parameter $\theta_t$. To properly measure this misalignment, we consider conditional adversarial distribution (CAD), denoted as $\mathcal{P}(\boldsymbol{\delta}_{\boldsymbol{x}}|\boldsymbol{x}, \theta)$. As CAD models the distribution of perturbations $\boldsymbol{\delta}_{\boldsymbol{x}}$ conditioned on clean examples $\boldsymbol{x}$ for each model with parameter $\theta$, minimal divergence between CADs of a surrogate-target model pair ensures a high transferability between the surrogate and target models. The rationale behind is that if $\mathcal{P}(\boldsymbol{\delta}_{\boldsymbol{x}}|\boldsymbol{x}, \theta_s)$ and $\mathcal{P}(\boldsymbol{\delta}_{\boldsymbol{x}}|\boldsymbol{x}, \theta_t)$ yield similar probabilities for perturbation $\boldsymbol{\delta}_{\boldsymbol{x}}$, then the perturbation craft by $f_s$ can effectively substitute the authentic perturbation crafted by $f_t$, thus misleading $f_t$. Building on this premise, BlackobxBench quantifies this divergence between $\mathcal{P}(\boldsymbol{\delta}_{\boldsymbol{x}}|\boldsymbol{x}, \theta_t)$ and $\mathcal{P}(\boldsymbol{\delta}_{\boldsymbol{x}}|\boldsymbol{x}, \theta_s)$ as Adversarial Divergence (AD): 
\bluetwo{\begin{equation}
    \text{AD} = 1-\text{Norm}_{[0,1]}\left(\frac{1}{\log_2 \hat{D}_{\mathrm{KL}}(s,t)}\right),
\end{equation}
where $\hat{D}_{\mathrm{KL}}(s,t)$ denotes the empirical estimation of KL divergence between $\mathcal{P}(\boldsymbol{\delta}_{\boldsymbol{x}}|\boldsymbol{x}, \theta_t)$ and $\mathcal{P}(\boldsymbol{\delta}_{\boldsymbol{x}}|\boldsymbol{x}, \theta_s)$, \ie, ${D}_{\mathrm{KL}}(s,t)=\left[\mathbb{E}_{\mathcal{P}\left(\boldsymbol{\delta}_{\boldsymbol{x}} \mid \boldsymbol{x}, \theta_s\right)}\left[\log \frac{\mathcal{P}\left(\boldsymbol{\boldsymbol{\delta}_{\boldsymbol{x}}} \mid \boldsymbol{x}, \theta_s\right)}{\mathcal{P}\left(\boldsymbol{\boldsymbol{\delta}_{\boldsymbol{x}}} \mid \boldsymbol{x}, \theta_t\right)}\right]\right]$. $\text{Norm}_{[0,1]}\left(\cdot\right)$ denotes the max normalization. Let $\text{sim}\left(s,t\right)=\frac{1}{\log_2 \hat{D}_{\mathrm{KL}}(s,t)}$, $\text{Norm}_{[0,1]}\left(\text{sim}\left(s,t\right)\right)=\frac{\text{sim}\left(s,t\right)}{\max_{s,t}\text{sim}\left(s,t\right)}$. AD normalizes the KL divergence to be bounded within $[0, 1]$.}
\bluetwo{In the following, we will detail two important parts for computing Adversarial Divergence. We will firstly introduce the modeling of CAD using the conditional generative flow model (c-Glow); then, with the CAD of two models, we will present the computation of KL divergence between them.}

\bluetwo{Following \cite{feng2022boosting}, we learned the c-Glow model for each classifier $f$ to approximate its CAD.}

\begin{enumerate}
    \item \bluetwo{\textit{Modeling CAD by c-Glow:} The c-Glow model learns an invertible mapping between latent variable $\boldsymbol{z}_0 \sim \mathcal{N}(\mathbf{0}, \mathbf{I})$ and perturbation $ \boldsymbol{\delta}_{\boldsymbol{x}}$ on benign input $\boldsymbol{x}$:}
    \bluetwo{\begin{equation}
    \boldsymbol{\delta}_{\boldsymbol{x}}=g_{\boldsymbol{x}, \boldsymbol{\theta}}(\boldsymbol{z}_0)=g_{\boldsymbol{x}, \boldsymbol{\phi}}(\boldsymbol{z}),
    \end{equation}
    where $\boldsymbol{\theta} = (\boldsymbol{\phi}, \boldsymbol{\mu}, \boldsymbol{\sigma})$ is the model parameter. $\boldsymbol{\phi}=\left(\boldsymbol{\phi}_1, \ldots, \boldsymbol{\phi}_M\right)$ is the $M$ layers mapping parameter of c-Glow model which takes $\boldsymbol{z}=\boldsymbol{\mu}+\boldsymbol{\sigma} \odot \boldsymbol{z}_0$ as input.}

    \bluetwo{Then, the conditional likelihood is given by:
    \begin{equation}
    \begin{aligned}
         \log \mathcal{P}_{\boldsymbol{\theta}}( \boldsymbol{\delta}_{\boldsymbol{x}} \mid \boldsymbol{x})&=\log \mathcal{P}_{\mathbf{0}, \mathbf{1}}\left(\boldsymbol{z}_0\right)\\&+\sum_{i=1}^{M+1} \log \left|\operatorname{det}\left(\frac{\partial g_{\boldsymbol{x}, \boldsymbol{\phi}_i}^{-1}\left(\boldsymbol{r}_{i-1}\right)}{\partial \boldsymbol{r}_{i-1}}\right)\right|,
    \end{aligned}
    \end{equation}
    where $\boldsymbol{r}_i=g_{\boldsymbol{x}, \boldsymbol{\phi}_i}^{-1}\left(\boldsymbol{r}_{i-1}\right)$, $ \boldsymbol{r}_0=\boldsymbol{\delta}_{\boldsymbol{x}}$, $\boldsymbol{r}_M=\boldsymbol{z}$ and $\boldsymbol{r}_{M+1}=\boldsymbol{z}_0$.}

    \item \bluetwo{\textit{Learning CAD with energy-based model:} Directly training c-Glow by maximum likelihood requires collecting a large number of actual adversarial perturbations $\boldsymbol{\delta}_{\boldsymbol{x}}$, which is computationally expensive. To address this, the CAD is instead approximated by an energy-based model:
    \begin{equation}
        \log \mathcal{P}_E(\boldsymbol{\delta}_{\boldsymbol{x}} \mid \boldsymbol{x}) \approx-\lambda \cdot \mathcal{L}_{a d v}(\boldsymbol{\delta}_{\boldsymbol{x}}, \boldsymbol{x}),
    \end{equation}
    where $\mathcal{L}_{a d v}(\cdot)$ is the adversarial loss function, $\lambda$ is a positive hyper-parameter. 
    \textit{Instead of generating adversarial examples, the method samples a large number of perturbations $\boldsymbol{\delta}_{\boldsymbol{x}}$ from the perturbation ball $\mathbb{B}_{\epsilon, \boldsymbol{x}}$ around each clean input $\boldsymbol{x}$.}
    These samples are then scored by the energy model to estimate their importance for learning the CAD.}

    \bluetwo{The c-Glow model is trained to approximate the energy-based distribution $\mathcal{P}_E$ by minimizing the KL divergence:
    \begin{equation}
    \mathcal{L}=\mathbb{E}_{\mathcal{P}_E(\boldsymbol{\delta}_{\boldsymbol{x}} \mid \boldsymbol{x})}\left[\log \frac{\mathcal{\mathcal { P }}_E(\boldsymbol{\delta}_{\boldsymbol{x}} \mid \boldsymbol{x})}{\mathcal{\mathcal { P }}_{\boldsymbol{\theta}}(\boldsymbol{\delta}_{\boldsymbol{x}} \mid \boldsymbol{x})}\right].
    \end{equation}}
\end{enumerate}
The training details can be found in the supplementary materials in \cite{feng2022boosting}.

After modeling the CADs of a pair of models, we estimate the KL divergence between them, \ie, $\hat{D}_{\mathrm{KL}}(s,t)$, as follows:
\begin{enumerate}
    \item \textit{Sampling}: Sample a set of values \(\{\boldsymbol{\delta}_{\boldsymbol{x}}^{(i)}\}\) from the distribution \(\mathcal{P}(\boldsymbol{\delta}_{\boldsymbol{x}} \mid \boldsymbol{x}, \theta_s)\).
    
    \item \textit{Compute the Log Ratio for Each Sample}:
    For each sampled \(\boldsymbol{\delta}_{\boldsymbol{x}}^{(i)}\), compute:
    \begin{equation}
        \log \frac{\mathcal{P}(\boldsymbol{\delta}_{\boldsymbol{x}}^{(i)} \mid \boldsymbol{x}, \theta_s)}{\mathcal{P}(\boldsymbol{\delta}_{\boldsymbol{x}}^{(i)} \mid \boldsymbol{x}, \theta_t)}.
    \end{equation}
    \item \textit{Average Over Samples}:
    Compute the average of the computed log ratios:
\begin{equation}
\begin{aligned}
    \hat{D}_{\mathrm{KL}}(s,t)= \frac{1}{N} \sum_{i=1}^{N} \log \frac{\mathcal{P}(\boldsymbol{\delta}_{\boldsymbol{x}}^{(i)} \mid \boldsymbol{x}, \theta_s)}{\mathcal{P}(\boldsymbol{\delta}_{\boldsymbol{x}}^{(i)} \mid \boldsymbol{x}, \theta_t)}
\end{aligned},
\end{equation}
    where \(N\) is the number of samples. Note that The accuracy of this approximation depends on the number of samples. We randomly sampled 100,000 examples in our evaluation. More samples generally provide a better approximation to the true expectation.
\end{enumerate}

A lower metric value indicates a lower divergence between models. With this metric, misalignment can be quantified, negative transfer can be understood, and more effective attack strategies can be devised.

\textbf{Decision boundary visualization tool} Decision Surface\cite{decisionsurface} measures the network's loss variations concerning the input perturbations and contains the explicit decision boundary, naturally aligning with our analytical requirements of measuring the distance between the benign images and the decision boundary.

Let $F(\cdot)$ be the loss function. To visualize the loss $F(x)$ \textit{w.r.t} the high-dimensional inputs in a 2D hyper-plane, Decision Surface~\cite{decisionsurface} selects two normalized basis vectors for $x-y$ hyper-plane and interpolates the start point $o$ with them, then the loss is calculated as :
\begin{equation}
    V(i, j, \alpha, \beta)=F(o+i \cdot \alpha+j \cdot \beta).
\end{equation}
For our purpose of assessing the network's loss variations \textit{w.r.t.} perturbations,  $\alpha$ represents an attack-specific perturbation direction, $\beta$ stands for a random direction vector, and $o$ denotes the benign image. The variables $i$ and $j$ quantify the degree of deviation from point $o$ along the $\alpha$ and $\beta$ directions, respectively.

Regarding the form the loss function $F(\cdot)$, \cite{decisionsurface} replaces the cross-entropy loss with a novel decision surface:
\begin{equation}
    S(x)=Z(x)_t-\max \left\{Z(x)_i, i \neq t\right\}.
\end{equation}
Here, $Z(x)_t$ is the logit output \textit{w.r.t} the ground truth label $t$. For correct predictions, $S(x)>0$, while for incorrect ones, $S(x) < 0$. The contour where $S(x)=0$ indicates an explicit decision boundary, which significantly aids our goal of visualizing the proximity of benign images to this boundary.

As shown in the figure of decision surface, with the loss for clean images serving as the central point, the decision surface shows the loss variation \textit{w.r.t} the perturbation density (shown in x-axis) and the noise density along (shown in y-axis). Evidently, it requires a shorter distance to traverse the decision boundary in the adversarial direction compared to the random direction. This observation indicates that adversarial attacks can identify directions toward the decision boundary within the immediate vicinity of clean images.

BlackboxBench also utilizes dbViz\cite{somepalli2022can} to visualize the decision boundaries of various DNN models. The authors propose plotting decision boundaries on the convex hull between data samples. This proposed method is simple, controllable, and capable of capturing significant portions of the decision space that are close to the data manifold. The following are some key mathematical concepts and formulas:
\begin{enumerate}
    \item \emph{Plotting Decision Boundaries}: To plot decision boundaries, dbViz first defines a plane spanned by three randomly chosen image samples \( x_1, x_2, x_3 \) and then constructs two vectors \( \tilde{v}_1 = x_2 - x_1 \) and \( \tilde{v}_2 = x_3 - x_1 \), which defines a plane. The decision boundaries are then plotted on this plane.
    
    \item \emph{Sampling Points on the Plane}: To plot the decision boundaries on this plane, it is necessary to sample inputs to the network. The coordinates of the sampling points can be expressed as \( \beta \tilde{v}_1 + \phi (\tilde{v}_2 - \text{proj}_{\tilde{v}_1} \tilde{v}_2) \), where \( 0 \leq \beta, \phi \leq 1 \), and \(\text{proj}_{\tilde{v}_1} \tilde{v}_2\) is the projection of \( \tilde{v}_2 \) onto \( \tilde{v}_1 \).

    \item \emph{Proof of Off-Manifold Behavior}: dbViz also mentions a lemma (Lemma 2.1) about off-manifold behavior. This lemma suggests that any neural network that varies smoothly as a function of its input will have nearly constant outputs over most of the input space. The mathematical statement of this lemma is: Let \( f: [0,1]^n \to [0,1] \) be a neural network satisfying \( |f(x) - f(y)| \leq L \|x - y\| \), where \( L \) is a constant. Then, for a uniformly random image \( x \) in the unit hypercube, the probability that \( |f(x) - \tilde{f}| \leq t \) is at least \( 1 - Le^{-\frac{2nt^2}{L^2}} \), where \( \tilde{f} \) is the median value of \( f \) on the unit hypercube.
\end{enumerate} 
These mathematical methods and formulas above conduct a detailed analysis of decision boundaries under different neural network architectures and training conditions. Through this approach, they are able to gain a deeper understanding and explanation of model behavior, especially in terms of the formation and changes of decision boundaries. More details can be found in \cite{somepalli2022can}.

\newpage
\begin{table*}[ht]
  \centering
\caption{Hyper-parameters of decision-based attacks.}
\begingroup
    \scalebox{0.83}{\begin{tabular}{c|l|cc|cc|cc|cc}
\toprule[2pt]
\multirow{2}{*}{Decision-based attacks} & \multicolumn{1}{c|}{\multirow{2}{*}{Hyperparameters}} & \multicolumn{2}{c|}{L\_inf, untargeted attack} & \multicolumn{2}{c|}{L\_2, untargeted attack} & \multicolumn{2}{c|}{L\_inf, targeted attack} & \multicolumn{2}{c}{L\_2,   targeted attack} \\ \cline{3-10} 
 & \multicolumn{1}{c|}{} & ImageNet & CIFAR-10 & ImageNet & CIFAR-10 & ImageNet & CIFAR-10 & ImageNet & CIFAR-10 \\ \hline \hline
\multirow{3}{*}{General    hyperparameters} & maximum perturbation & 0.03 & 0.03 & 3 & 1 & 0.03 & 0.03 & 3 & 1 \\
 & maximum queries & 10000 & 10000 & 10000 & 10000 & 100000 & 100000 & 100000 & 100000 \\
 & batch   size & 1 & 1 & 1 & 1 & 1 & 1 & 1 & 1 \\ \hline
\multirow{6}{*}{Boundary Attack \cite{brendel2017decision}} & steps & 25000 & 25000 & 25000 & 25000 & 25000 & 25000 & 25000 & 25000 \\
 & spherical\_step & 0.01 & 0.01 & 0.01 & 0.01 & 0.01 & 0.01 & 0.01 & 0.01 \\
 & source\_step & 0.01 & 0.01 & 0.01 & 0.01 & 0.01 & 0.01 & 0.01 & 0.01 \\
 & source\_step\_convergance & 1e-7 & 1e-7 & 1e-7 & 1e-7 & 1e-7 & 1e-7 & 1e-7 & 1e-7 \\
 & step\_adaptation & 1.5 & 1.5 & 1.5 & 1.5 & 1.5 & 1.5 & 1.5 & 1.5 \\
 & update\_stats\_every\_k & 10 & 10 & 10 & 10 & 10 & 10 & 10 & 10 \\ \hline
Evolutionray Attack \cite{dong2019efficient} & sub & 2 & 2 & 2 & 2 & 2 & 2 & 2 & 2 \\ \hline
\multirow{7}{*}{GeoDA \cite{rahmati2020geoda}} & sub\_dim & 10 & 10 & 10 & 10 & 10 & 10 & 10 & 10 \\
 & tol & 0.001 & 0.001 & 0.001 & 0.001 & 0.001 & 0.001 & 0.001 & 0.001 \\
 & alpha & 0.002 & 0.002 & 0.002 & 0.002 & 0.002 & 0.002 & 0.002 & 0.002 \\
 & mu & 0.6 & 0.6 & 0.6 & 0.6 & 0.6 & 0.6 & 0.6 & 0.6 \\
 & search\_space & ''sub'' & ''sub'' & ''sub'' & ''sub'' & ''sub'' & ''sub'' & ''sub'' & ''sub'' \\
 & grad\_estimator\_batch\_size & 40 & 40 & 40 & 40 & 40 & 40 & 40 & 40 \\
 & sigma & 0 & 0 & 0 & 0 & 0 & 0 & 0 & 0 \\ \hline
SFA \cite{chen2020boosting} & resize\_factor & 1.0 & 1.0 & 1.0 & 1.0 & 1.0 & 1.0 & 1.0 & 1.0 \\ \hline
Rays \cite{chen2020rays} & No other hyperpatameters & / & / & / & / & / & / & / & / \\ \hline
\multirow{2}{*}{OPT\cite{cheng2018query}} & alpha & 0.2 & 0.2 & 0.2 & 0.2 & 0.2 & 0.2 & 0.2 & 0.2 \\
 & beta & 0.001 & 0.001 & 0.001 & 0.001 & 0.001 & 0.001 & 0.001 & 0.001 \\ \hline
\multirow{6}{*}{Sign-OPT\cite{cheng2019sign}} & alpha & 0.2 & 0.2 & 0.2 & 0.2 & 0.2 & 0.2 & 0.2 & 0.2 \\
 & beta & 0.001 & 0.001 & 0.001 & 0.001 & 0.001 & 0.001 & 0.001 & 0.001 \\
 & svm & FALSE & FALSE & FALSE & FALSE & FALSE & 0.001 & FALSE & FALSE \\
 & momentum & 0 & 0 & 0 & 0 & 0 & 0 & 0 & 0 \\
 & k & 200 & 200 & 200 & 200 & 200 & 200 & 200 & 200 \\
 & sigma & 0 & 0 & 0 & 0 & 0 & 0 & 0 & 0 \\ \hline
\multirow{6}{*}{HSJA\cite{chen2020hopskipjumpattack}} & gamma & 1 & 1 & 1 & 1 & 1 & 1 & 1 & 1 \\
 & stepsize\_search & \makecell[c]{''geometric\_ \\ progression''} & \makecell[c]{''geometric\_ \\ progression''} & \makecell[c]{''geometric\_ \\ progression''} & \makecell[c]{''geometric\_ \\ progression''} & \makecell[c]{''geometric\_ \\ progression''} & \makecell[c]{''geometric\_ \\ progression''} & \makecell[c]{''geometric\_ \\ progression''} & \makecell[c]{''geometric\_ \\ progression''} \\
 & max\_num\_evals & 10000 & 10000 & 10000 & 10000 & 10000 & 10000 & 10000 & 10000 \\
 & init\_num\_evals & 100 & 100 & 100 & 100 & 100 & 100 & 100 & 100 \\
 & EOT & 1 & 1 & 1 & 1 & 1 & 1 & 1 & 1 \\
 & sigma & 0 & 0 & 0 & 0 & 0 & 0 & 0 & 0 \\ \hline
QEBA \cite{li2020qeba} & pgen & DCT & DCT & "DCT" & "DCT" & "DCT" & "DCT" & "DCT" & "DCT" \\ \hline
Nonlinear-BA \cite{li2021nonlinear} & pgen & AE9408 & AE9408 & AE9408 & AE9408 & AE9408 & AE9408 & AE9408 & AE9408 \\ \hline
PSBA \cite{zhang2021progressive} & pgen & AE9408 & AE9408 & AE9408 & AE9408 & AE9408 & AE9408 & AE9408 & AE9408 \\ \hline
\multirow{8}{*}{Triangle Attack \cite{wang2022triangle}} & ratio\_mask & 0.1 & 0.1 & 0.1 & 0.1 & 0.1 & 0.1 & 0.1 & 0.1 \\
 & dim\_num & 1 & 1 & 1 & 1 & 1 & 1 & 1 & 1 \\
 & max\_iter\_num\_in\_2d & 2 & 2 & 2 & 2 & 2 & 2 & 2 & 2 \\ \cline{2-10} 
 & init\_theta & 2 & 2 & 2 & 2 & 2 & 2 & 2 & 2 \\ \cline{2-10} 
 & init\_alpha & np.pi/2 & 2 & np.pi/2 & np.pi/2 & np.pi/2 & np.pi/2 & np.pi/2 & np.pi/2 \\
 & plus\_learning\_rate & 0.1 & 0.1 & 0.1 & 0.1 & 0.1 & 0.1 & 0.1 & 0.1 \\
 & minus\_learning\_rate & 0.005 & 0.005 & 0.005 & 0.005 & 0.005 & 0.005 & 0.005 & 0.005 \\
 & half\_range & 0.1 & 0.1 & 0.1 & 0.1 & 0.1 & 0.1 & 0.1 & 0.1 \\     \bottomrule[2pt]
\end{tabular}}
    \label{tab:hyperparameter_decision}%
\endgroup
\end{table*}

\begin{table*}[ht]
  \centering
\caption{Hyper-parameters of score-based attacks.}
\renewcommand\arraystretch{1.2}
\resizebox{0.97\textwidth}{!}{%
\begin{tabular}{c|l|cc|cc|cc|cc}
\toprule[2pt]
\multirow{2}{*}{Score-based attacks} & \multicolumn{1}{c|}{\multirow{2}{*}{Hyperparameters}} & \multicolumn{2}{c|}{L\_inf, untargeted attack} & \multicolumn{2}{c|}{L\_2, untargeted attack} & \multicolumn{2}{c|}{L\_inf, targeted attack} & \multicolumn{2}{c}{L\_2,   targeted attack} \\ \cline{3-10} 
 & \multicolumn{1}{c|}{} & ImageNet & CIFAR-10 & ImageNet & CIFAR-10 & ImageNet & CIFAR-10 & ImageNet & CIFAR-10 \\ \hline \hline
\multirow{3}{*}{General    hyperparameters} & maximum perturbation & 0.03 & 0.03 & 3 & 1 & 0.03 & 0.03 & 3 & 1 \\
 & maximum queries & 10000 & 10000 & 10000 & 10000 & 100000 & 100000 & 100000 & 100000 \\
 & batch   size & 1 & 1 & 1 & 1 & 1 & 1 & 1 & 1 \\ \hline
\multirow{3}{*}{SimBA \cite{guo2019simple}} & freq\_dims & 28 & 28 & 28 & 28 & 28 & 28 & 28 & 28 \\
 & order strided & TRUE & TRUE & TRUE & TRUE & TRUE & TRUE & TRUE & TRUE \\
 & stride & 7 & 7 & 7 & 7 & 7 & 7 & 7 & 7 \\ \hline
\multirow{4}{*}{Parsimonious Attack \cite{moon2019parsimonious}} & EOT & 1 & 1 & 1 & 1 & 1 & 1 & 1 & 1 \\
 & block\_size & 4 & 4 & 4 & 4 & 4 & 4 & 4 & 4 \\
 & block\_batch\_size & 64 & 64 & 64 & 64 & 64 & 64 & 64 & 64 \\
 & sigma & 0 & 0 & 0 & 0 & 0 & 0 & 0 & 0 \\ \hline
Square Attack \cite{andriushchenko2020square} & p\_init & 0.05 & 0.05 & 0.05 & 0.05 & 0.05 & 0.05 & 0.05 & 0.05 \\ \hline
\multirow{6}{*}{PPBA \cite{li2020projection}} & low-dim & 1500 & 1500 & 1500 & 1500 & 1500 & 1500 & 1500 & 1500 \\
 & mom & 1 & 1 & 1 & 1 & 1 & 1 & 1 & 1 \\
 & order & "strided" & "strided" & "strided" & "strided" & "strided" & "strided" & "strided" & "strided" \\
 & r & 2352 & 2352 & 2352 & 2352 & 2352 & 2352 & 2352 & 2352 \\
 & n\_samples & 1 & 1 & 1 & 1 & 1 & 1 & 1 & 1 \\
 & rho & 0.01 & 0.01 & 0.01 & 0.01 & 0.01 & 0.01 & 0.01 & 0.01 \\ \hline
\multirow{3}{*}{NES \cite{ilyas2018black}} & fd\_eta & 2.55 & 2.55 & 2.55 & 2.55 & 2.55 & 2.55 & 2.55 & 2.55 \\
 & lr & 2.55 & 2.55 & 2.55 & 2.55 & 2.55 & 2.55 & 2.55 & 2.55 \\
 & q & 15 & 15 & 15 & 15 & 15 & 15 & 15 & 15 \\ \hline
\multirow{5}{*}{Bandits \cite{ilyas2018prior}} & lr & 2.55 & 2.55 & 2.55 & 2.55 & 2.55 & 2.55 & 2.55 & 2.55 \\
 & fd\_eta & 2.55 & 2.55 & 2.55 & 2.55 & 2.55 & 2.55 & 2.55 & 2.55 \\
 & prior\_lr & 0.1 & 0.1 & 0.1 & 0.1 & 0.1 & 0.1 & 0.1 & 0.1 \\
 & prior\_size & 20 & 20 & 20 & 20 & 20 & 20 & 20 & 20 \\
 & prior\_exploraton & 0.1 & 0.1 & 0.1 & 0.1 & 0.1 & 0.1 & 0.1 & 0.1 \\ \hline
\multirow{8}{*}{AdvFlow \cite{mohaghegh2020advflow}} & high\_res\_blocks & 4 & 4 & 4 & 4 & 4 & 4 & 4 & 4 \\
 & low\_res\_blocks & 6 & 6 & 6 & 6 & 6 & 6 & 6 & 6 \\
 & channel\_hidden & 128 & 128 & 128 & 128 & 128 & 128 & 128 & 128 \\
 & batch\_norm & FALSE & FALSE & FALSE & FALSE & FALSE & FALSE & FALSE & FALSE \\
 & n\_blocks & 6 & 6 & 6 & 6 & 6 & 6 & 6 & 6 \\
 & internal\_width & 128 & 128 & 128 & 128 & 128 & 128 & 128 & 128 \\
 & fc\_dropout & 0 & 0 & 0 & 0 & 0 & 0 & 0 & 0 \\
 & clamping & 1.5 & 1.5 & 1.5 & 1.5 & 1.5 & 1.5 & 1.5 & 1.5 \\ \hline
\multirow{4}{*}{NP-Attack \cite{bai2020improving}} & N & 100 & 100 & 100 & 100 & 100 & 100 & 100 & 100 \\
 & I & 600 & 600 & 600 & 600 & 600 & 600 & 600 & 600 \\
 & E & 0.05 & 0.05 & 0.05 & 0.05 & 0.05 & 0.05 & 0.05 & 0.05 \\
 & LR & 0.05 & 0.05 & 0.05 & 0.05 & 0.05 & 0.05 & 0.05 & 0.05 \\ \hline
\multirow{3}{*}{ZO-signSGD\cite{liu2019signsgd}} & fd\_eta & 2.55 & 2.55 & 2.55 & 2.55 & 2.55 & 2.55 & 2.55 & 2.55 \\
 & lr & 2.55 & 2.55 & 2.55 & 2.55 & 2.55 & 2.55 & 2.55 & 2.55 \\
 & q & 30 & 30 & 30 & 30 & 30 & 30 & 30 & 30 \\ \hline
SignHunter \cite{al2020sign} & fd\_eta & 12.75 & 12.75 & 12.75 & 12.75 & 12.75 & 12.75 & 12.75 & 30 \\ \hline
\multirow{4}{*}{BABIES\cite{tran2022exploiting}} & order & "strided" & "strided" & "strided" & "strided" & "strided" & "strided" & "strided" & "strided" \\
 & interp & "store\_false" & "store\_false" & "store\_false" & "store\_false" & "store\_false" & "store\_false" & "store\_false" & "store\_false" \\
 & rho & 5 & 2.4 & 5 & 2.4 & 12 & 4 & 12 & 4 \\
 & eps & 2 & 2 & 2 & 2 & 3 & 2 & 3 & 2 \\ \hline
\multirow{19}{*}{CG-Attack \cite{feng2022boosting}} & x\_hidden\_channels & 64 & 64 & 64 & 64 & 64 & 64 & 64 & 64 \\
 & x\_hiddden\_sie & 128 & 128 & 128 & 128 & 128 & 128 & 128 & 128 \\
 & y\_hidden\_channels & 256 & 256 & 256 & 256 & 256 & 256 & 256 & 256 \\
 & flow\_depth & 8 & 8 & 8 & 8 & 8 & 8 & 8 & 8 \\
 & num\_levels & 3 & 3 & 3 & 3 & 3 & 3 & 3 & 3 \\
 & num\_epochs & 50 & 50 & 50 & 50 & 50 & 50 & 50 & 50 \\
 & checkpoints\_gap & 10000 & 10000 & 10000 & 10000 & 10000 & 10000 & 10000 & 10000 \\
 & nll\_gap & 1 & 1 & 1 & 1 & 1 & 1 & 1 & 1 \\
 & inference\_gap & 10000 & 10000 & 10000 & 10000 & 10000 & 10000 & 10000 & 10000 \\
 & lr & 0.0002 & 0.0002 & 0.0002 & 0.0002 & 0.0002 & 0.0002 & 0.0002 & 0.0002 \\
 & max\_grad\_clip & 0 & 0 & 0 & 0 & 0 & 0 & 0 & 0 \\
 & max\_grad\_norm & 10 & 10 & 10 & 10 & 10 & 10 & 10 & 10 \\
 & save\_gap & 10000 & 10000 & 10000 & 10000 & 10000 & 10000 & 10000 & 10000 \\
 & regularizer & 0 & 0 & 0 & 0 & 0 & 0 & 0 & 0 \\
 & learn\_top & FALSE & FALSE & FALSE & FALSE & FALSE & FALSE & FALSE & FALSE \\
 & tanh & FALSE & FALSE & FALSE & FALSE & FALSE & FALSE & FALSE & FALSE \\
 & only & TRUE & TRUE & TRUE & TRUE & TRUE & TRUE & TRUE & TRUE \\
 & clamp & TRUE & TRUE & TRUE & TRUE & TRUE & TRUE & TRUE & TRUE \\
 & margin & 20 & 20 & 20 & 20 & 20 & 20 & 20 & 20 \\ \hline
\multirow{7}{*}{MCG\cite{yin2023generalizable}} & generator\_path & / & / & / & / & / & / & / & / \\
 & surrogate\_model & / & / & / & / & / & / & / & / \\
 & down\_sample\_x & 1 & 1 & 1 & 1 & 1 & 1 & 1 & 1 \\
 & down\_sample\_y & 1 & 1 & 1 & 1 & 1 & 1 & 1 & 1 \\
 & finetune\_grow & TRUE & TRUE & TRUE & TRUE & TRUE & TRUE & TRUE & TRUE \\
 & finetune\_reload & TRUE & TRUE & TRUE & TRUE & TRUE & TRUE & TRUE & TRUE \\
 & finetune\_perturbation & TRUE & TRUE & TRUE & TRUE & TRUE & TRUE & TRUE & TRUE \\  \bottomrule[2pt]
\end{tabular}
    \label{tab:hyperparameter_score}%
}
\end{table*}

\begin{table*}[t]
  \centering
  \caption{Hyper-parameters of transfer-based attacks.}
\begingroup
    \scalebox{0.62}{\begin{tabular}{c|p{24.835em}|cccc|cccc}
\toprule[2pt]
    \multirow{3}[6]{*}{\textbf{Transfer-based attacks}} & \multicolumn{1}{c|}{\multirow{3}[6]{*}{\textbf{Hyper-parameters}}} & \multicolumn{4}{c|}{\textbf{ImageNet}} & \multicolumn{4}{c}{\textbf{CIFAR-10}} \bigstrut\\
\cline{3-10}          & \multicolumn{1}{c|}{} & \multicolumn{2}{c|}{\textbf{untargeted}} & \multicolumn{2}{c|}{\textbf{targeted}} & \multicolumn{2}{c|}{\textbf{untargeted}} & \multicolumn{2}{c}{\textbf{targeted}} \bigstrut\\
\cline{3-10}          & \multicolumn{1}{c|}{} & \multicolumn{1}{c}{\textbf{$\ell_\infty$}} & \multicolumn{1}{c|}{\textbf{$\ell_2$}} & \multicolumn{1}{c}{\textbf{$\ell_\infty$}} & \multicolumn{1}{c|}{\textbf{$\ell_2$}} & \multicolumn{1}{c}{\textbf{$\ell_\infty$}} & \multicolumn{1}{c|}{\textbf{$\ell_2$}} & \multicolumn{1}{c}{\textbf{$\ell_\infty$}} & \multicolumn{1}{c}{\textbf{$\ell_2$}} \bigstrut\\
    \hline
    \hline
    \multicolumn{1}{c|}{\multirow{2}[2]{*}{\textbf{General hyper-parameters}}} & maximum perturbation & \multicolumn{1}{c}{0.03}  & \multicolumn{1}{c|}{3} & \multicolumn{1}{c}{0.06}  & \multicolumn{1}{c|}{6}     & \multicolumn{1}{c}{0.03}  & \multicolumn{1}{c|}{1} & \multicolumn{1}{c}{0.03}  & \multicolumn{1}{c}{1} \bigstrut[t]\\
          & step size & 0.003 & \multicolumn{1}{c|}{0.3} & 0.006 & 0.6   & 0.003 & \multicolumn{1}{c|}{0.1} & 0.003 & 0.1 \bigstrut[b]\\
    \hline
    \textbf{I-FGSM\cite{ifgsm}} & iterations & \multicolumn{2}{c}{100} & \multicolumn{2}{c|}{300} & \multicolumn{2}{c}{100} & \multicolumn{2}{c}{300} \bigstrut\\
    \hline
    \multirow{2}[2]{*}{\textbf{PGD\cite{pgd}}} & iterations & \multicolumn{2}{c}{100} & \multicolumn{2}{c|}{300} & \multicolumn{2}{c}{100} & \multicolumn{2}{c}{300} \bigstrut[t]\\
          & bound of random start of perturbation & \multicolumn{4}{c|}{0.03}     & \multicolumn{4}{c}{1e-7} \bigstrut[b]\\
    \hline
    \multirow{2}[2]{*}{\textbf{MI-FGSM\cite{mifgsm}}} & iterations & \multicolumn{2}{c}{10} & \multicolumn{2}{c|}{300} & \multicolumn{2}{c}{100} & \multicolumn{2}{c}{300} \bigstrut[t]\\
          & decay factor & \multicolumn{4}{c|}{1}        & \multicolumn{4}{c}{1} \bigstrut[b]\\
    \hline
    \multirow{2}[2]{*}{\textbf{NI-FGSM\cite{nisifgsm}}} & iterations & \multicolumn{2}{c}{10} & \multicolumn{2}{c|}{300} & \multicolumn{2}{c}{100} & \multicolumn{2}{c}{300} \bigstrut[t]\\
          & decay factor / step size of looking ahead & \multicolumn{4}{c|}{1 / 1/255} & \multicolumn{4}{c}{1 / 1e-7} \bigstrut[b]\\
    \hline
    \multirow{2}[2]{*}{\textbf{PI-FGSM\cite{pifgsm}}} & iterations & \multicolumn{2}{c}{100} & \multicolumn{2}{c|}{300} & \multicolumn{2}{c}{100} & \multicolumn{2}{c}{300} \bigstrut[t]\\
          & decay factor / step size of looking ahead & \multicolumn{4}{c|}{1 / 1/255} & \multicolumn{4}{c}{1 / 0.001} \bigstrut[b]\\
    \hline
    \textbf{VT\cite{vt}} & iterations & \multicolumn{2}{c}{100} & \multicolumn{2}{c|}{300} & \multicolumn{2}{c}{100} & \multicolumn{2}{c}{300} \bigstrut[t]\\
          & number of samples to calculate variance & \multicolumn{4}{c|}{5}        & \multicolumn{4}{c}{5} \bigstrut[b]\\
    \hline
    \multirow{2}[2]{*}{\textbf{RAP\cite{rap}}} & iterations & \multicolumn{4}{c|}{400}      & \multicolumn{4}{c}{400} \bigstrut[t]\\
          & late start / iteration number of inner maximization & \multicolumn{4}{c|}{100 /10}  & \multicolumn{4}{c}{100 / 5} \bigstrut[b]\\
    \hline
    \multirow{2}[2]{*}{\textbf{LinBP\cite{linbp}}} & iterations & \multicolumn{2}{c}{100} & \multicolumn{2}{c|}{300} & \multicolumn{2}{c}{100} & \multicolumn{2}{c}{300} \bigstrut[t]\\
          & position to perform linear backpropagation & \multicolumn{4}{p{22em}|}{ResNet-50: the 2nd bottleneck block at the 3rd stage;\newline{}VGG-19$^{\dagger}$: the conv layer at the 12nd VGG block;\newline{}Inception-V3: \newline{}DenseNet-121: \newline{}ViT-B/16: } & \multicolumn{4}{p{22em}}{ResNet-50: the 1st bottleneck block at the 4th stage;\newline{}VGG-19$^{\dagger}$: the conv layer at the 8th VGG block;\newline{}Inception-V3: \newline{}DenseNet-BC: } \bigstrut[b]\\
    \hline
    \multirow{2}[2]{*}{\textbf{SGM\cite{sgm}}} & iterations & \multicolumn{2}{c}{100} & \multicolumn{2}{c|}{300} & \multicolumn{2}{c}{100} & \multicolumn{2}{c}{300} \bigstrut[t]\\
          & decay parameter & \multicolumn{4}{c|}{ResNet-50: 0.2; DenseNet-121: 0.5; ViT-B/16: 0.6} & \multicolumn{4}{c}{ResNet-50: 0.5; DenseNet-BC: 0.6} \bigstrut[b]\\
    \hline
    \multirow{2}[2]{*}{\textbf{DI2-FGSM\cite{difgsm}}} & iterations & \multicolumn{2}{c}{100} & \multicolumn{2}{c|}{300} & \multicolumn{2}{c}{100} & \multicolumn{2}{c}{300} \bigstrut[t]\\
          & resized dimension / transformation probability & \multicolumn{4}{c|}{299 / 0.5} & \multicolumn{4}{c}{40 / 0.5} \bigstrut[b]\\
    \hline
    \multirow{2}[2]{*}{\textbf{SI-FGSM\cite{nisifgsm}}} & iterations & \multicolumn{2}{c}{100} & \multicolumn{2}{c|}{300} & \multicolumn{2}{c}{100} & \multicolumn{2}{c}{300} \bigstrut[t]\\
          & number of the scale copies / scale factor & \multicolumn{4}{c|}{5 / 2}    & \multicolumn{4}{c}{3 / 1.5} \bigstrut[b]\\
    \hline
    \multirow{3}[2]{*}{\textbf{Admix\cite{admix}}} & iterations & \multicolumn{2}{c}{100} & \multicolumn{2}{c|}{300} & \multicolumn{2}{c}{100} & \multicolumn{2}{c}{300} \bigstrut[t]\\
          & strength of sampled image / number of sampled images & \multicolumn{4}{c|}{0.2 / 3}  & \multicolumn{4}{c}{0.2 / 3} \\
          & number of admixed copies / scale factor & \multicolumn{4}{c|}{5 / 2}    & \multicolumn{4}{c}{3 / 1.5} \bigstrut[b]\\
    \hline
    \multirow{2}[2]{*}{\textbf{TI-FGSM\cite{tifgsm}}} & iterations & \multicolumn{2}{c}{100} & \multicolumn{2}{c|}{300} & \multicolumn{2}{c}{100} & \multicolumn{2}{c}{300} \bigstrut[t]\\
          & kernel size & \multicolumn{4}{c|}{5}        & \multicolumn{4}{c}{3} \bigstrut[b]\\
    \hline
    \multirow{3}[2]{*}{\textbf{FIA\cite{fia}}} & iterations & \multicolumn{2}{c}{100} & \multicolumn{2}{c|}{\multirow{3}[2]{*}{FIA could't perform targeted attack.}} & \multicolumn{2}{c}{100} & \multicolumn{2}{c}{\multirow{3}[2]{*}{FIA could't perform targeted attack.}} \bigstrut[t]\\
          & intermediate layer to extract feature maps & \multicolumn{2}{p{11em}}{ResNet-50: the 2nd stage;\newline{}VGG-19$^{\dagger}$: the conv layer at the 10th VGG block;\newline{}Inception-V3: the 2nd Inception module in the 2nd Inception block;\newline{}DenseNet-121: the 3rd dense block;\newline{}ViT-B/16: the 3rd transformer encoder layer;} & \multicolumn{2}{c|}{} & \multicolumn{2}{p{11em}}{ResNet-50: the 2nd stage;\newline{}VGG-19$^{\dagger}$: the conv layer at the 7th VGG block;\newline{}Inception-V3: the 3rd Inception module in the 2nd Inception block;\newline{}DenseNet-BC: the 2nd transition layer;} & \multicolumn{2}{c}{} \\
          & ensemble number in aggregate gradient / drop probability & \multicolumn{2}{c}{30 / 0.3} & \multicolumn{2}{c|}{} & \multicolumn{2}{c}{30 / 0.3} & \multicolumn{2}{c}{} \bigstrut[b]\\
    \hline
    \multirow{3}[2]{*}{\textbf{NAA\cite{naa}}} & iterations & \multicolumn{2}{c}{100} & \multicolumn{2}{c|}{\multirow{3}[2]{*}{NAA could't perform targeted attack.}} & \multicolumn{2}{c}{100} & \multicolumn{2}{c}{\multirow{3}[2]{*}{NAA could't perform targeted attack.}} \bigstrut[t]\\
          & target layer & \multicolumn{2}{p{11em}}{ResNet-50: the 2nd stage;\newline{}VGG-19$^{\dagger}$: the conv layer at the 10th VGG block;\newline{}Inception-V3: the 2nd Inception module in the 2nd Inception block;\newline{}DenseNet-121: the 3rd dense block;\newline{}ViT-B/16: the 3rd transformer encoder layer;} & \multicolumn{2}{c|}{} & \multicolumn{2}{p{11em}}{ResNet-50: the 2nd stage;\newline{}VGG-19$^{\dagger}$: the conv layer at the 7th VGG block;\newline{}Inception-V3: the 3rd Inception module in the 2nd Inception block;\newline{}DenseNet-BC: the 2nd transition layer;} & \multicolumn{2}{c}{} \\
          & integrated step & \multicolumn{2}{c}{30} & \multicolumn{2}{c|}{} & \multicolumn{2}{c}{30} & \multicolumn{2}{c}{} \bigstrut[b]\\
    \hline
    \multirow{2}[2]{*}{\textbf{ILA\cite{ila}}} & iterations of attack / baseline attack & \multicolumn{2}{c}{100 /10} & \multicolumn{2}{c|}{300 / 30} & \multicolumn{2}{c}{100 / 10} & \multicolumn{2}{c}{300 / 30} \bigstrut[t]\\
          & intermediate layer & \multicolumn{4}{p{22em}|}{ResNet-50: the 2nd stage;\newline{}VGG-19$^{\dagger}$: the conv layer at the 10th VGG block;\newline{}Inception-V3: the 2nd Inception module in the 2nd Inception block;\newline{}DenseNet-121: the 3rd dense block;\newline{}ViT-B/16: the 3rd transformer encoder layer;} & \multicolumn{4}{p{22em}}{ResNet-50: the 2nd stage;\newline{}VGG-19$^{\dagger}$: the conv layer at the 7th VGG block;\newline{}Inception-V3: the 3rd Inception module in the 2nd Inception block;\newline{}DenseNet-BC: the 2nd transition layer;} \bigstrut[b]\\
    \hline
    \multirow{3}[2]{*}{\textbf{RD\cite{lgv}}} & iterations & \multicolumn{2}{c}{100} & \multicolumn{2}{c|}{300} & \multicolumn{2}{c}{100} & \multicolumn{2}{c}{300} \bigstrut[t]\\
          & scale of noise added on DNN weights & \multicolumn{4}{p{22em}|}{ResNet-50: 0.005; VGG-19$^{\dagger}$: 0.01; Inception-V3: 0.01; DenseNet-121: 0.01; ViT-B/16: 0.005;} & \multicolumn{4}{p{22em}}{ResNet-50: 0.005; VGG-19$^{\dagger}$: 0.01; Inception-V3: 0.005; DenseNet-BC: 0.01;} \\
          & number of ensemble models & \multicolumn{4}{c|}{10}       & \multicolumn{4}{c}{10} \bigstrut[b]\\
    \hline
    \multirow{3}[2]{*}{\textbf{GhostNet\cite{ghostnet}}} & iterations & \multicolumn{2}{c}{100} & \multicolumn{2}{c|}{300} & \multicolumn{2}{c}{100} & \multicolumn{2}{c}{300} \bigstrut[t]\\
          & dropout probability & \multicolumn{4}{c|}{VGG-19$^{\dagger}$: 0.05;\newline{}Inception-V3: 0.01;} & \multicolumn{4}{c}{VGG-19$^{\dagger}$: 0.1;\newline{}Inception-V3: 0.2;} \\
          & parameter of uniform distribution that modulating scalars are drawn from & \multicolumn{4}{c|}{ResNet-50: 0.22;\newline{}DenseNet-121: 0.6;\newline{}ViT-B/16: 0.3;} & \multicolumn{4}{c}{ResNet-50: 1;\newline{}DenseNet-BC: 0.2} \bigstrut[b]\\
    \hline
    \multirow{3}[2]{*}{\textbf{DRA\cite{dra}}} & iterations & \multicolumn{2}{c}{100} & \multicolumn{2}{c|}{300} & \multicolumn{4}{c}{\multirow{3}[2]{*}{No officially released code.}} \bigstrut[t]\\
          & batch size / epochs / learning rate / regularization constant & \multicolumn{4}{c|}{32 / 20 / 0.001 / 6} & \multicolumn{4}{c}{} \\
          & learning rate schedule & \multicolumn{4}{c|}{decay by10 at epochs 10} & \multicolumn{4}{c}{} \bigstrut[b]\\
    \hline
    \multirow{3}[2]{*}{\textbf{IAA\cite{iaa}}} & iterations & \multicolumn{2}{c}{100} & \multicolumn{2}{c|}{300} & \multicolumn{4}{c}{\multirow{3}[2]{*}{No officially released code.}} \bigstrut[t]\\
          & shape-related hyper-parameter of Softplus & \multicolumn{4}{p{22em}|}{ResNet-50: 25\newline{}others: no officially released code} & \multicolumn{4}{c}{} \\
          & decay factor on residual module & \multicolumn{4}{p{22em}|}{ResNet-50: [1.0, 0.85, 0.65, 0.15]\newline{}others: no officially released code} & \multicolumn{4}{c}{} \bigstrut[b]\\
    \hline
    \multirow{3}[2]{*}{\textbf{LGV\cite{lgv}}} & iterations & \multicolumn{2}{c}{100} & \multicolumn{2}{c|}{300} & \multicolumn{2}{c}{100} & \multicolumn{2}{c}{300} \bigstrut[t]\\
          & batch size / epochs / learning rate schedule & \multicolumn{4}{c|}{512 / 10 / constant} & \multicolumn{4}{c}{256 / 10 / constant} \\
          & learning rate/weight decay/momentum & \multicolumn{4}{p{22em}|}{ResNet-50: 0.05/1e-4/0.9; VGG-19$^{\dagger}$: 0.1/7e-4/0.9;\newline{}Inception-V3: 0.1/5e-4/0.9; DenseNet-121: 0.05/1e-4/0.9; ViT-B/16: 0.05/1e-4/0.9;} & \multicolumn{4}{p{22em}}{ResNet-50: 0.05/5e-4/0.9; {}VGG-19$^{\dagger}$: 0.05/5e-4/0.9;\newline{}Inception-V3: 0.1/5e-4/0.9; {}DenseNet-BC: 0.5/1e-4/0.9;} \bigstrut[b]\\
    \hline
\multirow{4}[2]{*}{\textbf{SWA\cite{bayesian}}} & iterations & \multicolumn{2}{c}{100} & \multicolumn{2}{c|}{300} & \multicolumn{2}{c}{100} & \multicolumn{2}{c}{300} \bigstrut[t]\\
          & batch size / epochs / learning rate schedule / momentum / $\sigma$ & \multicolumn{4}{c|}{512 / 10 / constant / 0.9 / 0.002} & \multicolumn{4}{c}{512 / 10 / constant / 0.9 / 0.002} \\
          & learning rate / weight decay / $\lambda_{\varepsilon, \sigma}$ / $\gamma$ & \multicolumn{4}{p{22em}|}{ResNet-50: 0.05 / 5e-4 / 1 / {$0.1 /\left\|\boldsymbol{\Delta} \mathbf{w}^*\right\|_2$};\newline{}VGG-19$^{\dagger}$: 0.005 / 1e-4 / 1 / {$0.15 /\left\|\boldsymbol{\Delta} \mathbf{w}^*\right\|_2$}/;\newline{}Inception-V3: 0.002 / 1e-4 / 1 / {$0.1 /\left\|\boldsymbol{\Delta} \mathbf{w}^*\right\|_2$};\newline{}DenseNet-121: 0.05 / 1e-4 / 1 / {$0.1 /\left\|\boldsymbol{\Delta} \mathbf{w}^*\right\|_2$};\newline{}ViT-B/16: 0.001 / 1e-4 / 1 / {$0.05 /\left\|\boldsymbol{\Delta} \mathbf{w}^*\right\|_2$};} & \multicolumn{4}{p{22em}}{ResNet-50: 0.05 / 1e-4 / 0.2 / {$0.1 /\left\|\boldsymbol{\Delta} \mathbf{w}^*\right\|_2$};\newline{}VGG-19$^{\dagger}$: 0.1 / 1e-4 / 1 / {$0.02 /\left\|\boldsymbol{\Delta} \mathbf{w}^*\right\|_2$};\newline{}Inception-V3: 0.05 / 1e-4 / 0.2 / {$0.1 /\left\|\boldsymbol{\Delta} \mathbf{w}^*\right\|_2$};\newline{}DenseNet-BC: 0.1 / 1e-4 / 1 / {$0.1 /\left\|\boldsymbol{\Delta} \mathbf{w}^*\right\|_2$};} \bigstrut[b]\\
    \hline
    \multirow{6}[2]{*}{\textbf{Bayesian Attack\cite{bayesian}}} & iterations & \multicolumn{2}{c}{100} & \multicolumn{2}{c|}{300} & \multicolumn{2}{c}{100} & \multicolumn{2}{c}{300} \bigstrut[t]\\
          & batch size / epochs / learning rate schedule / momentum / $\sigma$ & \multicolumn{4}{c|}{512 / 10 / constant / 0.9 / 0.002} & \multicolumn{4}{c}{512 / 10 / constant / 0.9 / 0.002} \\
          & learning rate / weight decay / $\lambda_{\varepsilon, \sigma}$ / $\gamma$ & \multicolumn{4}{p{22em}|}{ResNet-50: 0.05 / 5e-4 / 1 / {$0.1 /\left\|\boldsymbol{\Delta} \mathbf{w}^*\right\|_2$};\newline{}VGG-19$^{\dagger}$: 0.005 / 1e-4 / 1 / {$0.15 /\left\|\boldsymbol{\Delta} \mathbf{w}^*\right\|_2$}/;\newline{}Inception-V3: 0.002 / 1e-4 / 1 / {$0.1 /\left\|\boldsymbol{\Delta} \mathbf{w}^*\right\|_2$};\newline{}DenseNet-121: 0.05 / 1e-4 / 1 / {$0.1 /\left\|\boldsymbol{\Delta} \mathbf{w}^*\right\|_2$};\newline{}ViT-B/16: 0.001 / 1e-4 / 1 / {$0.05 /\left\|\boldsymbol{\Delta} \mathbf{w}^*\right\|_2$};} & \multicolumn{4}{p{22em}}{ResNet-50: 0.05 / 1e-4 / 0.2 / {$0.1 /\left\|\boldsymbol{\Delta} \mathbf{w}^*\right\|_2$};\newline{}VGG-19$^{\dagger}$: 0.1 / 1e-4 / 1 / {$0.02 /\left\|\boldsymbol{\Delta} \mathbf{w}^*\right\|_2$};\newline{}Inception-V3: 0.05 / 1e-4 / 0.2 / {$0.1 /\left\|\boldsymbol{\Delta} \mathbf{w}^*\right\|_2$};\newline{}DenseNet-BC: 0.1 / 1e-4 / 1 / {$0.1 /\left\|\boldsymbol{\Delta} \mathbf{w}^*\right\|_2$};} \\
          & number of sampled models & \multicolumn{4}{c|}{50}       & \multicolumn{4}{c}{50} \\
          & rescaling factor of the covariance matrix & \multicolumn{4}{c|}{1.5}      & \multicolumn{4}{c}{1.5} \bigstrut[b]\\
    \bottomrule[2pt]
    \end{tabular}}
    \label{tab:hyperparameter}%
\endgroup
\end{table*}%



 




\vfill


\bibliographystyle{IEEEtran}
\bibliography{reference}

\vfill

\end{document}